\documentclass[12pt,letterpaper]{article}

\usepackage{amsfonts,natbib,hyperref,url}
\usepackage{graphicx}
\usepackage{latexsym,color,lscape,subcaption,placeins}

\usepackage{ee}

\usepackage[latin1]{inputenc}
\usepackage{amssymb,amsmath,amsfonts}
\usepackage{multirow}
\usepackage{extarrows}
\usepackage{delarray}

\usepackage{setspace}

\numberwithin{equation}{section}
\newcounter{result}[section]

\newtheorem{theo}{\textbf{Theorem}}

\newtheorem{ass}{\textbf{Assumption}}

\newtheorem{lemma}[result]{\textbf{Lemma}}

\newcommand{\inp}{\stackrel{p}{\rightarrow}}
\newcommand{\ind}{\stackrel{d}{\rightarrow}}
\newcommand{\Op}{\mathcal{O}_P}
\newcommand{\op}{o_P}
\newcommand{\ur}{u.r}
\renewcommand{\vec}{\mbox{vect }}
\newcommand{\deff}{\xlongequal{\text{def}}}
\newcommand{\avar}{\mbox{AVar}}
\newcommand{\acov}{\mbox{ACov}}

\renewcommand{\var}{\mbox{Var}}

\renewcommand{\H}{\mathcal H}

\renewcommand{\S}{\mathcal S}

\title{Efficient two-sample instrumental variable estimators with change points and near-weak identification}

\author{Bertille Antoine \\ Simon Fraser University \\ \small{Bertille\_Antoine@sfu.ca}
\and
Otilia Boldea \\ Tilburg University \\ \small{O.Boldea@uvt.nl}
\and
Niccol\`{o} Zaccaria \\ Tilburg University \\ \small{N.Zaccaria@tilburguniversity.edu}
}

\begin{document}

\maketitle

\begin{abstract}
We consider estimation and inference in a linear model with endogenous regressors where the parameters of interest change across  two samples. If the first-stage is common, we show how to use this information to obtain more efficient two-sample GMM estimators than the standard split-sample GMM, even in the presence of near-weak instruments.
We also propose two tests to detect change points in the parameters of interest, depending on whether the first-stage is common or not. We derive the limiting distribution of these tests and show that they have non-trivial power even under weaker and possibly time-varying identification patterns. The finite sample properties of our proposed estimators and testing procedures are illustrated in a series of Monte-Carlo experiments, and in an application to the open-economy New Keynesian Phillips curve. Our empirical analysis using US data provides strong support for a New Keynesian Phillips curve with incomplete pass-through and reveals important time variation in the relationship between inflation and exchange rate pass-through.

\end{abstract}

\newpage
\section{Introduction}


The evidence for change points is pervasive in macroeconomic applications estimating key structural parameters for monetary policy rules: \cite{CGG1999}, \cite{SW2002} and \cite{ALW}, or when estimating the relationship between inflation and unemployment via open or close-economy New Keynesian Phillips curves - \cite{HHB}, \cite{MagnussonMavroeidis}, \cite{AB17}, \cite{syed2022}, \cite{RossiInoue2024}. There is also strong evidence that many key macroeconomic variables responses to fiscal and monetary policies change over the sample - \cite{inoue2024a} or depend on the state of the economy
- \cite{auerbachgorodnichenko2013}, \cite{owyangetal2013}, \cite{rameyzubairy2018}, \cite{Barnichon:2018}, \cite{Jorda:2020}, \cite{Alpanda:2021}, \cite{Alloza:2022}, \cite{Jo:2022}, \cite{Klepacz}, and \cite{Cloyne2023}, and the state dependence in many of these models can be viewed as a change point upon reordering the data.

A model with endogenous regressors and  a parameter change at a given point in time is usually estimated by GMM separately over the two samples before and after the change (we refer to this estimator as the \textit{split-sample estimator}). This practice can lead to large efficiency losses, because it ignores the fact that a change in parameters over two samples need not imply that everything is changing across these two samples. For example, external instruments  - such as the ones used in the local projection literature or for estimating structural parameters - are constructed as shocks that are only correlated with the endogenous regressors contemporaneously, and there is often no reason to suspect that this correlation also changes over time. \textit{If, in fact, the first-stage linking instruments and endogenous regressors has common parameters}, this information can be used to obtain more efficient estimators for the parameters that do change across the two samples. Additionally, since in most macroeconomic models, instruments are relatively weak - \cite{Mavro}, \cite{DKK2006}, \cite{Nason}, \cite{KMavro}, \cite{rameyzubairy2018}, \cite{AB17}, \cite{BarnichonMesters2020} - splitting the sample may make instruments appear even weaker than they are, especially when ignoring information that is common across the two samples.

We start by showing how to use the information from a common first-stage to increase asymptotic efficiency of the parameter estimates that do change across samples. While a 2SLS estimator with common first-stage is available for strong instruments in \cite{HHB}, we show that it need not be more efficient than our estimator or than the split-sample GMM estimator. We instead develop a \textit{two-sample GMM estimator} with a common first-stage, and prove that it is asymptotically at least as efficient as the split-sample estimator, even when the instruments are near-weak. The asymptotic variance we derive allows us to construct a variance estimator that is guaranteed to be no larger than that of the split-sample estimator in finite samples.\footnote{Note that the estimator we derive is not a second-step GMM estimator with the 2SLS estimator in \cite{HHB} as a first-step, because in the case of change points, the relationship between the two breaks down - \cite{HHB}.} Because we allow for autocorrelated errors, our framework can directly be applied to local projection estimators - \cite{stock2018}.

While our focus is on macroeconometric applications, \textit{when the change point is known}, our estimator can also be viewed as a new two-sample estimator where both the first and the second-stages are observed, and therefore we also contribute to the microeconometric literature on two-sample estimators.  In the original formulation of the two-sample estimators - \cite{angrist1992}, \cite{inoue2010} - the endogenous regressors and the instruments are observed in the first sample, the outcome and the instruments in the second sample, and the first-stage of the first sample is used as proxy for the first-stage in the second sample to obtain IV and 2SLS estimators, assuming that the two first-stages are common. In our case, both first-stages and second-stages are observed, yielding new two-sample estimators.

Additionally, \textit{when the change point is unknown}, we show that the implied change fraction can be estimated consistently even when the instruments are near-weak, and at the fastest available rate in the presence of exogenous regressors such as an intercept. We also provide two Wald tests for a change point in the presence of near-weak instruments. The first test assumes a common first-stage: its asymptotic distribution is pivotal, and has non-trivial power even under the alternative, although the implied change point estimator need not be consistent under the alternative and this depends on how much heterogeneity is allowed in the second moments of the data. The second test assumes that the first-stage has a change point, and checks whether the second-stage exhibits the same change point. Plugging in the implied OLS change point estimator yields a chi-squared asymptotic distribution under the null, with non-trivial power under the alternative.

We also discuss \textit{a general procedure for multiple change points}, and show that \textit{a two-sample estimator for the first-stage} using information about the second-stage is also more efficient than its OLS counterpart over the full-sample or over sub-samples, generalizing \cite{AB17}.

A simulation study compares the performance of the split-sample GMM estimators, the two-sample 2SLS estimators in \cite{HHB}, and the proposed two-sample GMM estimators in the case of a known change point, an unknown change point, and pre-testing for a change point. The two-sample 2SLS estimator has larger mean squared-error compared to the split-sample estimator in several designs with heteroskedastic errors, while our two-sample GMM estimator consistently displays both lower variance and lower mean squared error than both.

We illustrate our method by estimating an open-economy New Keynesian Phillips Curve (NKPC) with monthly US data. Our empirical results provide strong support for imperfect exchange-rate pass-through - \cite{monacelli2005}. We find that this pass-through is significant and comparable in magnitude to that of the unemployment coefficient. Our results also suggest that the relationship between inflation and exchange rate pass-through exhibits important time variation.

The rest of the paper is organized as follows. Section \ref{section:framework} motivates our framework. Section \ref{section:stable-RF} introduces our two-sample GMM estimator and derives its asymptotic properties in presence of a known change point, a common first-stage and near-weak instruments. Section \ref{section:unstable-RF} presents results for the change point estimator and the two Wald tests for a change point. It also proposes a general empirical approach for multiple change points. Section \ref{section:MC} contains the simulations, and Section \ref{section:NKPC} the empirical application. Section \ref{section:conclusion} concludes. The proofs of the theoretical results are in the Appendix.

\section{Framework and examples} \label{section:framework}

To motivate our framework, consider first the standard linear regression model,
\begin{eqnarray*}
	y_t  &=& Z_{1t}' \theta_{z}^0  + X_t'\theta_{x}^0 + u_t  \\
	X_t' &=& Z_t' \Pi^0 + v_t' \, ,
\end{eqnarray*}
where $Z_{1t}$ are $p_1$ exogenous variables, $X_t$ are $p_2$ endogenous variables, $Z_t$ are $q$ valid instruments (with $q\geq p_2$) that include the exogenous regressors $Z_{1t}$ as the first elements, and $\Pi^0$ is a full-rank matrix of size $(q,p_2)$. The parameter of interest is the $p \times 1$-vector $\theta^{0}=[\theta_z^{0'} \ \theta_x^{0'}]'$ (with $p= p_1+p_2$).

We extend this model in three directions. First, we allow the parameters of interest $\theta^0$ to change abruptly across sub-samples. Second, we allow the second moments of the data  - the variance of the moment conditions, the second moment of the instruments, and the variance of the structural errors - to change over the sample and therefore to exhibit substantial time heterogeneity. Third, we allow for the instruments to be near-weak. For all three extensions, we first consider the case when the relationship between the instruments and endogenous regressors does not change over time.\footnote{In the general procedure in Section \ref{section:unstable-RF}, we allow for this relationship to also change.}

Three examples below motivate the need for such an extension and, in particular, why the first-stage may be common while the parameters of interest may change over the sample.

\textbf{New Keynesian Phillips curve.} Consider a model where inflation depends on current inflation predictions and unemployment gap, which are typically endogenous  and often instrumented by using lags of these variables - \cite{HHB}. As discussed in the introduction, there is strong evidence that the parameter linking inflation to the unemployment gap changed over time, while there is no reason to believe that the dynamics of measured inflation expectations or unemployment gap changed at the same time as the unemployment gap parameter. As shown in our empirical application, as well as in \cite{syed2022} - \textit{there are sample periods over which the first-stage is common}; that is, the parameter $\Pi^0$ remains the same before and after the change point. Even when external instruments are used instead of lags, as in \cite{RossiInoue2024}, the same reasoning applies.


\textbf{Local projections.} Suppose we are interested in whether government spending multipliers are different when interests rates are at the zero lower bound - Section 5 in \cite{rameyzubairy2018}. Since interest rates were close to the zero lower bound from 2008 until recently, we can write this model as a regression of output at different horizons on government spending before and after 2008. Since government spending is endogenous, we can use external instruments such as the government spending shocks in  \cite{rameyzubairy2018}. Since the \cite{blanchard2002} shock used in \cite{rameyzubairy2018} as instrument is constructed to measure an unexpected  discretionary change in fiscal policy over the sample, it is \textit{unclear why the relationship of the unexpected component of the fiscal policy with fiscal policy itself should change exactly at the time the interest rates hit the lower bound}.

\textbf{Two-sample IV estimators in microeconometrics.} Let $t$ above refer to individuals instead of time, and suppose we are interested in the effect of age at school entry on educational attainment - \cite{angrist1992}. Because some parents may enroll children earlier, the age at school entry is endogenous, and \cite{angrist1992} use quarter of birth dummies as instruments. They employed data on quarter of birth and age at school entry for one sample of individuals, data on quarter of birth and educational attainment on another sample, and introduced a two-sample IV estimator where the first-stage is estimated in one sample, and plugged into the IV estimator of the second sample. This plug-in approach is employed in all two-sample IV estimators that we are aware of - \cite{inoue2010}, \cite{sanderson2022} - and assumes that the first-stage is common for the two populations. However, if we observed the age at school entry and educational attainment in both samples - which is likely the case for new administrative data - \textit{it is conceivable that the relationship between educational attainment and age at school entry would be heterogeneous across the two samples (and that the parameters of interest would therefore change across the two samples), yet the first-stage linking quarter of birth to age at school entry would have constant parameters across the same two samples}. In general, the causal effect of a treatment on an outcome can be heterogeneous across groups of populations, while the relationship between treatment uptake and intent to treat could be homogeneous over certain groups.


\section{Two-sample GMM estimators with a known change point}\label{section:stable-RF}

To simplify the exposition, consider first the case when the first-stage is common; the case of change points in the first-stage is discussed in Section \ref{section:unstable-RF}. For the remainder of this section, the reference model is:
\begin{eqnarray}
	y_{t} &=& \left\{ \begin{array}{lclclccr}
			Z_{1t}' \theta_{z,1}^0  &+& X_t'\theta_{x,1}^0 &+& u_t & & , & t\leq \lfloor T \lambda^0 \rfloor \\
			Z_{1t}' \theta_{z,2}^0 &+& X_t'\theta_{x,2}^0 &+& u_t & & , & t > \lfloor T \lambda^0 \rfloor
		\end{array} \right. \, , \label{eq:1change SE}\\
	X_t' &=& Z_t'\frac{ \Pi^0}{r_T} + v_t' \, , t=1,\ldots, T\label{eq:stable RF}
\end{eqnarray} \normalsize
where $\lfloor T\lambda_0\rfloor$ is a known change point with $0<\lambda_0<1$, $Z_{1t}$ are $p_1$ exogenous variables, $X_t$ are $p_2$ endogenous variables, $Z_t$ are $q$ valid instruments (with $q\geq p_2$) that include the exogenous regressors $Z_{1t}$, and $\Pi^0$ is a full-rank matrix of size $(q,p_2)$. The instruments share the same identification strength over the whole sample, which can range from strong ($r_T=1$) to near-weak ($\lim_{T\rightarrow \infty} r_T = \infty$ and $r_T=o(\sqrt{T})$). The parameters of interest are the $p \times 1$ vectors $\theta_i^0 =[\theta_{z,i}^{0'} \ \theta_{x,i}^{0'}]', i=1,2$, with $p= p_1+p_2$. For simplicity, we assume that all elements of $\theta_1^0-\theta_2^0$ are different than zero, so all parameters of interest change. To streamline notations,
let $\theta_{vec}^0 =\vec(\theta_{z,1}^0, \theta_{x,1}^0,\theta_{z,2}^0,\theta_{x,2}^0)$, $W_t =\vec(Z_{1t,}, X_t)$, $T_1^0=T^0 =\lfloor T\lambda_0 \rfloor$, and $T_2^0=T-T_1^0$.\footnote{$\vec(\cdot)$ is a generalization of the usual $vec(\cdot)$ operator, in the sense that for $S$ general matrices $A_1,\ldots, A_S$,  $\vec(A_1,\ldots, A_s) = ( (vec(A_1))',\ldots, (vec(A_S))')'$.}
This model can be estimated with the split-sample GMM estimator $\hat \theta_{GMM,vec}(\lambda_0)$ in \cite{Andrews1993}, which minimizes:
\begin{eqnarray}\nonumber
O_{GMM}(\theta_{vec}; \lambda_0) &=&  g_T'(\theta_{vec}; \lambda_0) \hat S_u^{-1} g_T(\theta_{vec}; \lambda_0)   ,\\
\text{where} && \theta_{vec} = \vec(\theta_1,\theta_2) \, ,  \text{with }  \theta_i=\vec(\theta_{z,i},\theta_{y,i}) \, ,i=1,2 \, ,  \nonumber \\
&& g_T(\theta_{vec}; \lambda_0) = \left[ \begin{array}{l}
(T_1^0)^{-1} \sum_{t=1}^{T_1^0}   Z_t( y_t - W_t'\theta_1) \\
(T_2^0)^{-1} \sum_{t=T_1^0+1}^{T} Z_t\left( y_t - W_t'\theta_2 \right)
\end{array} \right] \, , \label{eq:gT}\\
&& \hat S_{u} = diag \ [\hat S_{u,1}, \hat S_{u,2}] \inp \avar \left[T^{1/2} g_T(\theta_{vec}^0; \lambda_0)\right] \, . \nonumber
\end{eqnarray}

\cite{Andrews1993} introduced the split-sample GMM estimator to study the properties of a change point test. Here, we derive its asymptotic distribution under a constant first-stage and near-weak instruments: the structural parameters associated with the endogenous variables are asymptotically normally distributed at rate $\sqrt{T}/r_T$, whereas those associated with the exogenous regressors are asymptotically normally distributed at the standard rate $\sqrt{T}$, as shown in Theorem  \ref{theo:consistent}.\footnote{See \cite{AntoineRenault2009} who derive asymptotic properties of 2SLS and GMM in stable models with near-weak instruments.} However, this GMM estimator is not the most efficient estimator of \eqref{eq:1change SE}  -\eqref{eq:stable RF} as it ignores the constancy of the first-stage across the two samples.

An estimator that does take into account that the first-stage is common is the two-sample 2SLS estimator (TS2SLS) estimator in \cite{HHB}, which estimates the first-stage on the full sample. This estimator  $\hat \theta_{TS2SLS,vec}(\lambda_0)$ minimizes:
\begin{eqnarray*}
	O_{2SLS}(\theta_{vec};\lambda_0) &=& \textstyle \sum_{t=1}^{T_1^0} ( y_t - \hat W_t' \theta_{1} )^2 + \textstyle \sum_{t=T_1^0+1}^{T} ( y_t - \hat W_t' \theta_{2})^2  \\
	\text{where}  && \hat W_t' =[ Z_{1t}'\,,\, Z_t' \hat\Pi], \mbox{ and } \hat \Pi = \textstyle(\sum_{t=1}^{T} Z_t Z_t')^{-1}\textstyle (\sum_{t=1}^{T} Z_t X_t').
\end{eqnarray*}

With $\Pi^{z} = \begin{bmatrix} I_{p_1} \\ O_{(q-p_1) \times p_1} \end{bmatrix}$, and $\Pi^a = [\Pi_z, \Pi^0]$, such that $ W_t'\frac{\Pi^a}{r_T} = (Z_{1t}', Z_t'\frac{\Pi^0}{r_T})$,  the implied moment conditions for this estimator are\footnote{A more efficient two-step GMM estimator for the moment conditions \eqref{eq:ts2slsmom} exists; however, because the moment conditions are nonlinear, it will suffer from the curse of dimensionality even with moderate number of instruments $Z_t$. We therefore do not pursue it in this paper.}
\begin{eqnarray} \label{eq:ts2slsmom}
\begin{bmatrix}
\E [Z_t(y_t - W_t'\frac{\Pi^a}{r_T}\theta_{1}^0 ) ]= 0 & t\leq T^0 \\
\E [Z_t(y_t - W_t'\frac{\Pi^a}{r_T}\theta_{2}^0 )] = 0 & t> T^0\\
\E [\vec (Z_t(X_t' - Z_t'\frac{\Pi^0}{r_T})) ]= 0 & t=1,\ldots T
\end{bmatrix}.
\end{eqnarray}

To mimic the design of the TS2SLS estimator above and still obtain a quadratic objective function, one may be tempted to add the $qp_2$ full-sample OLS moment condition of the first-stage to the split-sample GMM estimator. However, \cite{AhnSchmidt} showed that adding a just-identified moment condition with its own nuisance parameter does not change the asymptotic distribution of the original estimator. Instead, we propose adding $2qp_2$ first-stage moments, for the samples before and after the change point, with the restriction that $\Pi^0$ is the same across the two sub-samples. Let $\Pi^0_{vec}=\vec(\Pi^0)$, and the new moment conditions be:
\begin{eqnarray*}
\breve g_T(\theta_{vec}, \Pi_{vec}; \lambda_0) = \left[ \begin{array}{c} g_{T}(\theta_{vec}; \lambda_0) \\ g_{T,2}(\Pi_{vec}; \lambda_0) \end{array} \right] \, ,
\end{eqnarray*}
where $g_{T,2}(\Pi_{vec}; \lambda_0)$ stacks the OLS moment conditions from the first-stage:
\begin{align}\label{eq:addmom}
g_{T,2}(\Pi_{vec}; \lambda_0) = \begin{bmatrix}
(T_1^0)^{-1} \sum_{t=1}^{T_1^0} \vec (Z_t(X_t' - Z_t'\Pi))  \\
 (T_2^0)^{-1} \sum_{t=T_1^0+1}^{T} \vec (Z_t(X_t' - Z_t'\Pi))
\end{bmatrix}.
\end{align}
The two-sample GMM (TSGMM) estimators are defined as:
\begin{eqnarray*}
\left[\begin{array}{c} \hat{\theta}_{TSGMM,vec}(\lambda_0) \\ \hat{\Pi}_{TSGMM,vec}(\lambda_0) \end{array} \right]
&=& \arg \min_{\theta_{vec},\Pi} \left[ \breve g_T'(\theta_{vec}, \Pi_{vec}; \lambda_0)  \ \hat{\mathcal  S}^{-1} \ \breve g_T(\theta_{vec}, \Pi_{vec}; \lambda_0)  \right] \, , \\
\text{where } \ \hat{\mathcal S} &\inp& \avar [\sqrt{T} \breve g_T(\theta_{vec}^0, \Pi_{vec}^0; \lambda^0)] \, .
\end{eqnarray*}
In Theorem \ref{theo:eff} below, we show that the additional moment conditions \eqref{eq:addmom}  - that now overidentify $\Pi^0$ - are not redundant and deliver an asymptotically more efficient estimator of $\theta^0_{vec}$. It is worth pausing first to compare these estimators to the original two-sample IV estimator in \cite{angrist1992} and the two-sample 2SLS estimator in \cite{inoue2010}. These estimators were developed when $W_t, Z_t$ are observed in the first sample, and $y_t, Z_t$ in the second sample. The two-sample IV estimator uses the following moment condition:
$$
E(Z_t y_t 1[t> T^0] - Z_t W_t'\theta^0 1[t\leq T^0] )=0,
$$
which is very different from the moment conditions above. This estimator plugs in the first-stage of the first sample $Z_t W_t' 1[t\leq T^0]$  for the unavailable first-stage of the second sample $Z_t W_t' 1[t> T^0]$. In contrast, the two-sample 2SLS estimator in \cite{inoue2010} uses:
$$
\begin{bmatrix}
	\E [Z_t(y_t - W_t'\frac{\Pi^a}{r_T}\theta_{2}^0 ) ]= 0 & t>T^0 \\
	\E [\vec(Z_t(X_t' - Z_t'\frac{\Pi^a}{r_T})) ]= 0 & t \leq T^0
\end{bmatrix}.
$$
 \cite{inoue2010} also show that it is more efficient than the two-sample IV in \cite{angrist1992}. Note that a first-step estimator of the TSGMM estimator we proposed - based on two-stage least-squares - has four sets of moment conditions, two of which are equivalent to the moment conditions above, and two of which are the equivalent moment conditions in the other samples which are observed in our paper.

Next, we derive the asymptotic distributions of the split-sample GMM, TS2SLS and TSGMM. To do so, we impose the assumptions below.
\begin{ass} (Regularity of the error terms and first-stage equation \eqref{eq:stable RF}) \label{ass:reg1} \\
(i) Let $h_t=Z_t \otimes \vec(u_t, v_t)$ with $i^{th}$ element $h_{t,i}$. \\
- $E(h_t)=0$.\\
- The eigenvalues of $S = \avar\left(T^{-1/2} \sum_{t=1}^Th_t\right)$ are $\mathcal{O}(1)$. \\
- $\E (h_{t,i})=0$ and for some $d>2$, $\|h_{t,i}\|_d < \infty$ for $t=1,\cdots,T$ and $i=1,\cdots,(p_2+1)q$.\\
- $\{h_{t,i}\}$ is near epoch dependent with respect to some process $\{\xi_t\}$, $\|h_t-E(h_t|\mathcal{G}_{t-m}^{t+m})\|_2 \leq \nu_m$ with $\nu_m=\mathcal{O}(m^{-1/2})$ where $\mathcal{G}_{t-m}^{t+m}$ is a $\sigma$-algebra based on $(\xi_{t-m},\cdots,\xi_{t+m})$. \\
- $\{\xi_t\}$ is either $\phi$-mixing of size $m^{-d/[2(d-1)]}$ or $\alpha$-mixing of size $m^{-d/(d-2)}$. \\
(ii) $\Pi^0$ is full column-rank equal to $p_2$.
\end{ass}
Assumption \ref{ass:reg1} part (i) states that the instruments are valid, and allows for general patterns of weak dependence in the data, including autocorrelated errors. Therefore, our paper also applies to local projection IV estimators. Assumption \ref{ass:reg1} part (ii) ensures that the instruments are not redundant.

\begin{ass} (Regularity of the change fraction and second moment matrices) \label{ass:reg-change} \\
(i)  $0<\lambda^0<1$. \\
(ii)  Let $\sum_1 = \sum_{t=1}^{T_1^0} $ and  $\sum_2 = \sum_{t=T_1^0+1}^{T}$. Then $Q_i(\lambda_0)$ and $S_i(\lambda_0)$ are positive definite for $i=1,2$, where
\begin{eqnarray*}
 T^{-1} \textstyle \sum_{i} Z_tZ_t' \stackrel{p}{\rightarrow} Q_i(\lambda_0) =Q_i\, ; \,
S_i(\lambda_0) = AVar \left[ T^{-1/2}\textstyle \sum_i h_t \right] = \left[ \begin{array}{cc} S_{u,i}(\lambda_0) & S_{uv,i}'(\lambda_0) \\
S_{uv,i}(\lambda_0) & S_{u,i}(\lambda_0)
\end{array} \right]
\, = S_i.
\end{eqnarray*}
\end{ass}
Assumption \ref{ass:reg-change}(i) ensures that there are enough observations in each of the two samples to estimate parameters consistently. Note that while we defined the two samples based on a known change point, they could also be defined in terms of a state variable splitting the two samples based on a known threshold value, as in the local projection literature - \cite{rameyzubairy2018}. Additionally, they could just be two groups of observations with heterogeneous parameters - \cite{bonhomme2015}. Assumption \ref{ass:reg-change}(ii) allows for general time variation (heterogeneity) in the second moments of the data.

\begin{ass} (Regularity of the identification strength) \label{ass:reg-identif} \\
 $r_T = o(\sqrt{T})$.
\end{ass}
Assumption \ref{ass:reg-identif} ensures that the identification is sufficiently strong for the structural parameters to be consistently estimated. Under these assumptions, Theorem \ref{theo:consistent} derives the asymptotic distribution of the split-sample GMM, TS2SLS and TSGMM estimators.
\begin{theo} (Asymptotic distributions of the split-sample GMM, TS2SLS and TSGMM estimators) \label{theo:consistent} \\
Let $\Lambda_T = \diag(T^{1/2} I_{p_1}, T^{1/2}  r_T^{-1} I_{p_2})$. Under Assumptions \ref{ass:reg1} to \ref{ass:reg-identif},

\noindent (i) $[I_2 \otimes \Lambda_T]  \, (\hat \theta_{TS2SLS,vec}(\lambda_0) - \theta_{vec}^0) \ind \mathcal{N}(0,V_{TS2SLS,vec})$\\

\noindent (ii) $[I_2 \otimes \Lambda_T]\left(\hat{\theta}_{GMM,vec}(\lambda_0)-\theta_{vec}^0\right) \ind \mathcal N(0, V_{GMM,vec})$\\

\noindent (iii) $
\diag(I_2 \otimes \Lambda_T,  T^{1/2} I_{p_2q})\begin{bmatrix} \hat \theta_{TSGMM,vec}(\lambda_0) - \theta_{vec}^0 \\
\hat \Pi_{TSGMM,vec}(\lambda_0) - \Pi_{vec}^0/r_T \end{bmatrix} \ind \mathcal{N} (0 , V_{TSGMM,vec}^a),
 $
 \\

and  $ [I_2 \otimes \Lambda_T]\left(\hat{\theta}_{TSGMM,vec}(\lambda_0)-\theta_{vec}^0\right) \ind \mathcal N(0, V_{TSGMM,vec})$, with\\
$ V_{TSGMM,vec} = (V_{GMM,vec}^{-1} +  \mathcal G'\mathcal G)^{-1}$.\\
The variances $V_{TS2SLS,vec}$,  $V_{GMM,vec}$, $V_{TSGMM,vec}^a$ and $V_{TSGMM,vec}$ and $\mathcal G$ are in Appendix Definition 1.

\end{theo}

The estimators of the endogenous regressor parameters $\theta_{x,i}^0$ converge at rate $\sqrt{T}/r_T$, which is slower than usual: this rate is due to the presence of near-weak instruments. The other parameter estimators are not affected by the near-weak identification rate, and their estimators are asymptotically normally distributed at the standard rate $\sqrt{T}$.\footnote{See \cite{AR2009} for related results without change points.} More importantly, the formula for the variance of the TSGMM estimator, containing the positive semi-definite matrix $\mathcal G'\mathcal G$, reveals that this variance is at least as small as that of the split-sample GMM estimator asymptotically. When plugging in a consistent estimator for $\mathcal G$, this guarantees lower variance of the TSGMM estimator in finite samples as well.

The next assumptions consider additional regularity conditions on homogeneity of second moments and conditional homoskedasticity, which are key to derive conditions under which the TSGMM estimators are \textit{strictly more efficient} than the TS2SLS estimators, and the TS2SLS estimators are strictly more efficient than the split-sample GMM.
\begin{ass} (Homogeneity of the second moments) \label{ass:reg-secondmom}\\
(i) $Q_1(\lambda_0)=\lambda_0 Q$ and $Q_2(\lambda_0)= (1-\lambda_0) Q$; (ii) $S_1(\lambda_0)= \lambda_0S$ and $S_2(\lambda_0)= (1-\lambda_0)S$.
\end{ass}
The homogeneity assumption prevents changes in the second moments of the instruments and in their correlation with the error terms of the first-stage. For example, Assumption \ref{ass:reg-secondmom}(i) is violated when there is a change in $E(Z_t Z_t')$, while Assumption \ref{ass:reg-secondmom}(ii) is violated when there is a change in $\var(Z_t u_t)$, for example.

\begin{ass} (Conditional homoskedasticity) \label{ass:cond-hom}\\
$$S_{i} = \Phi \otimes Q_i,  \mbox{ where } \Phi= \E \left[\left( \begin{array}{cc}
                                                                                      u_t^2 & u_t v_t'\\
                                                                                      u_t v_t & v_t v_t'\\
                                                                                    \end{array}
                                                                                  \right)
 | \mathcal{F}_{t} \right] = \left(
           \begin{array}{cc}
             \Phi_u & \Phi_{uv}' \\
             \Phi_{uv} & \Phi_v \\
           \end{array}
         \right),
 $$
$Q_1 = Q_1(\lambda_0)$, $Q_2= Q_2(\lambda_0)$, with $\mathcal{F}_{t}$ the $\sigma$-algebra generated by $\{Z_t, Z_{t-1}, \ldots \}$, and $i=1,2$.
\end{ass}

The following theorem compares the asymptotic variance of the proposed estimators with and without Assumptions \ref{ass:reg-secondmom} and \ref{ass:cond-hom}.\footnote{For two square symmetric matrices $A,B$ of size $p \times p$, we write $A\leq B$ iff $B-A$ is positive semidefinite, and $A<B$ if $B-A$ is positive definite.}

\begin{theo} (Efficiency comparison) \label{theo:eff}\\
(i) Under Assumptions \ref{ass:reg1} to \ref{ass:reg-identif}, $V_{TSGMM,vec} \leq V_{GMM,vec}$, with the difference being of rank $min(p_2q, 2p)$. If, additionally, $S_{uv,1}(\lambda_0)$, respectively $S_{uv,1}(1)-S_{uv,1}(\lambda_0)$ , the covariances between the moment conditions in the first- and second-stages of both samples, are of full column rank, and  $p_2q\geq 2p$, then $V_{TSGMM,vec} < V_{GMM,vec}$.\\
(ii) Under Assumptions \ref{ass:reg1} to \ref{ass:cond-hom}, for $i=1,2$,
\begin{eqnarray}
V_{TS2SLS,i} < V_{GMM,i}  &\Leftrightarrow& 2 \Phi_{uv}' \theta_{x,i}^0 + \theta_{x,i}^{0'} \Phi_v \theta_{x,i}^0 < 0 \label{eq:in1} \\
V_{TSGMM,i} < V_{TS2SLS,i}&\Leftrightarrow& 2 \Phi_{uv}' \theta_{x,i}^0 + \theta_{x,i}^{0'} \Phi_v \theta_{x,i}^0  > -\frac{\delta \Phi_u^2}{1+\delta \Phi_u} \, , \label{eq:in2}\\
&& \text{with $\delta = \Phi_u^{-2} \Phi_{uv}' (\Phi_v - \Phi_{uv} \Phi_u^{-1} \Phi_{uv}')^{-1} \Phi_{uv}$}, \nonumber
\end{eqnarray}
with equalities instead of inequalities if \eqref{eq:in1}, respectively \eqref{eq:in2}, hold with equality.
\end{theo}

Theorem \ref{theo:eff} part \textit{(i)} shows that even when the first sample moments of the first-stage are redundant for the second sample moments of the first-stage - as it is the case under homogeneity Assumptions \ref{ass:reg-secondmom} and \ref{ass:cond-hom} -  they are not redundant \textit{overall} for the estimation of $\theta_{vec}^0$. The intuition is similar to Theorem 4 in \cite{Breusch1999}, where it is shown that with three moment conditions, say $(g_1, g_2, g_3)$, the redundancy of $g_3$ given $g_2$ does not imply that $g_3$ is redundant given $(g_1,g_2)$. This non-redundancy is also related to recent results by \cite{AR2017} who extend \cite{Breusch1999} to frameworks that allow for heterogenous identification strengths.

The proof relies on determining the rank of $\mathcal G$ defined in Theorem \ref{theo:consistent}. We intentionally allowed in part \textit{(i)} for moments of the data to be heterogeneous over time for two reasons. First, because the second moments of the data may also have change points at the same time as the parameters of interest. Second, because the sample split need not be due to a change point, but rather a threshold variable as in the literature on state-dependent local projections - \cite{rameyzubairy2018}, \cite{Alpanda:2021}. Reordering the data according to whether the threshold variable is below or above a certain known value can induce heterogeneity in the second moments of the data. Nonetheless, with this reordering, we obtain the change point model in \eqref{eq:1change SE}.

We also give conditions for strict efficiency of TSGMM over split-sample GMM estimators in part \textit{(i)}.
The first condition is natural if a first-stage linking regressors and instruments exists: if the moment conditions of the first and second-stage were not correlated, then $X_t$ would not be endogenous according to that first-stage. The second condition is automatically satisfied if we have two or more endogenous regressors, as is the case in our empirical application, because then $p_2 q \geq 2 q \geq 2p$. If we have only one endogenous regressor, then the condition becomes $q \geq 2 p$, meaning we need twice as many instruments as regressors in each of the two samples.

Theorem \ref{theo:eff} part \textit{(ii)} compares the TS2SLS with the split-sample and the TSGMM estimators, a comparison that is not possible in the general case due to the moment conditions of the TS2SLS not being a special case of the moment conditions of the other estimators.

We first show that TS2SLS estimators can be more efficient than the split-sample GMM estimators. Condition (\ref{eq:in1}),
$$ 2 \Phi_{uv}' \theta_{x,i}^0 + \theta_{x,i}^{0'} \Phi_v \theta_{x,i}^0 < 0
\ \Leftrightarrow \
\Phi_u + 2 \Phi_{uv}' \theta_{x,i}^0 + \theta_{x,i}^{0'} \Phi_v \theta_{x,i}^0  < \Phi_u$$
is equivalent to
$ \var(u_t + v_t'\theta_{x,i}^0 | \mathcal F_t) < \var( u_t | \mathcal F_t) $.
Heuristically, it states that TS2SLS estimators are more efficient when the second-stage error - obtained after plugging in the full-sample first-stage  - has smaller variance than that of the structural error. This condition \textit{should induce caution when using the TS2SLS estimators in practice}, because it is not satisfied, for example,  if $Cov(u_t, v_t)$ is of the same sign as $\theta_{x,i}^0$.

Additionally, the TS2SLS estimators are not a special case of the TSGMM estimators either, for which a first-step estimator based on 2SLS can be constructed.\footnote{The first-stage 2SLS estimator for TSGMM based on moment conditions \eqref{eq:addmom} is different.}. Nevertheless, the TSGMM estimators are strictly more efficient than the TS2SLS estimators despite second moment homogeneity and conditional homoskedasticity, as long as (\ref{eq:in2}) holds.
This condition is harder to interpret, but it is automatically satisfied for at least one of the two TSGMM estimators in the case of a single endogenous regressor and no exogenous regressors, that is $p_2=1$ and $p_1=0$. In such a case, $\Phi_u \delta = \Phi_{uv}^2 (\Phi_v \Phi_u - \Phi_{uv}^2)^{-1}$, and the above condition becomes,
\begin{align*}
2 \Phi_{uv} \theta_{x,i}^0 + (\theta_{x,i}^{0})^2 \Phi_v > -\frac{\Phi_{uv}^2}{\Phi_v}
\ \Leftrightarrow  \
\left(\theta_{x,i}^0 + \frac{\Phi_{uv}}{\Phi_v} \right )^2 > 0,
\end{align*}
which has to hold for at least one of $i=1,2$ since we assumed there the coefficients of the endogenous regressors change over the sample ($\theta_{x,1}^0 \neq \theta_{x,2}^0$).

To interpret this condition, assume that the regressors are fixed. Then $\displaystyle \beta^0=\frac{\Phi_{uv}}{\Phi_v}$ is the limiting coefficient of a regression of $u_t$ on $v_t$, suggesting that the TSGMM estimator ``purges'' $u_t$ of the true correlation with $v_t$. On the other hand, the TS2SLS estimator transforms the error $u_t$ into $(u_t+ v_t'\theta_{x,i}^0)$ through an orthogonal projection, so that $-\theta_{x,i}^0$ plays the role of $\beta^0$ for each sub-sample. As a result, when these two are equal, say over the first sample, the two associated estimators TS2SLS and TSGMM are asymptotically equivalent over this sample, but the TSGMM estimator will be more efficient over the second sample.

\section{Two-sample GMM estimators with unknown change points} \label{section:unstable-RF}

In this section, we first explain how to deal with an unknown change point in the equation of interest. First, we propose a consistent estimator of the associated change fraction, and a new test to detect its presence. Second, we propose a new test to detect whether the change point in the first also shows up in the second-stage, and extend our inference procedure to the presence of multiple change points.

\subsection{Unknown change point estimation} \label{subsection:SF change}

\subsubsection{Estimation of the change fraction $\lambda^0$}

We extend the results in \cite{HHB} to show that minimizing a 2SLS criterion delivers consistent estimators of both the change fraction $\lambda^0$ and the parameters $\theta_{vec}^0$ in the presence of near-weak instruments. In the first-stage, $X_t$ is estimated over the full-sample by OLS to get $\hat{X}_t = \hat \Pi' Z_t$, and the augmented projected regressors $\hat W_t = \vec (Z_{1t}, \hat X_t)$. In the second-stage, consider the following 2SLS criterion given a candidate change point $\lfloor T\lambda \rfloor$ and parameters $\theta_{vec}$,
\begin{eqnarray*}
&& Q_{2SLS}(\lambda,\theta_{vec} ) = \sum_{t=1}^{\lfloor T\lambda\rfloor} \left( y_t - \hat W_t' \theta_{1} \right)^2 + \sum_{t=\lfloor T\lambda\rfloor+1}^{T} \left( y_t - \hat W_t' \theta_{2} \right)^2, \\
\text{with} && \theta_i=\vec(\theta_{z,i},\ \theta_{x,i}) \, , i=1,2 \, .
\end{eqnarray*}

To get the 2SLS estimators for each candidate change fraction $\lambda$, first concentrate with respect to $\theta_{vec}$ to get $\hat{\theta}_{vec}(\lambda)$ and, then, minimize $Q_{2SLS}(\lambda,\hat \theta_{vec} (\lambda) )$ over all partitions of the sample defined by $\lfloor T\lambda\rfloor$ to get
\begin{eqnarray} \label{eq:2sls-lambda-estim}
\hat{\lambda} &=& \arg \min_{\lambda \in \Lambda_{\epsilon}} Q_{2SLS}(\lambda,\hat{\theta}_{vec}(\lambda))
\quad \text{with} \quad \hat{\theta}_{vec}(\lambda)\equiv \vec (\hat{\theta}_1(\lambda),\hat{\theta}_2(\lambda)),
\end{eqnarray}
and $\Lambda_{\epsilon} = [\epsilon, 1-\epsilon]$ for some $\epsilon>0$,  a cut-off to ensure enough observations in each of the two samples.
The associated 2SLS estimators of the change point $\hat{T}$ and parameters $\hat{\theta}_{vec}$ are then defined as:
\begin{eqnarray*}
\hat{T}= \lfloor T\hat{\lambda} \rfloor \hspace{.25cm} \text{and} \hspace{.25cm} \hat{\theta}_{vec} \equiv \hat{\theta}_{vec} (\hat \lambda) \equiv \vec (\hat{\theta}_1(\hat{\lambda}),\hat{\theta}_2(\hat{\lambda})) \equiv \vec(\hat \theta_1,\hat \theta_2).
\end{eqnarray*}
To derive the asymptotic properties of $\hat{\lambda}$, impose the following regularity assumptions, which extend the known change-point assumptions in Section \ref{section:stable-RF} to all change point candidates.

\begin{ass} (Regularity of the change fraction) \label{ass:reg} \\
The candidate change points satisfy $ \max (\lfloor T\lambda\rfloor,T-\lfloor T\lambda\rfloor) > \max (p-1, \epsilon T)$ for some $\epsilon>0$ such that $\epsilon < \min(\lambda^0,1-\lambda^0)$. Accordingly,  $\lambda \in \Lambda_{\epsilon} = [\epsilon, 1-\epsilon]$.
\end{ass}

\begin{ass} (Regularity of the instrumental variables) \label{ass:reg-IV} \\
Let $\hat Q_1(r) = T^{-1} \sum_{t=1}^{\lfloor Tr \rfloor} Z_t Z_t'$. Then $\hat Q_1(r) \inp Q_1(r)$, uniformly in $r\in[0,1]$ where $Q_1(r)$ is positive definite and strictly increasing in $r$.
\end{ass}

\begin{ass} (Regularity of the variances)\label{ass:reg-var}
$$
\avar\left[T^{-1/2} \sum_{t=1}^{\lfloor Tr\rfloor} h_t\right] = S_1(r) = \left(
                                                                  \begin{array}{cc}
                                                                    S_{u,1}(r) & S_{uv,1}'(r) \\
                                                                    S_{uv,1}(r) & S_{v,1}(r) \\
                                                                  \end{array}
                                                                \right)
,
$$
uniformly in $r\in[0,1]$, where $S_1(r)$ is positive definite and strictly increasing in $r$, with $S_{u,1}(r)$, $S_{v,1}(r)$ of size $q \times q$, respectively $(p_2 q) \times (p_2 q)$.
\end{ass}

Assumptions \ref{ass:reg}, \ref{ass:reg-IV} and \ref{ass:reg-var} are typical for the change point literature. Assumption \ref{ass:reg} ensures that there are enough observations in each sub-sample to identify the true change point. Assumption \ref{ass:reg-IV} ensures that there is enough variation in the instruments to identify the change point. It also allows for the second moment of instruments to change over the sample. Assumption \ref{ass:reg-var} allows for (unconditional) heteroskedasticity in the sample moments of the first and the second-stage. It also allows for a change in the variance of the moment conditions.

\begin{theo} (Asymptotic properties of $\hat{\lambda}$ and associated estimators) \label{theo:change-order} \\
Under Assumptions \ref{ass:reg1} to \ref{ass:reg-identif} and \ref{ass:reg} to \ref{ass:reg-var}, \\
(i) $\| \hat{\lambda}-\lambda^0 \| = \mathcal{O}_P(T^{-1})$. \\
(ii)
the asymptotic distribution of the split-sample GMM, TS2SLS and TSGMM estimators is as if the change point was known, hence as described in Theorem \ref{theo:consistent}.
\end{theo}

Note that the estimator of the change fraction $\hat\lambda$ converges at the fastest available rate $T$. This stems from the presence of fixed changes in the parameters of the exogenous regressors; intuitively, the exogenous regressors act as their own strong instruments, and the strongest instruments determine the fast convergence rate of $\hat{\lambda}$.

\subsubsection{Detection of the change point} \label{subsection:sup-wald test}

In practice, the existence of the change point in the second-stage equation often needs to be established. To that end, we consider the sup-Wald test of \cite{HHB} for which the null and alternative hypotheses are: $H_0:\,\mathcal R_p \theta^0=0$ versus $H_A:\,\mathcal R_p \theta^{0}\neq 0$, with $\mathcal R_p = (1,-1) \otimes I_p$. The associated test statistic is defined as:
\begin{equation} \label{supwald}
Sup-Wald_T\;=\;\underset{\lambda \in\Lambda_{\epsilon}}{\sup} Wald_T(\lambda),
\end{equation}
\begin{eqnarray*}
\text{where} && Wald_T(\lambda) =  T  \hat{\theta}_{vec}'(\lambda) \mathcal R_p^\prime  \left[\mathcal R_p \, \hat G(\lambda) \,\mathcal R_p' \right]^{-1} \mathcal R_p \hat{\theta}_{vec}(\lambda) \, , \\
&& \hat G=\diag[ \ \hat G_{1}(\lambda) ,  \hat G_{2}(\lambda)] \quad
\text{with } \   \hat G_i(\lambda) = \hat{\mathcal{A}}_{i}^{-1}(\lambda) \hat H_{i}(\lambda) \hat{\mathcal{A}}_{i}^{-1}(\lambda) \, , \\
&& \hat{\mathcal{A}}_{i}(\lambda) =  T^{-1}\sum_{t\in I_i} \hat{X}_t \hat{X}_t^\prime \quad \, ,
I_1 = \{1,\ldots, [T\lambda]\} \, , I_2 = \{[T\lambda]+1,\ldots, T\} \, , \\
\text{and} && \text{$\hat H_{i}(\lambda)$ is a HAC estimator such that} \\
&& \Lambda_T^{-1} \hat H_{i}(\lambda) \Lambda_T^{-1} \inp H_i(\lambda) = \avar \left[ \sum_{t\in I_i} \Lambda_T^{-1}  \hat{X}_t (u_t + v_t'\theta_{x,i}^0) \right] \, .
\end{eqnarray*}

We impose the following additional assumption under the null hypothesis, which maintains the same homogeneity assumption as Assumption \ref{ass:reg-secondmom}, but with respect to all candidate change points. Without this assumption, the asymptotic distribution of the test is not pivotal and needs to be simulated or bootstrapped - \cite{Boldea2019}.

\begin{ass} (Homogeneity of the second moments) \label{ass:reg-secondmom all candidates}\\
	For any $r\in[0,1]$, we have: (i) $Q_1(r)=r Q$; (ii) $S_1(r)= rS$.
\end{ass}

The following theorem provides the limiting distribution of the sup-Wald test statistic.

\begin{theo} (Test for a change in the equation of interest) \label{theo:wald} \\
(i) Under $H_0: \theta_1^0=\theta_2^0$, and Assumptions \ref{ass:reg1} and \ref{ass:reg} to \ref{ass:reg-secondmom all candidates},
$$
Sup-Wald_T\;\Rightarrow\; \sup_{\lambda\in \Lambda_{\epsilon}} \frac{||\mathcal B_{p} (\lambda)-\lambda \mathcal B_{p}(1)||^2}
{\lambda(1-\lambda)},
$$
where $\mathcal B_{p}(\lambda)$ is a $p \times 1$ vector of independent standard Brownian motions defined on $[0,1]$.\\
(ii) Under $H_A: \theta_{\Delta}^0 = \theta_1^0 -\theta_2^0 \neq 0$, and Assumptions \ref{ass:reg1} to \ref{ass:reg-identif} and \ref{ass:reg} to \ref{ass:reg-var}, the test statistic diverges such that 
$$
Sup-Wald_T = \left\{ \begin{array}{ll}
\mathcal{O}_p(T/r_T^2) & \text{without exogenous regressors ($p_1=0$)} \\
\mathcal{O}_p(T) & \text{in presence of exogenous regressors ($p_1\neq0$)}
\end{array} \right.
$$
In addition, if Assumption \ref{ass:reg-secondmom all candidates} holds and either $p_1=0, p_2=1$ (only one endogenous regressor, no exogenous regressors) or $\theta_{x,1}^0=\theta_{x,2}^0$, then the implicit change fraction estimator is consistent,
$\hat \lambda^{W} =\arg \sup_{\lambda\in\Lambda_{\epsilon}} Wald_T(\lambda) \inp \lambda^0$. Otherwise, $\hat \lambda^{W} \inp \lambda^0$ is not guaranteed.
\end{theo}

We recommend using the sup-Wald test to test for a change point because of its robustness to conditional heteroskedasticity and autocorrelation. We also note that if Assumption \ref{ass:reg-secondmom all candidates} is violated, then the implicit change fraction estimator may not be consistent, a result which to our knowledge was not previously available and may be of independent interest.

Note that the global power of the test depends on the presence of exogenous regressors: without exogenous regressors, the rate of divergence is affected by the identification strength of the instruments and is equal to $T/r_T^2$; in presence of exogenous regressors, the rate is standard and equal to $T$, and thus is not affected by the identification strength.

\subsection{General framework}

\noindent We now discuss a generalization of the model considered in Section \ref{section:stable-RF} to allow for changes in the parameters of first and second-stages,
\begin{eqnarray}
y_{t} &=& \left\{ \begin{array}{lclclccr}
    Z_{1t}' \theta_{z,1}^0 &+&X_t'\theta_{x,1}^0  &+& u_t & & , & t\leq \lfloor T \lambda^0 \rfloor \\
    Z_{1t}' \theta_{z,2}^0 &+& X_t'\theta_{x,2}^0  &+& u_t & & , & t > \lfloor T \lambda^0 \rfloor
    \end{array} \right. \label{eq:one-change3} \\
X_t' &=& \left\{ \begin{array}{lclccl}
\frac{Z_t' \Pi_1^0}{r_{1T}} &+& v_t' & & , & t\leq \lfloor T \nu^0 \rfloor \\
\frac{Z_t' \Pi_2^0}{r_{2T}} &+& v_t' & & , & t> \lfloor T \nu^0 \rfloor \\
\end{array} \right. \, , \label{eq:unstable-RF}
\end{eqnarray}
where $r_{iT}=1$, or $r_{iT} \rightarrow \infty, r_{iT} = o(\sqrt(T))$, with $i=1,2$, and $Z_t$ is not correlated with $v_t$ or $u_t$. When there is no change in the identification strength (that is, $r_{1T} \propto r_{2T}$),  (\ref{eq:unstable-RF}) naturally extends the RF models considered in \cite{HHB} to weaker identification patterns. Otherwise, (\ref{eq:unstable-RF}) captures changes in identification strength that may or may not be concomitant to those in the parameter of interest $\theta^0_{vec}$. Our goal is to detect and locate both parameter instability and changes in the identification strength, as well as to provide correct and sharp inference on $\theta^0_{vec}$.

\subsubsection{Test for common change} \label{subs:test-common-change}

This section proposes a Wald test for a common change. Consider the case where the RF change $T^*=\lfloor T\nu^0\rfloor$ has been detected and estimated consistently by $\hat{T}^*$ (for example, using the methods developed in \cite{AB17} or \cite{baiperron1998}). To test whether $T^*$  is also a change point in the equation of interest, we test whether the 2SLS parameter estimates defined over each corresponding sub-sample are equal to each other using a standard Wald test. These parameter estimates are defined as follows for $i=1,2$, and $\hat{I}_1^*=\{1, \ldots, \hat{T}^*\}$ and $\hat{I}_2^*=\{\hat{T}^*+1, \ldots, T\}$,
\begin{eqnarray*}
&& \hat \theta_i^c = \left(\sum_{\hat{I}_i^*} \hat{W}_{t,i} \hat{W}_{t,i}^{\prime} \right)^{-1} \left( \sum_{\hat{I}_i^*} \hat{W}_{t,i} y_t \right) \, , \\
\text{with} &&  \hat W_{t,i}= \vec (Z_{1t},\hat X_{t,i}) \, , \quad \text{and} \quad \hat X_{t,i}'= Z_t' \hat \Pi_i \, .
\end{eqnarray*}
The Wald test statistic for a common change is defined as:
\begin{equation*}
Wald_T^{c}   \, =  \, T \, \hat  \theta_{vec}^{c\prime} \,  \mathcal R_p^\prime  \,  (\mathcal R_p  \,  \hat G^c  \,  \mathcal R_p^\prime )^{-1}  \,  \mathcal R_p^\prime   \,  \hat \theta^c_{vec},
\end{equation*}
\begin{eqnarray*}
\text{with}
&& \hat  \theta_{vec}^c =\vec(\hat \theta^c_1, \hat \theta^c_2) \, , \quad  \mathcal R_p = (1,-1) \otimes I_p \, , \\
&& \hat G^c = \diag \left[ \hat G_{1}^c, \hat G_{2}^c \right] \, , \quad
\hat G_{i}^c = (\hat A_{i}^c)^{-1} \hat B_{i}^c  (\hat A_{i}^c)^{-1} \, , \\
&& \hat A_{i}^c = T^{-1} \sum_{\hat{I}_i^*} \hat X_{t,i} \hat X_{t,i}^{\prime} \, ,  \quad
\Lambda_{iT} = \diag(T^{1/2} I_{p_1}, T^{1/2}  r_{iT}^{-1} I_{p_2}) \, , \\
\text{and} && \Lambda_{iT}^{-1} \hat B_i^c \Lambda_{iT}^{-1} \inp B_i^c = \avar[ T^{-1/2} \sum_{\hat I_i^*} \Lambda_{iT}^{-1} \hat X_{t,i} u_t ] \, .
\end{eqnarray*}

The following theorem provides the limiting distribution of the above Wald test statistic.

\begin{theo} (Wald test for common change) \label{theo:common} \\
	If $\hat T - [T\nu^0] = \Op(1)$, then:\\
(i) Under $H_0: \theta_1^0=\theta_2^0$, Assumptions \ref{ass:reg1} to \ref{ass:reg-identif} and \ref{ass:reg} to \ref{ass:reg-secondmom all candidates}, $Wald_T^c \ind \chi^2_p$. \\
(ii) Under $H_A: \theta_1^0 \neq \theta_2^0$, Assumptions \ref{ass:reg1} to \ref{ass:reg-identif} and \ref{ass:reg} to \ref{ass:reg-secondmom all candidates},
$$ Wald_T^c =
\left\{ \begin{array}{ll}
\mathcal{O}_p(T) & \text{when $p_1\neq0$ (with exogenous regressors)} \\
\mathcal{O}_p(T/\overline{r}_T^2) & \text{when $p_1=0$ (no exog. regressors) and $\overline{r}_T=\max_i(r_{iT})$.}
\end{array} \right.
$$
\end{theo}

Unlike the test in the previous section, the common change point test is simpler, because it is computed directly at the estimated change point coming from the first-stage; however, note that its the rate of divergence is different in absence of exogenous regressors, since it depends on the weakest identification rate across the two samples $\overline{r}_T$.

\subsubsection{Extension to multiple changes and inference procedure} \label{subs:3stage-inf-procedure}

Both the estimation and the change point detection methods can be extended to multiple changes. For detecting the number of change points, we recommend the sequential testing methods in \cite{HHB}. Since we have shown that the sup-Wald test of zero versus one change in \cite{HHB} remains valid in the presence of near-weak instruments, we conjecture that the sequential tests remain valid as well. Similarly, the TSGMM estimators can be extended to multiple changes in a straightforward way.

We now present a four-stage inference procedure that builds on results developed in Sections \ref{section:stable-RF} and \ref{section:unstable-RF}.

\begin{enumerate}
\item \textbf{First stage} \\
Find the number changes in the first-stage via the methods in \cite{HHB}. Find their location via the methods in \cite{HHB} or \cite{AB17} and collect them in the set $\mathcal B_{S1}$.  Use  $\mathcal B_{S1}$ to partition the sample, and construct $\hat{X}_t$ by OLS (over sub-samples, if $\mathcal B_{S1}$ is not empty, and over the full-sample, if $\mathcal B_{S1}$ is empty).

\item \textbf{Second stage} \\
Partition the sample using $\mathcal B_{S1}$, and work over the associated sub-samples. Find the number of changes in the equation of interest for each of these sub-samples, via a sequential version of the sup-Wald test presented in Section \ref{subsection:sup-wald test}. Estimate them via the change point estimator of Section \ref{section:stable-RF}, extended to multiple changes, and collect these changes in the set $\mathcal B_{S2,NC}$.

\item \textbf{Third stage: common changes.} \\
Partition the sample into sub-samples according to $\mathcal B_{S1,NC}$ again.  Test in each of these for common changes to both the first and the second-stages, via the Wald test in Section \ref{subs:test-common-change}. Collect the common changes in the set $\mathcal B_{S2,C}$, and obtain $B_{S2} = B_{S2,NC} \ \cup B_{S2,C}$.

\item \textbf{Forth stage: estimation.} \\
Partition the sample according to $B_{S1}$: each resulting sub-sample first-stage is stable. If there are second-stage changes in any of these sub-samples according to $\mathcal B_{S2}$, use the TSGMM estimators as described in Section \ref{section:stable-RF}; otherwise, use the standard full-sample GMM estimator.
\end{enumerate}

Of course, the multiple pre-testing steps can affect the procedure, and to safeguard against this, we suggest using a Bonferroni correction.

\section{Efficient estimators of the first-stage parameters} \label{section:efficient RF}

This section shows that, in the presence of second-stage change, the TSGMM estimators of the first-stage $\hat \Pi_{TSGMM,vec}$ introduced in Section \ref{section:stable-RF} are also more efficient than the full-sample OLS estimates and the split-sample OLS estimates that ignore the second-stage change point. To formalize this result, consider the following stable first-stage, where we are interested in estimating $\Pi^0$:
\begin{equation*}
X_t = Z_t' \Pi^0+ v_t \, .
\end{equation*}
For simplicity, we consider one endogenous regressor $X_t$, no additional exogenous regressor ($p_1=0, p_2=1$), and we impose strong identification, i.e. $r_T=1$. Therefore, we drop the ``vec'' subscript on all estimators.

The parameter $\Pi$ is stable, but we allow for second-stage changes, as well as potential changes in $\verb'Var'(v_t Z_t)$, $\verb'E'(Z_t Z_t')$, or in all, at $T^0$, which is assumed known for simplicity. Notice that a change in $\verb'Var'(v_t Z_t)$ at $T^0$ means that $\verb'Var'(v_t Z_t)$ changes once at $t=T^0$: as a result, Assumption \ref{ass:reg-secondmom}(ii) is violated. A change in $\verb'E'(Z_t Z_t')$ implies that $\verb'E'(Z_t Z_t')$ changes once as $t=T^0$, so Assumption \ref{ass:reg-secondmom}(i) is violated.

We  introduce a third estimator that ignores the second-stage moment conditions and uses the OLS sub-sample moment conditions before and after the change - $g_{T,2}(\Pi;\lambda_0)$, defined in Section \ref{section:stable-RF}. It corresponds to the optimal (two-step) GMM estimator that uses the $2q$ moments $g_{T,2}(\Pi;\lambda_0)$ to estimate $q$ parameters and we denote it $\hat \Pi_{GMM}$. The theorem below shows that $\hat \Pi_{TSGMM}$ is at least as efficient as the other two estimators.

\begin{theo} (Efficiency of first-stage estimators) \label{theo:eff-ls}\\
(i) Under Assumptions \ref{ass:reg1} to \ref{ass:reg-identif},
$\sqrt{T}(\hat \Pi - \Pi^0)$,
$\sqrt{T}(\hat \Pi_{GMM} - \Pi^0)$,
and
$\sqrt{T}(\hat \Pi_{TSGMM} - \Pi^0)$ are asymptotically normally distributed with mean zero and respective asymptotic variance-covariance matrices:
\begin{eqnarray*}
V_{OLS,\Pi} &=& (Q_1+Q_2)^{-1} (S_{v,1}+S_{v,2}) (Q_1+Q_2)^{-1} \\
V_{GMM,\Pi} &=& (Q_1 S_{v,1}^{-1} Q_1 +Q_2 S_{v,2}^{-1} Q_2)^{-1} \\
V_{TSGMM, \Pi} &=& (V_{GMM,\Pi}^{-1} +  \H_*' \mathcal E_*^{-1/2} \mathcal M_{\mathcal J_*} \mathcal E_*^{-1/2} \H_*)^{-1},
\end{eqnarray*}
where $\mathcal E_* =  \diag(S_{u,1} - S_{uv,1}' S_{v,1}^{-1} S_{uv,1} \ , \ S_{u,2} - S_{uv,2}' S_{v,2}^{-1} S_{uv,2})$, \\
and $\mathcal M_{\mathcal J_*} = I - \mathcal J_* (\mathcal J_*' \mathcal J_*)^{-1} \mathcal J_*',  \mathcal J_* =\mathcal E_*^{-1/2} \Gamma_1, \mathcal H  = \diag( S_{uv,1}' S_{v,1}^{-1},  S_{uv,2}' S_{v,2}^{-1})  \ \Gamma_2$. \\
(ii) Under Assumptions  \ref{ass:reg1} to \ref{ass:reg-secondmom}: $V_{TSGMM,\Pi} \leq V_{GMM,\Pi} \leq V_{OLS, \Pi}$. \\
(iii) Under Assumptions  \ref{ass:reg1} to \ref{ass:cond-hom}: $V_{TSGMM, \Pi} \leq V_{GMM,\Pi} = V_{OLS, \Pi}$.
\end{theo}

The restriction of a common first-stage delivers not only more efficient GMM estimators of the parameters of interest, but also of the first-stage parameters. Since the first-stage is part of a system of equations linked through correlation between $u_t$ and $v_t$, the OLS estimators of the first-stage are no longer the most efficient in presence of changes in the second-stage, or in second moments of the instruments or in the variance of the OLS moment conditions. It is important to note that OLS estimators remain the most efficient among the class of estimators that ignore the second-stage, and as long as no second moments change over the sample. Consequently, our results do not conflict with the Gauss-Markov Theorem or classical results on asymptotic efficiency of OLS.

\section{Monte-Carlo study} \label{section:MC}
In this section, we examine by simulation the small sample behavior of the split-sample GMM estimators, TS2SLS and TSGMM estimators when the change point is known, when it is unknown and needs to be estimated, and when pre-testing for a change point.
\subsection{Data generating process}
We consider the following model:
\begin{align}
	y_t&=\begin{cases}
		Z_{1t}'\theta_{1,z}^0+X_t'\theta_{x,1}^0+\sigma_t u_t,\quad,\quad t\leq \lfloor T\lambda^0\rfloor\\
		Z_{1t}'\theta_{2,z}^0+X_t'\theta_{x,2}^0+\sigma_t u_t,\quad,\quad t> \lfloor T\lambda^0\rfloor
	\end{cases} \label{eq:MC1}\\
	X_t&'=Z_t'\Pi+v_t, \label{eq:MC2}
\end{align}
with $p$ explanatory variables $W_t$, $p_1$ exogeneous variables $Z_{1t}$, $p_2$ endogeneous variables $X_t$, and $p$ parameters of interest $\theta^{0'}$. $Z_t$ is a vector of $q$ instruments, including the exogeneous regressors $Z_{1t}$. $\sigma_t$ determines whether the structural errors are homoskedastic or heteroskedastic.

We set $p=2$ with $p_1=p_2=1$. The exogenous regressor $Z_{1t}$ is a constant, and the number of external instruments $n_{IV}=q-1$ are set to 1 or 4. The sample sizes are $T=400$ and $T=800$, and we implement $N=1,000$ repetitions. The true change fraction is $\lambda^0=0.4$, thus the (true) change location is either $\lfloor T\lambda^0\rfloor=160$ or 320 depending on the sample size. The change size is assumed to be equal to one, and the true parameters are
\begin{align*}
	\theta^{0}_{vec}=\text{vect}(\theta_{1,z}^0,\theta_{1,x}^0,\theta_{2,z}^0,\theta_{2,x}^0)=(0,0,1,1)',\quad\Pi=(1,1),
\end{align*}
where $\theta_2^0-\theta_1^0=1$ is referred to as the ```change size'' in the Appendix Section A Results.
The errors are jointly generated as follows:
\begin{align*}
	\left(\begin{array}{c}u_t\\v_t\end{array}\right) \sim \mathcal{N}\left( \left(\begin{array}{c}0\\0\end{array}\right),\left[\begin{array}{cc}1&\rho\\\rho & 1\end{array}\right]\right)
\end{align*}
with correlation coefficient $\rho=-0.5$. The RMSE is computed using the sum of the average Monte carlo bias and of the average asymptotics-based standard deviation. \footnote{This standard deviation is calculated by plugging in the Appendix, Definition 1, consistent and heteroskedasticity-robust estimators of the quantities that enter the asymptotic variance.} as
\[
RMSE=\sqrt{\left(\frac{1}{N}\sum_{n=1}^N \text{Bias}(\hat\theta^n_{j,i})\right)^2+\left(\frac{1}{N}\sum_{i=1}^N\text{Asympt.Std}(\hat\theta^n_{j,i})\right)^2}
\]
for $n=1,\dots,N$ repetitions, where $\hat \theta^n_{i,j}$ is short-hand for the estimators in repetition $n$ of the coefficients on the endogenous regressors in samples $i=1,2$ and of type $j \in \{GMM, TS2SLS, TSGMM\}$. All GMM estimators are computed with a 2SLS first-step estimator based on their respective moment conditions.

HOM corresponds to conditional homoskedasticity and thus $\sigma_t=1$. HET1 corresponds to conditional heteroskedasticity given by $\sigma_t^2=(1+(\sum_t^T Z_t)^2)/2$, and HET2 to a GARCH(1,1) model with $\sigma_t^2=0.1+0.6\epsilon_{t-1}^2+0.3\sigma_{t-1}^2$, where $\epsilon_t=\sigma_t u_t$. Under HET2, we employ a burn-in sample of 100 observations for $T=400$, and 200 observations for $T=800$.

The data generating process satisfies Assumptions \ref{ass:reg1}-\ref{ass:reg-secondmom} and \ref{ass:reg}-\ref{ass:reg-secondmom all candidates}, and also Assumption \ref{ass:cond-hom} in the HOM case. Condition \ref{eq:in1} in Theorem \ref{theo:eff} holds with equality, so the asymptotic distributions of the TS2SLS and of the split-sample GMM are the same under HOM, as shown in Theorem \ref{theo:eff}, and the ranking of the two estimators based on asymptotic variances is unclear under HET1-HET2. However, condition \ref{eq:in2} holds as stated in Theorem \ref{theo:eff}, so TSGMM is asymptotically strictly more efficient than both the TS2SLS and the split-sample GMM estimators under HOM, and at least as efficient as the other two estimators under HET1-HET2.

\subsection{Results with a known change point}

We first consider the case where the DGP displays one change point which is fully known: that is, both the existence of the change and its location are known. Estimation results for GMM, TS2SLS, and TSGMM are reported in Tables \ref{tab:hom}, \ref{tab:het1}, and \ref{tab:het2} with sample sizes $T=400$ and $800$ under HOM, HET1 and HET2, respectively. We report average Monte-Carlo bias, Monte-Carlo standard deviations, as well as asymptotics-based standard deviations and RMSE of estimators $\theta^n_{i,j}$ (for simplicity we drop the superscript $n$ in all tables). In addition, we report the average length of a 95\% confidence interval computed with the asymptotics-based standard deviations, and the associated coverage rates.

Overall, the three estimation procedures are well-behaved, and fairly close to each other: all display small biases, and coverage rates close to the nominal 95\%. In addition, Monte-Carlo standard deviations are close to their asymptotics-based counterparts, but they are slightly higher, so there are benefits to using the asymptotic formulas in Appendix Definition 1. The TS2SLS estimator does not uniformly dominate the split-sample GMM estimators in terms of efficiency: the two asymptotics-based standard deviations are close as predicted by Theorem \ref{theo:eff} in the HOM case, but with more instruments, the asymptotic standard deviations of the TS2SLS becomes larger than that of the split-sample GMM in both HOM and HET cases. In contrast, the TSGMM consistently display lower standard deviations compared to both TS2SLS and split-sample GMM, as predicted by Theorem \ref{theo:eff} and implied by the asymptotic variance formulas. The bias of the TSGMM is also small, and as a result, TSGMM has also consistently lower RMSE than the other two procedures, while the empirical coverage of the confidence intervals is still close to the nominal size in most cases.

\subsection{Results with an unknown change point}

Next, we assume the researcher knows that there is one change point, but doesn't know its location and estimates it via the method proposed in Section \ref{section:unstable-RF}. To ensure there are enough observations in each sub-sample, the candidate change points are taken to be between $[0.15T]$ and $[0.85T]$.

\begin{figure}[h]
\centering
\includegraphics[trim=0cm 0cm 0cm 0cm, clip=false, width=0.49\textwidth]{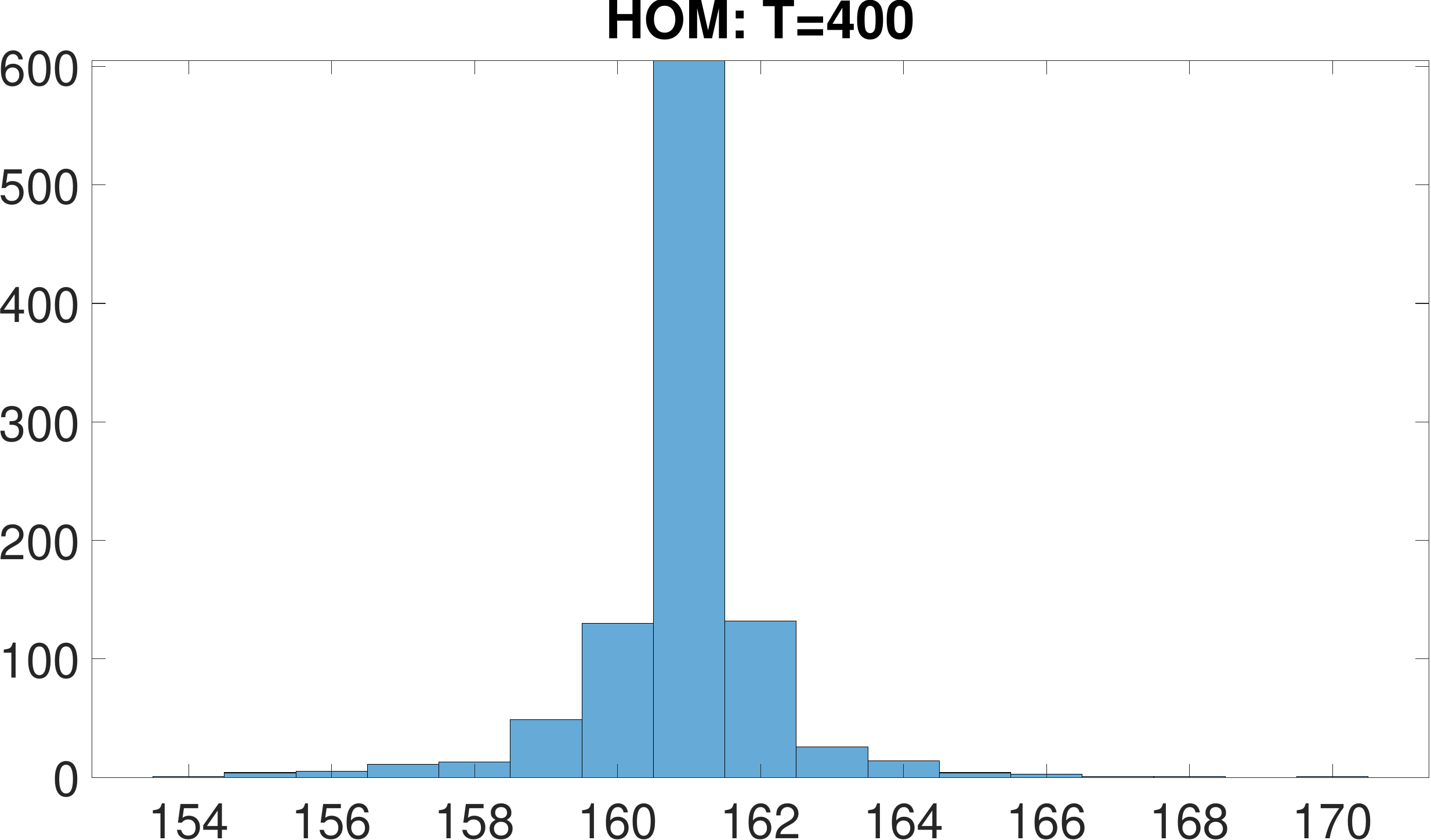}
\includegraphics[trim=0cm 0cm 0cm 0cm, clip=false, width=0.49\textwidth]{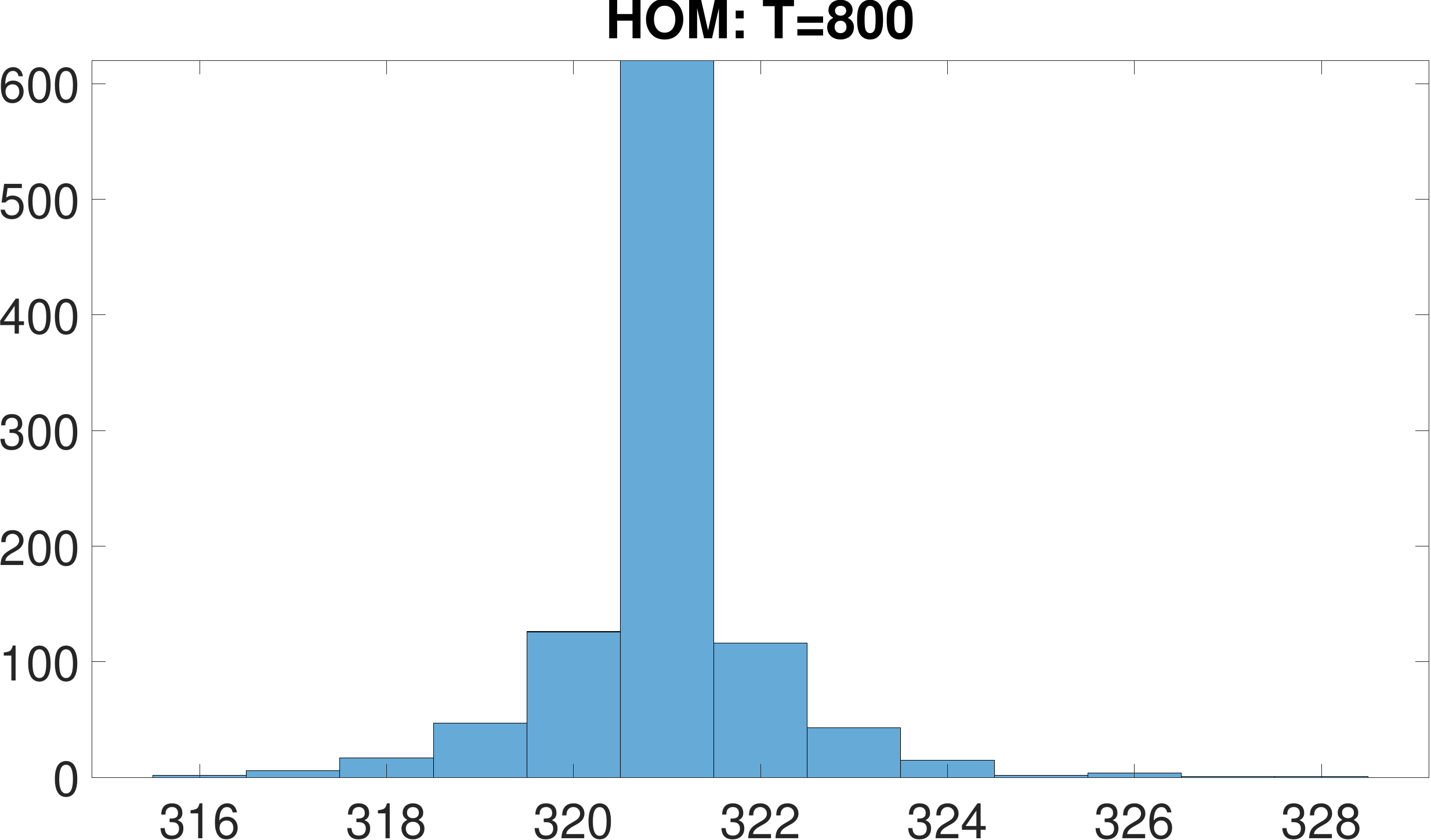}\\
\includegraphics[trim=0cm 0cm 0cm 0cm, clip=false, width=0.49\textwidth]{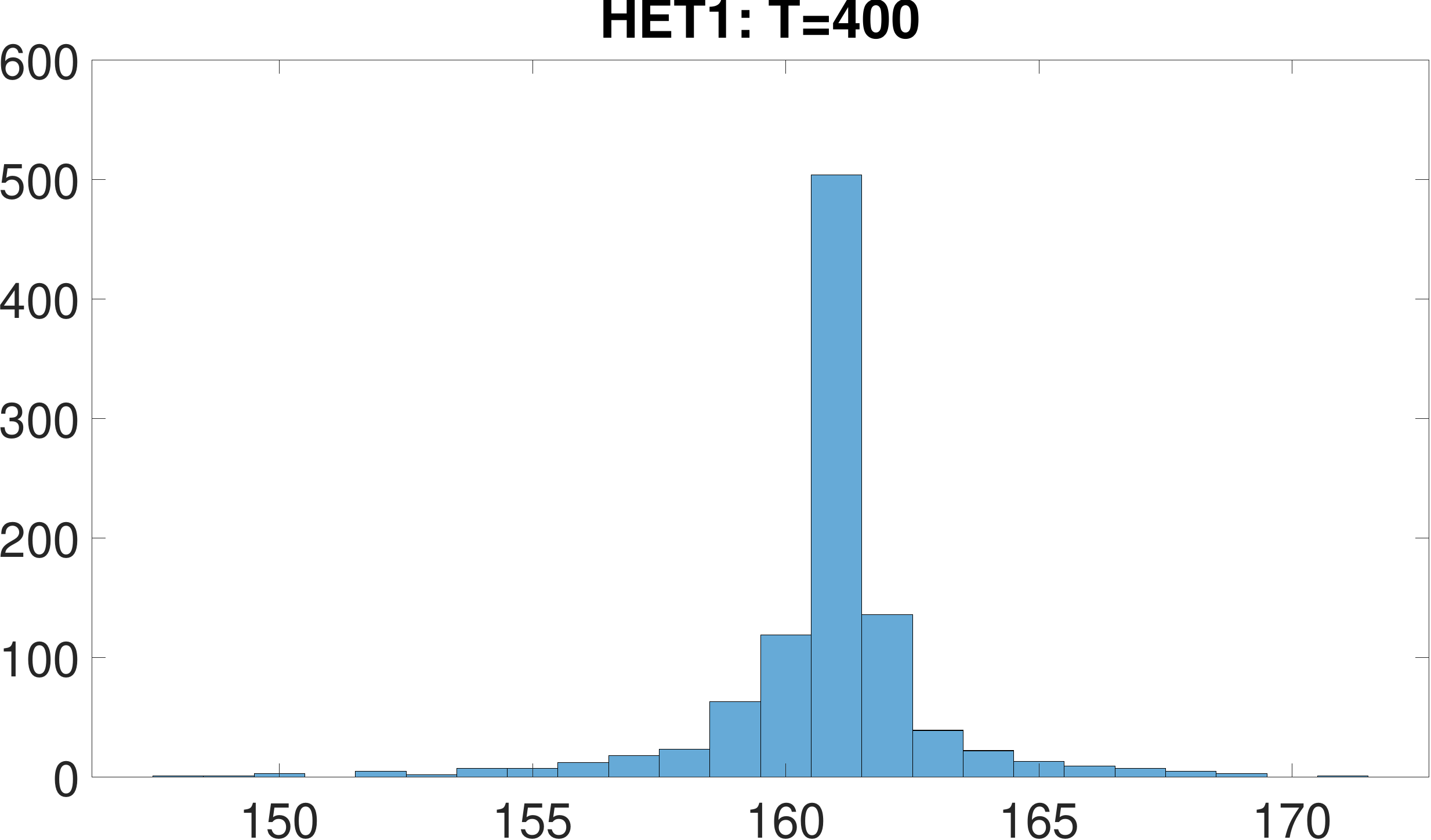}
\includegraphics[trim=0cm 0cm 0cm 0cm, clip=false, width=0.49\textwidth]{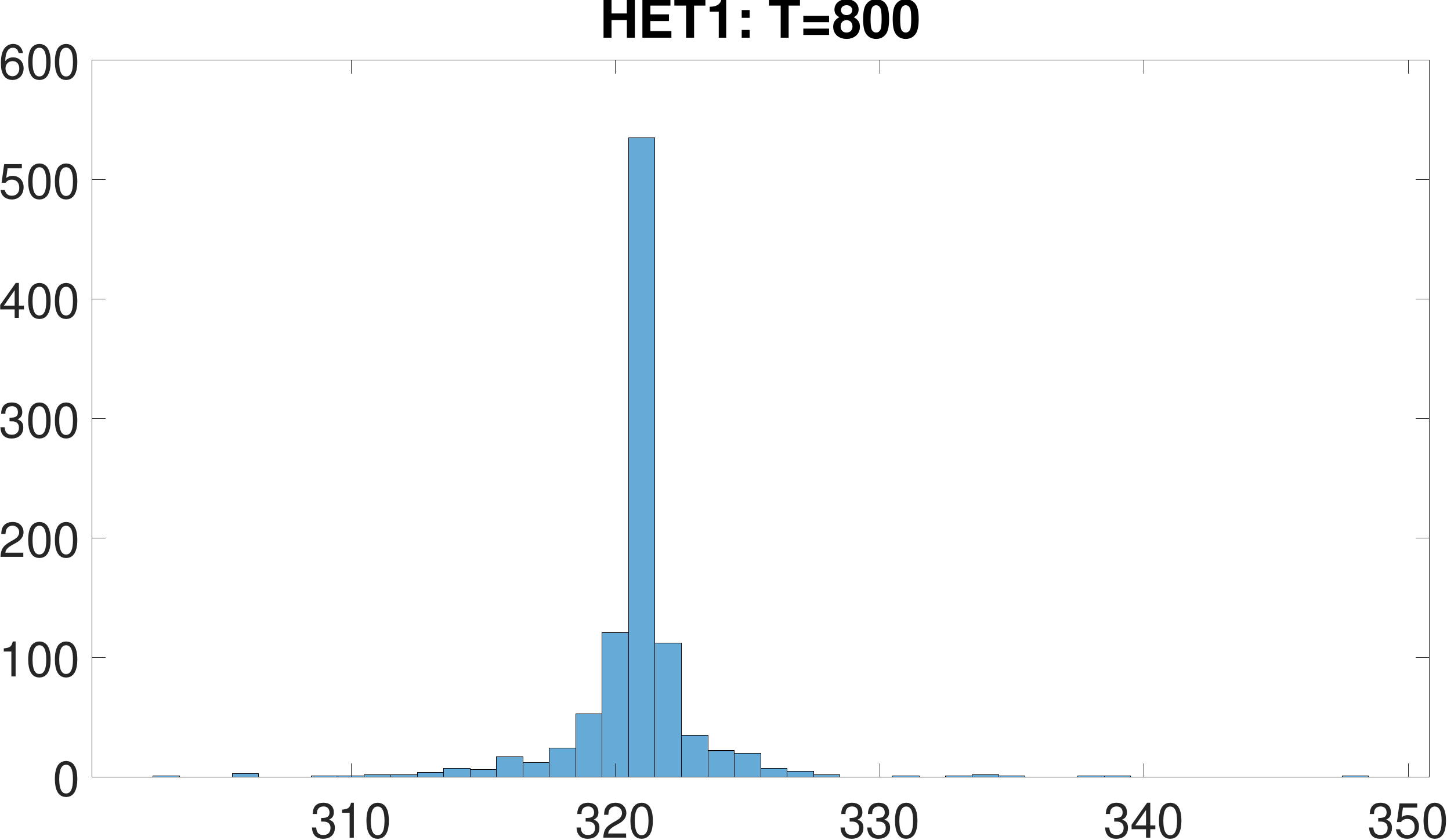}\\
\includegraphics[trim=0cm 0cm 0cm 0cm, clip=false, width=0.49\textwidth]{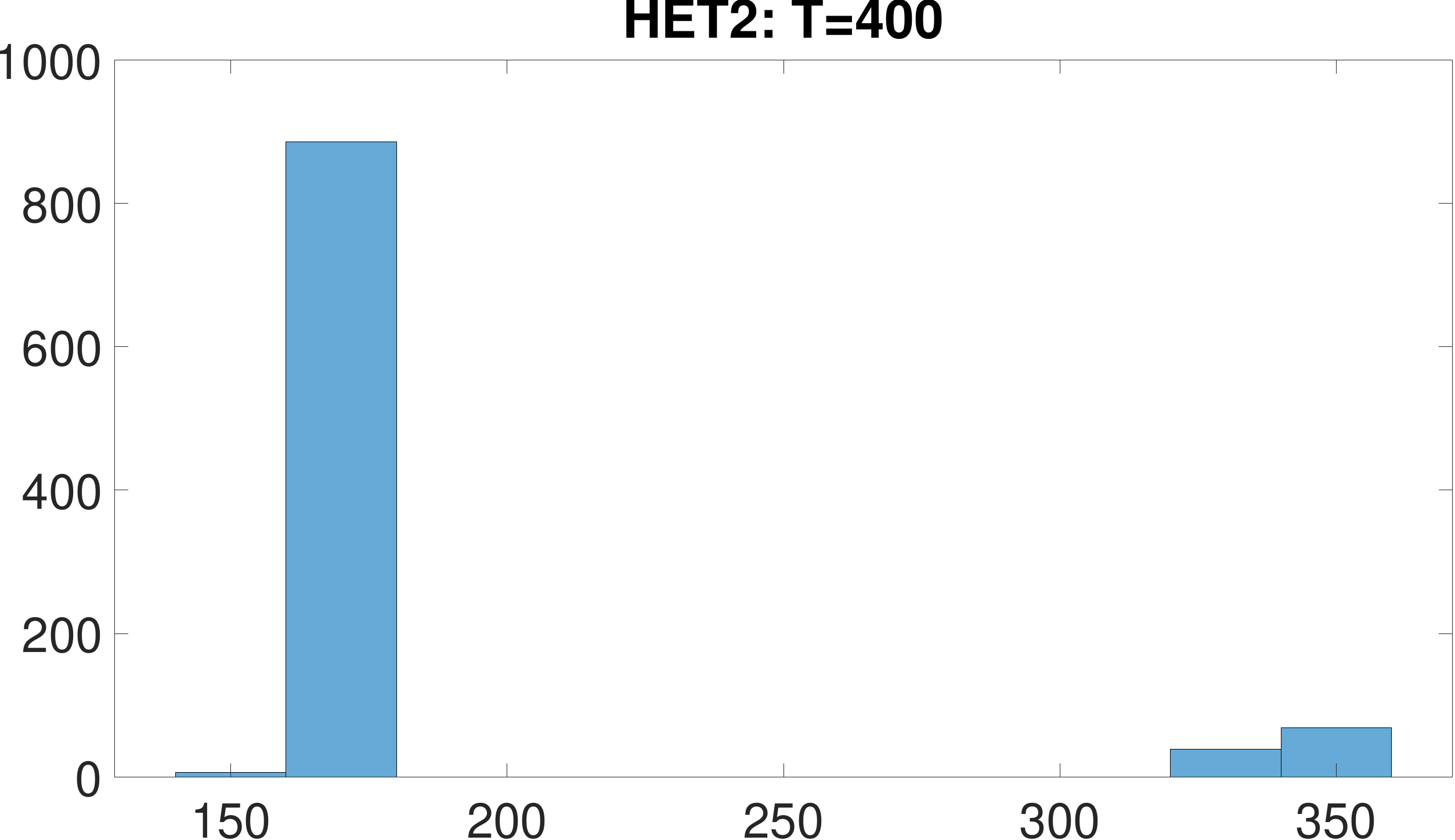}
\includegraphics[trim=0cm 0cm 0cm 0cm, clip=false, width=0.49\textwidth]{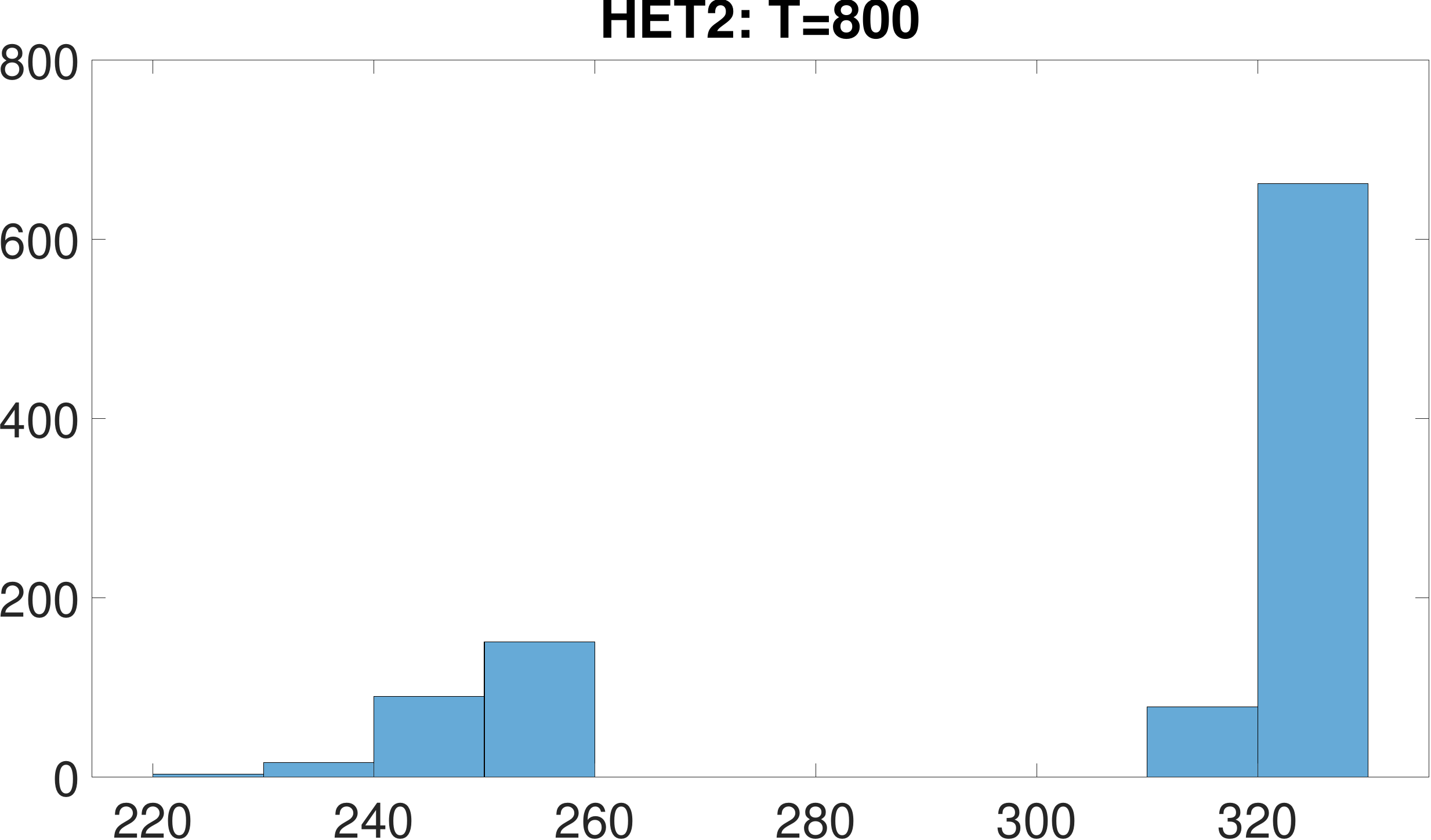}
		\caption{Histograms of the estimated change point for $\lambda^0=0.4$ and $\text{$n_{IV}$}=4$.}
		\label{fig:esthet1}
\end{figure}

In Figure \ref{fig:esthet1}, we display the histograms of estimated change points with sample sizes $T=400$ and $800$ under HOM (top row), HET1 (middle row), and HET2 (bottom row). Note that the change points are estimated accurately in the HOM and HET1 cases, and are less accurate in the HET2 case due to the GARCH(1,1) process chosen to be close to a unit root.

Estimation results are reported in Tables \ref{tab:homb}, \ref{tab:het1b}, and \ref{tab:het2b} with sample sizes $T=400$ and $800$ under HOM, and HET1-HET2. We also report the Monte-Carlo mean of the associated estimated change points. Due to estimation error from the change points, the differences between the three estimators are smaller, but the TSGMM continues to consistently display lower standard deviations and RMSE than the other two estimates.

\begin{figure}[h]

\centering
\includegraphics[trim=0cm 0cm 0cm 0cm, clip=false, width=0.65\textwidth]{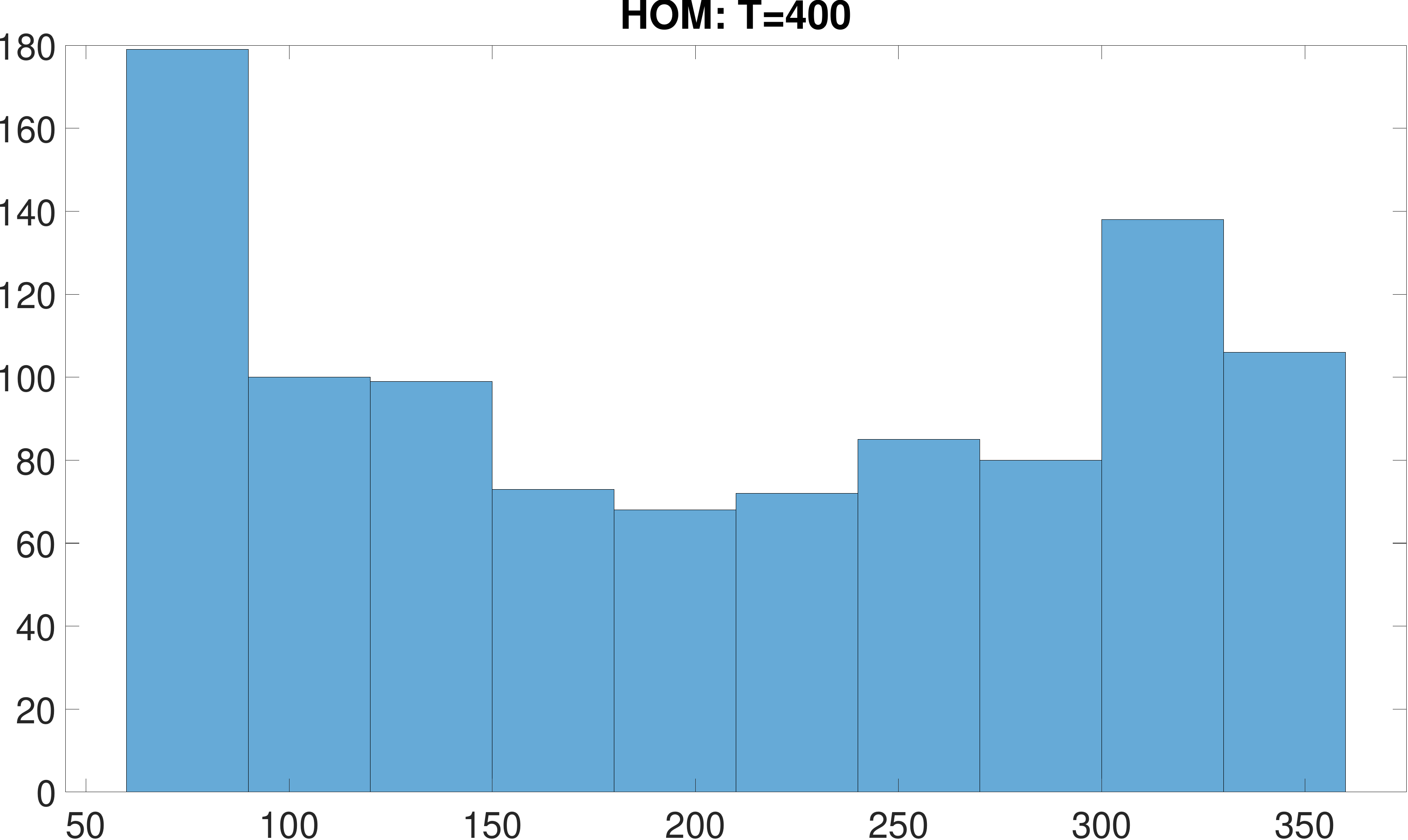}\\
\includegraphics[trim=0cm 0cm 0cm 0cm, clip=false, width=0.65\textwidth]{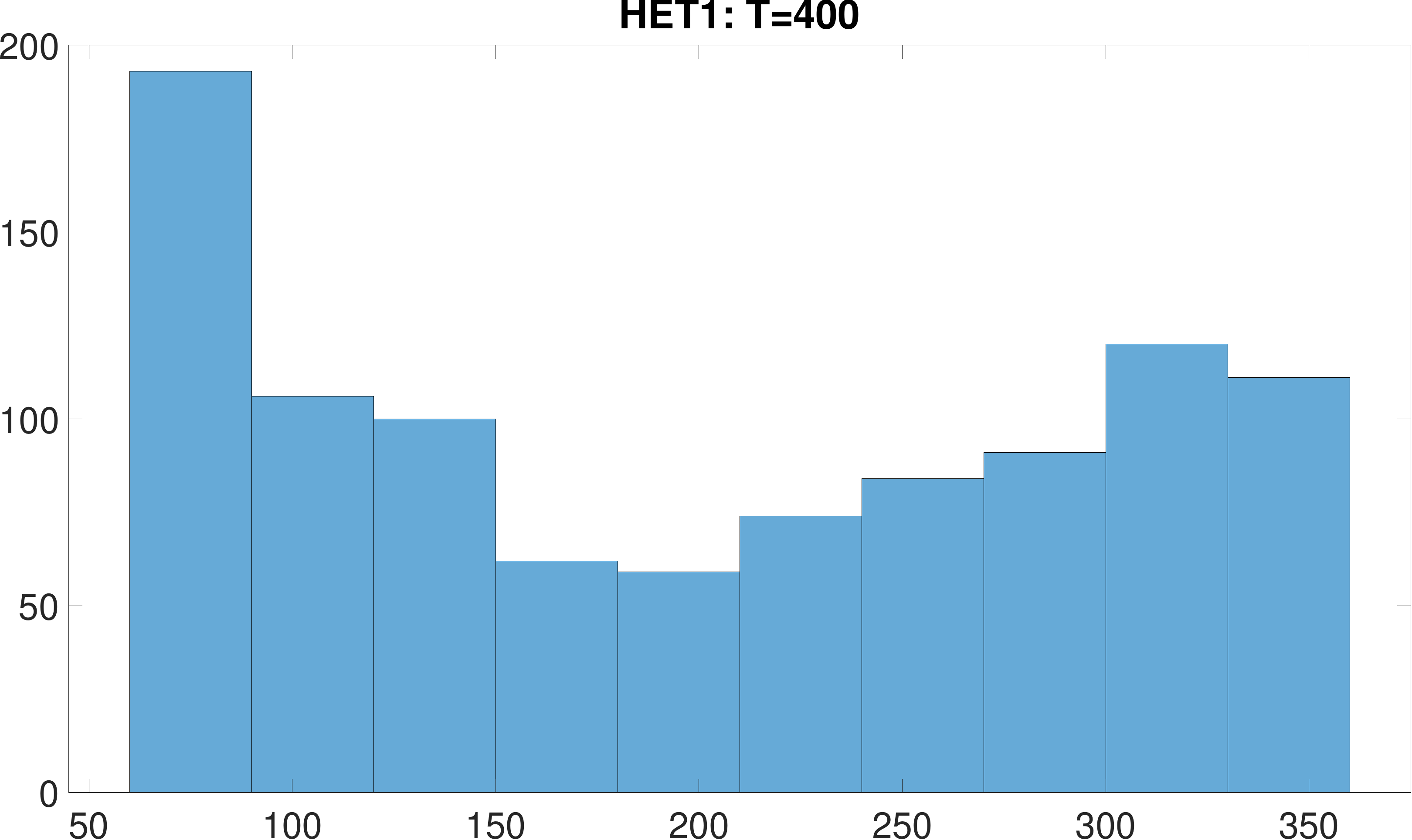}\\
\includegraphics[trim=0cm 0cm 0cm 0cm, clip=false, width=0.65\textwidth]{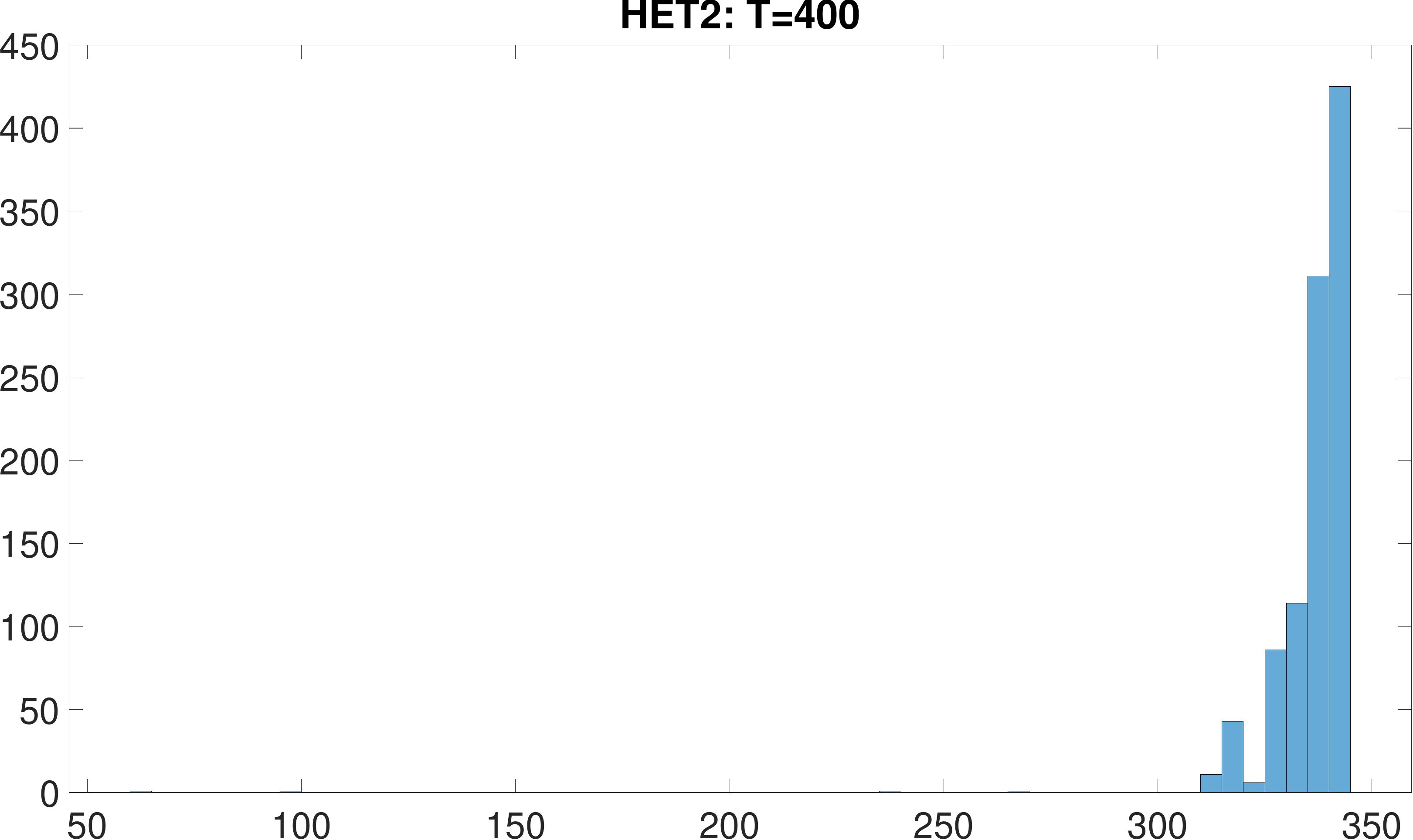}
	\caption{Histograms of the estimated change point, when the true DGP has no change points. Here $T=400$ and $n_{IV}=4$}
	\label{fig:histbob}
\end{figure}

As a robustness check, we also consider the case where there is no change point in the DGP, but we estimate a change point, and we set $T=400$. Estimation results are reported in Table \ref{tab:bob}, while histograms of associated estimated change points are displayed in Figure \ref{fig:histbob}. In all cases, the TSGMM still has lower standard errors and RMSE than the other estimates, and its coverage slightly deteriorates in the HET cases, which is to be expected due to more moment conditions than the other estimators and a small sample size.

\subsection{Detection of change point}

Finally, we present results related to the detection of the change point. Table \ref{tab:ttest} displays the probability of detecting a change, given that the DGP has one change point, using the sup-Wald test in Section \ref{subsection:sup-wald test}. The null hypothesis of the test is that there is no change point in the data. The critical values of limiting distribution of the test are from \cite{baiperron2004}. As can be seen from Table \ref{tab:ttest}, the empirical size of the test is close to the nominal size, and the power is also very good as long as the magnitude of the change in both coefficients is large enough.

In Table \ref{tab:avgspec}, we pre-test for a change point, and subsequently estimate the parameters. If the test rejects the null, we estimate the location of the change point, and the parameters over the two sub-samples. If the test fails to reject, we estimate the parameters over the whole sample. We report Monte-Carlo bias, standard deviation, as well as asymptotic standard deviation (computed using the asymptotic theory), and RMSE: each measure is obtained as a weighted average of the corresponding quantity computed over each sub-sample. We see that pre-testing does not affect the efficiency of the TS2GMM estimators, which still displays lower standard deviations and RMSE. However, it affects coverage for all estimators in small samples such as $T=400$, especially in the case HET2, where the change point is estimated less accurately.

\section{Estimation of the NKPC in an open economy} \label{section:NKPC}

The New Keynesian Phillips Curve (NKPC) played an important role in monetary policy analysis: in an open economy, it relates inflation to the movements of business cycles and expected future exchange rates. In this context, one usually assumes instantaneous import price adjustment and full transformation of exchange rate movements into consumer prices. However, it has been widely documented that the exchange rate pass-through - that is, the real marginal cost of importers - into consumer and import prices is incomplete, at least for US and other developed countries -\cite{campa2005}. Accordingly, we focus here on estimating the NKPC under imperfect exchange rate pass-through (as in \cite{monacelli2005}) which relates inflation to a domestic driving force variable (e.g. unemployment rate) in addition to the pass-through rate measured as the difference between the exchange rate and the terms of trade:
\begin{equation}
	\pi_t = \gamma_f \pi_{t+1}^e+\kappa_u u_t +\kappa_\Psi \Psi_t +\beta_0
\end{equation}
where $\pi_t$ is the consumer price inflation, $\pi_{t+1}^e$ is the expected future inflation, $u_t$ is the unemployment rate, and $\Psi_t$ is the pass-through rate.

Empirically, there does not seem to be a consensus on the exact definition of the exchange rate pass-through, and whether - and how - it might affect the Phillips curve. In the context of an incomplete pass-through, an increase in the law of one price gap, i.e. $\kappa_\Psi$, causes an increase in inflation, similar to the one caused by a cost-push shock. Hence, an incomplete exchange rate pass-through could have relevant consequences for monetary policy. At the same time, there is also some evidence suggesting that the exchange rate pass-through has been declining, which brings into question whether it should be taken into account in any policy consideration (see e.g. \cite{taylor2000}).

Using monthly US data from 1990 to 2020, we find strong support for the NKPC with imperfect pass-through. The results in the next section suggest that the exchange rate pass-through plays a statistically significant role which is comparable, in magnitude, to that of unemployment - at least over part of the sample. We also find that relationship between inflation and exchange rate pass-through exhibits important time variation.

\subsection{Main empirical analysis} \label{subsection:estim procedure}
Our main specification considers the following NKPC model with imperfect pass-through,
\begin{equation}\label{eq:1}
	\pi_t = \gamma_f \pi_{t+1}^e+\gamma_1 \pi_{t-1}+\gamma_2\pi_{t-2}+\kappa_u u_t +\kappa_\Psi \Psi_t +\beta_0
\end{equation}
Notice that two lags of inflation are included as controls. To control for potential endogeneity of (future) expected inflation and the unemployment rate, we consider two first-stages with the following four additional instruments: one lag of expected inflation, unemployment rate, interest rate spread, and terms of trade. All variables are first de-trended using the Hodrick-Prescott filter, and we only consider their cyclical component. Our sample size consists of monthly observations over the period 1983m9-2020m4, for a total of 440 observations. The data is:
\begin{itemize}
	\item inflation $\pi_t$: \href{https://fred.stlouisfed.org/series/CPIAUCSL}{Consumer Price Index for All Urban Consumers: All Items in U.S. City Average} , Percent Change from Year Ago, Monthly, Seasonally Adjusted.
	\item expected inflation $\pi_{t+1}^e$: \href{https://fred.stlouisfed.org/series/MICH}{Michigan Consumer Survey}, expected inflation one-year ahead.
	\item unemployment rate $u_t$: \href{https://fred.stlouisfed.org/series/UNRATE}{Unemployment Rate}, Percent Change from Year Ago, Monthly, Seasonally Adjusted.
	\item pass-through rate $\Psi_t$: $q_t-(1-\gamma)s_t$, where $\gamma$ is assumed to be 0.25, following \cite{gali2005} and \cite{syed2022}. The pass-through rate is the difference between the real exchange rate $q_t$ and the terms of trade $s_t$, with $\gamma$ indicating the degree of openness of the economy.
	\item real exchange rate $q_t$: \href{https://stats.oecd.org/index.aspx?queryid=6779#}{US Exchange Rate: Relative Consumer Price Indices}, index with year 2010 as base 100, monthly.
	\item terms of trade $s_t$ (partially interpolated): ratio between \href{https://fred.stlouisfed.org/series/IQ}{Export Price Index (All Commodities)} and \href{https://fred.stlouisfed.org/series/IR}{Import Price Index (All Commodities)}. Both variables are available monthly, starting from 1982-09-01 and 1983-09-01, respectively. However, they both have missing values for the first few years, hence we use cubic splines to interpolate the missing values.
	\item interest rate spread: difference between \href{https://fred.stlouisfed.org/series/DTB3}{3-Months Treasury Bill Secondary Market} and \href{https://fred.stlouisfed.org/series/DGS5}{Market Yield on U.S. Treasury Securities at 5-Year Constant Maturity}.
\end{itemize}

\begin{figure}[h!]	

\includegraphics[trim=0cm 0cm 0cm 0cm, clip=false, width=0.49\textwidth]{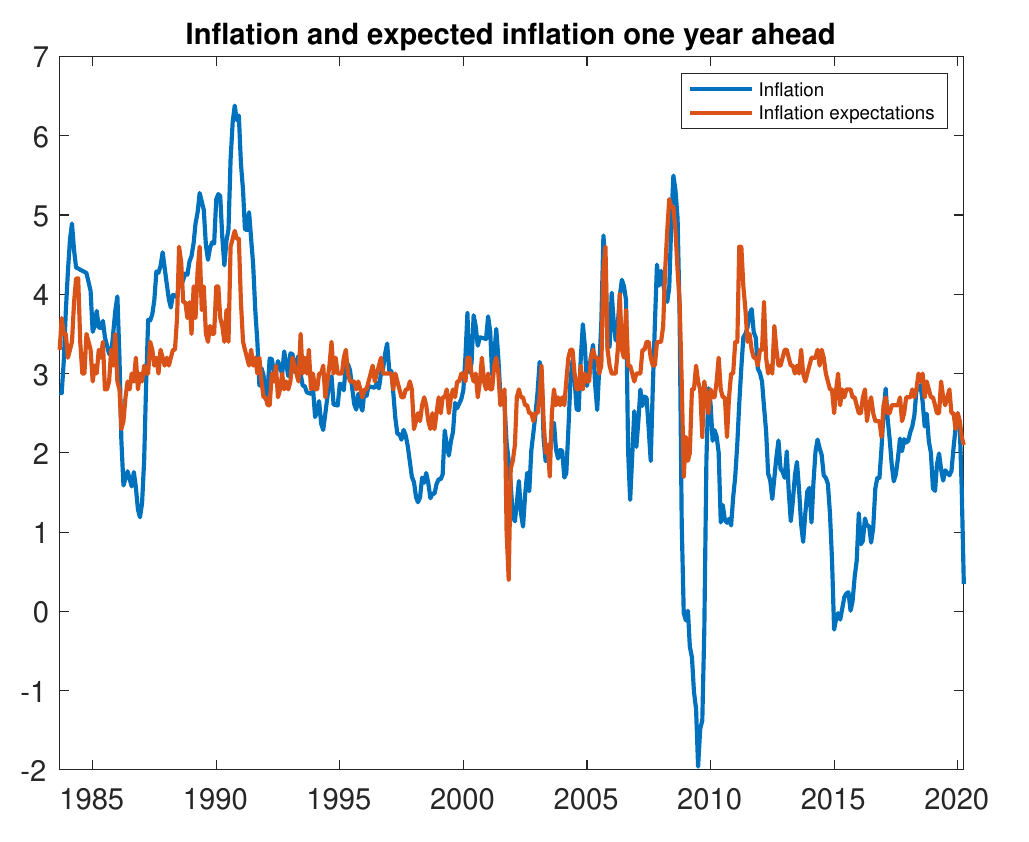}
\includegraphics[trim=0cm 0cm 0cm 0cm, clip=false, width=0.51\textwidth]{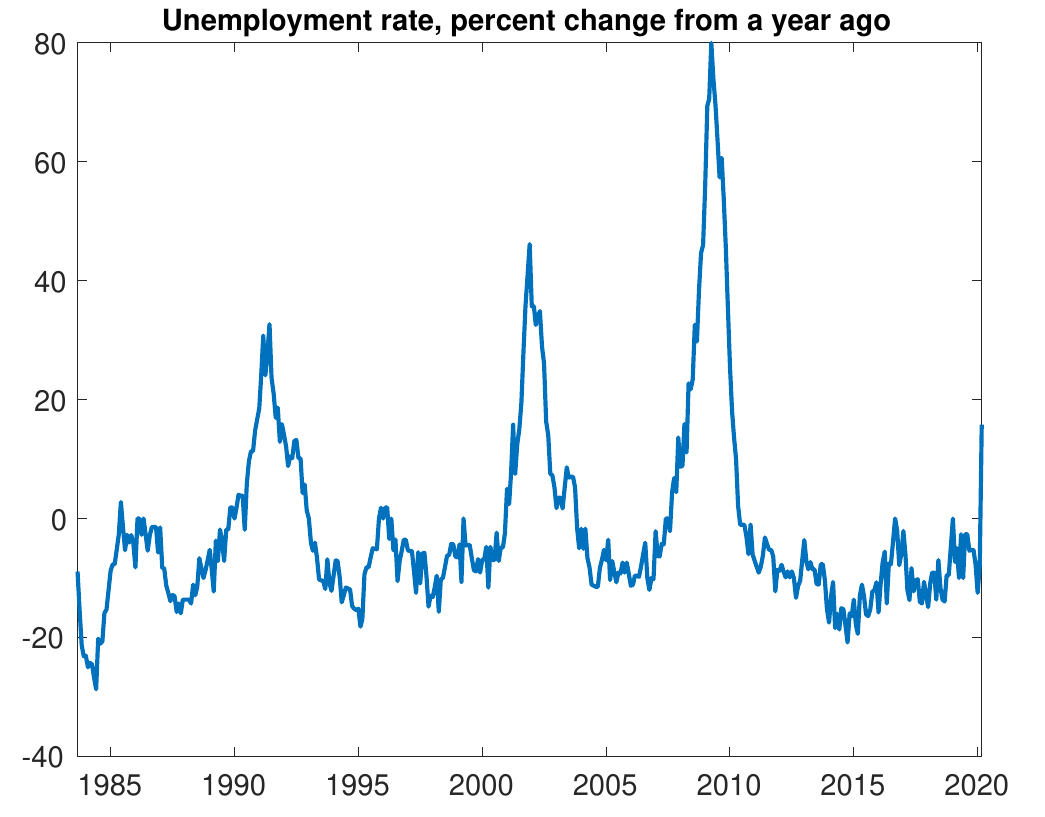}\\
\includegraphics[trim=0cm 0cm 0cm 0cm, clip=false, width=0.50\textwidth]{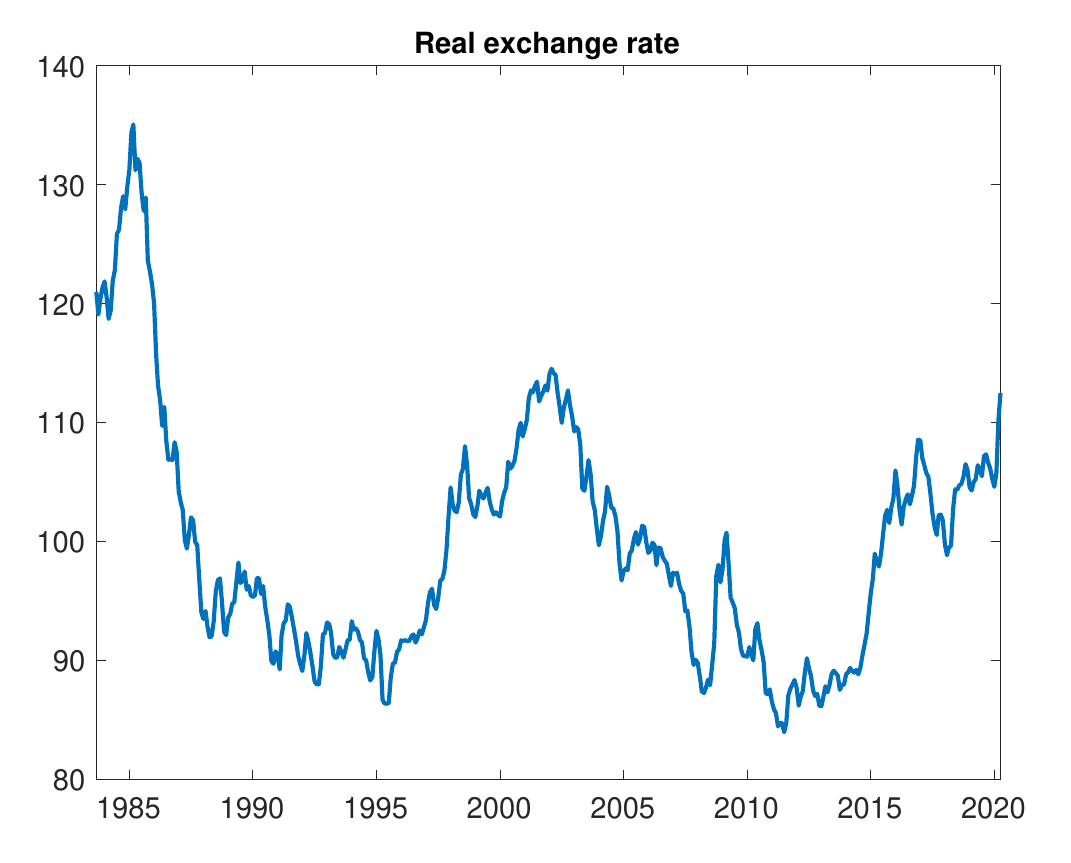}
\includegraphics[trim=0cm 0cm 0cm 0cm, clip=false, width=0.50\textwidth]{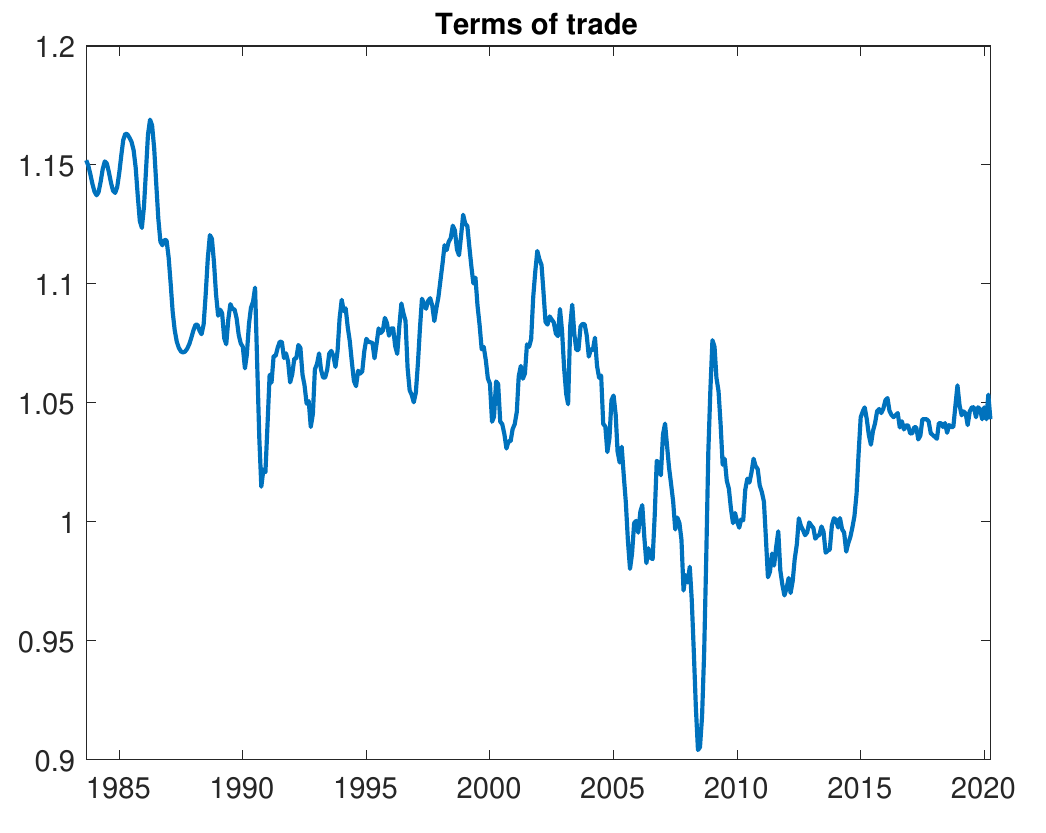}
		\caption{Time series of inflation and expected inflation (top left), unemployment (top right), real exchange rate (bottom left), and terms of trade (bottom right).}
	\label{fig:plot}
\end{figure}

Figure \ref{fig:plot} plots the time series of inflation and expected inflation, unemployment rate, terms of trade, and real exchange rate.

We first estimate the change points in the two first-stage equations which we call for simplicity reduced forms henceforth (RFs), i.e. expected inflation and unemployment rate, using the Stata package \cite{ditzen2021} that implements \cite{baiperron2004}'s algorithm. Then, we consider the segments over which both RFs are stable, and test in each for a change point in the equation of interest using the sup-Wald test described in Section \ref{subsection:SF change}; if the test rejects, wee estimate the change points in \eqref{eq:1} over each such segment using the estimator described in Section 3.\footnote{Recall that both the sup-Wald test employed and the change-point estimator is the same as in \cite{HHB}, and in this paper we showed that they are valid in the presence of near-weak identification.} Once the change location has been estimated, we estimate the model parameters in \eqref{eq:1} using either TS2SLS, split-sample GMM, or TSGMM. The results are displayed in Tables \ref{tab:RF_changes} to \ref{tab:1983-2001}.

In Table \ref{tab:RF_changes}, we report the change point estimation results obtained over each RF, respectively. The RF for expected inflation has one change at 2009m3 (significant at 1\%), while the RF for unemployment rate has two changes, respectively, at 2001m10 and 2008m12. Given the proximity of the first change points obtained in each reduced form, it seems plausible and reasonable to interpret them as the very same change. Accordingly, we consider the following three segments over which both reduced forms are stable: \\
(i) 1983m9 -- 2001m10; (ii) 2001m11 -- 2009m3; and (iii) 2009m4 -- 2020m4.

Table \ref{tab:sf_change} reports the change point estimation results obtained in the structural equation (equation of interest) using the proposed sup-Wald test in Section \ref{subsection:SF change}; this analysis is conducted over each of the three aforementioned (stable) RF segments. We find one change point in 1983m9 -- 2001m10 significant at $1\%$, one in 2001m11 -- 2009m3 significant only at $5\%$, and no change over the last period 2009m4 -- 2020m4. Given the small sample size of the second segment 2001m11 -- 2009m3 (less than 90 observations) - which should then be further split into two sub-periods - we limit our remaining analysis to the first segment 1983m9 -- 2001m9.\footnote{The remaining results are presented in the Supplementary appendix.} The corresponding change point in the SF is estimated at 1992m4 which is significant at 1\%.

Finally, Table \ref{tab:1983-2001} collects the estimation results of the structural parameters obtained by TS2SLS, GMM, and TSGMM over the two corresponding sub-samples, (i) 1983m9-1992m4 and (ii) 1992m5-2001m10, as well as GMM estimation results obtained over the whole segment 1983m9-2001m10 (which effectively ignores the change point discovered in the SF). These results are discussed in the next subsection.

\subsection{Discussion of the empirical results}

The empirical analysis of the open-economy NKPC reveals three important features with key policy implications: (i) the detection of structural change, (ii) more precise estimation of the parameters of interest, and (iii) significant time variation in these parameters of interest. If the structural change discovered by our test was ignored, traditional GMM estimates - computed over the whole sample - would imply a \textit{flat} Phillips curve with respect to unemployment rate, and limited importance of the exchange rate pass-through: indeed, as can be seen from column (7) in Table \ref{tab:1983-2001}, the estimate of $\kappa_u$ is not found to be statistically significant, while the estimate of $\kappa_{\Psi}$ is quite small.
Instead, once the (structural) change point is detected and estimated, and the sample is split accordingly, both $\kappa_u$ and $\kappa_{\Psi}$ become statistically significant over the second interval 1992m5--2001m9. Importantly, both estimates are economically meaningful, e.g. $\kappa_u<0$ and $\kappa_{\Psi}>0$, and somewhat comparable in terms of magnitude; see columns (5) and (6) in Table \ref{tab:1983-2001} for estimation results obtained with split-sample GMM and TSGMM respectively. In addition, the estimate of $\kappa_{\Psi}$ is also significantly positive over the first sample, though its magnitude is more modest. Noticeably, the magnitude of each $\kappa_{\Psi}$ estimate obtained over a sub-sample remains much larger than the (misleading) estimate obtained by GMM over the whole sample: TSGMM estimates are respectively 0.0175 over 1983m9--1992m4 and 0.0266 over 1992m5-2001m9, while the full-sample GMM estimate is 0.0111.

Our analysis also suggests time-variation in the relationship between inflation and unemployment, and this is not surprising, as it is a prevalent finding in the NKPC literature. We also provide evidence of time-variation between inflation and the exchange rate pass-through - confirming \cite{syed2022}'s findings - and the magnitude of both parameters increases from the first sub-sample to the second. 
When comparing sub-sample estimates obtained by TSGMM and GMM, both sets of estimates are relatively close to each other: for example, over the second sub-sample, $\kappa_{\Psi}$ is estimated at 0.0266 with TSGMM and 0.0267 with GMM, while corresponding estimates for $\kappa_u$ are -0.0427 and -0.0434, respectively; see columns (2), (3), (5) and (6) in Table \ref{tab:1983-2001}. One key difference between these two sets of estimates relates to their precision: TSGMM delivers more precise estimates throughout. For example, $\kappa_{\Psi}$ and $\kappa_u$ are significant at 1\% with TSGMM, and not with GMM. This is in line with our theoretical results which show that TSGMM is more efficient than GMM as long as there is \textit{any change} in the structural equation over a stable reduced form. Intuitively, TSGMM exploits (useful) information that is neglected by other estimators such as split-sample GMM. Interestingly, with TSGMM, the 95\%-confidence intervals obtained for both $\kappa_{\Psi}$ and $\kappa_u$ over the two samples do not intersect, suggesting significant changes in the relationship between inflation and the exchange rate pass-through and between inflation and unemployment. Specifically, the relation between inflation and unemployment and inflation and exchange rate becomes more important in the second sample, where the estimate of the exchange rate slope increases by over 50\%.

Overall, the empirical analysis provides strong support for the NKPC with imperfect pass-through.

\subsection{Robustness checks}

$\bullet$ \textbf{Concerns about weak identification.} \\
In Table \ref{tab:kp}, we report results obtained by the Kleibergen-Paap rank test\footnote{The traditional test of weak identification based on the F-test and the rule-of-thumb only applies when there is one endogenous variable.} for different sub-samples. The matrix of reduced form coefficients is of size $(2,4)$ - since we have two reduced forms and four instruments - and we test the null of a rank equal to 1 over each stable RF sub-sample: the associated critical values are from the $\xi^2$ distribution with 3 degrees of freedom. The null of under-identification is rejected at 5\% over the first two sub-samples, 1983m9 -- 2001m10 and 2001m11 -- 2009m3, while it is borderline at 10\% over the last sub-sample, 2009m4 -- 2020m4. Overall, the identification seems to be sufficiently strong to support our estimation results.

\noindent $\bullet$ \textbf{Alternative specification.}\\
We consider alternative choices of instruments: specifically, we report results obtained with (i) only two instruments (one lag of unemployment rate and expected inflation), as well as with (ii) five instruments (one lag of unemployment rate, expected inflation, real exchange rate, terms of trade, interest rate spread). These results are reported Tables \ref{tab:rf_2IV} to \ref{tab:1983-2001 2IV} and Tables \ref{tab:rf change 5iv} to \ref{tab:1983-2001_ex}, respectively. Overall, these confirm our main results: the estimates obtained with TSGMM are more precise and the parameters of interest are subject to significant changes over the sample.

\section{Conclusion} \label{section:conclusion}

In this paper, we considered a linear model with endogenous regressors.
We first showed  how to exploit the stability of the first-stage to deliver more efficient estimators of the parameters of interest, even in the presence of near-weak instruments. We also derived the limiting distributions of two tests that detect parameter changes in the equation of interest: the first one applies when the first-stage is stable, and the second one when the first-stage exhibits a change point that is common to the equation of interest. Third, we also showed how to exploit information from the equation of interest to obtain more efficient estimators of the (stable) first-stage parameters than the standard full-sample OLS and showed how using the TSGMM estimator can provide tighter confidence intervals in an application to the open-economy NKPC.

While the paper is focused on the efficiency of TSGMM, the moment conditions of the TSGMM estimator could also be used for identification purposes, and therefore our paper contributes to a large literature tackling identification through stability restrictions - \cite{rigobon2003}, \cite{MagnussonMavroeidis}, \cite{angelini2024a}, \cite{angelini2024b}.

\section*{Acknowledgement}

We would like to thank conference and seminar participants for valuable comments. B. Antoine acknowledges financial support from Social Sciences and Humanities Research Council of Canada (SSHRC). A previous version of this paper was circulated under the title ``Inference in Linear Models with Structural Changes and Mixed Identification Strength''.

\bibliographystyle{chicago}
\bibliography{biblio-companion}

\newpage
\appendix

\begin{center}
	\LARGE{\textbf{Appendix}}
\end{center}
\section{Results from the Monte-Carlo study}
\subsection{Known change point}
\FloatBarrier
\begin{table}[h]
	\scalebox{1}{
		\begin{tabular}{lcccccc}
			\hline
			Estimator & Bias & MC Std & As. Std. &  RMSE  & Length & Coverage\\\hline\hline
			$T=400$, $\text{$n_{IV}$}=1$\\
			\hline
			$\theta_{GMM,1}$  &    0.0035  &  0.0818  &  0.0792 &   0.0793  &  0.3106  &  0.9400\\
			$\theta_{TS2SLS,1}$  &    0.0015 &   0.0808  &  0.0786 &   0.0786   & 0.3080&   0.9390\\
			$\theta_{TSGMM,1}$ &    -0.0008  &  0.0749  &  0.0724  &  0.0724  &  0.2838  &  0.9330\\
			$\theta_{GMM,2}$  &   0.0009  &  0.0649  &  0.0646  &  0.0646  &  0.2532  &  0.9520\\
			$\theta_{TS2SLS,2}$  &    -0.0004 &   0.0634  &  0.0644  &  0.0644 &   0.2525   & 0.9550\\
			$\theta_{TSGMM,2}$ &    -0.0012 &   0.0607   & 0.0609  &  0.0609  &  0.2388  &  0.9570\\
			
			\hline
			$T=400$, $\text{$n_{IV}$}=4$\\
			\hline
			$\theta_{GMM,1}$  &    -0.0008 &   0.0393  &  0.0384 &   0.0384  &  0.1506  &  0.9310\\
			$\theta_{TS2SLS,1}$  &    -0.0000 &   0.0385   & 0.0391  &  0.0391 &   0.1532  &  0.9420\\
			$\theta_{TSGMM,1}$ & -0.0004 & 0.0371 & 0.0351 & 0.0351 & 0.1376 & 0.9250 \\
			$\theta_{GMM,2}$& -0.0010 & 0.0329 & 0.0318 & 0.0318 & 0.1245 & 0.9320 \\
			$\theta_{TS2SLS,2}$ & -0.0009 & 0.0325 & 0.0321 & 0.0321 & 0.1260 & 0.9360 \\
			$\theta_{TSGMM,2}$ & -0.0010 & 0.0315 & 0.0299 & 0.0299 & 0.1173 & 0.9320 \\

			\hline
			$T=800$, $\text{$n_{IV}$}=1$\\
			\hline
			$\theta_{GMM,1}$   &     0.0016 &   0.0573 &   0.0560  &  0.0561  &  0.2197  &  0.9440\\
			$\theta_{TS2SLS,1}$  &     0.0004  &  0.0570  &  0.0559  &  0.0559  &  0.2190  &  0.9440\\
			$\theta_{TSGMM,1}$   &     0.0003   & 0.0514  &  0.0514  &  0.0514  &  0.2013  &  0.9530\\
			$\theta_{GMM,2}$   &     0.0034  &  0.0469  &  0.0459  &  0.0460  &  0.1800  &  0.9460\\
			$\theta_{TS2SLS,2}$  &     0.0021  &  0.0471  &  0.0457  &  0.0458  &  0.1793  &  0.9480\\
			$\theta_{TSGMM,2}$   &     0.0022  &  0.0441  &  0.0434  &  0.0435  &  0.1701  &  0.9440\\
			\hline
			
			$T=800$, $\text{$n_{IV}$}=4$\\
			\hline
			$\theta_{GMM,1}$   &    -0.0016 &   0.0292 &   0.0275  &  0.0276 &   0.1078  &  0.9340\\
			$\theta_{TS2SLS,1}$  &    -0.0012  &  0.0289  &  0.0278  &  0.0278  &  0.1089  &  0.9400\\
			$\theta_{TSGMM,1}$   &    -0.0011  &  0.0272  &  0.0252  &  0.0252  &  0.0989  &  0.9340\\
			$\theta_{GMM,2}$   &     0.0004  &  0.0229  &  0.0226  &  0.0226  &  0.0885  &  0.9510\\
			$\theta_{TS2SLS,2}$  &     0.0004  &  0.0231  &  0.0227  &  0.0227  &  0.0891 &   0.9440\\
			$\theta_{TSGMM,2}$   &     0.0003  &  0.0219  &  0.0213  &  0.0213  &  0.0836 &   0.9340\\
			
			\hline
		\end{tabular}
	}
	\caption{Performance of the GMM, TS2SLS and TSGMM estimators under homoskedasticity (HOM) when the change point is known. We consider either 1 or 4 instruments with a sample size of either 400 or 800.}
	\label{tab:hom}
\end{table}

\newpage
\begin{table}[h]
	\scalebox{1}{
		\begin{tabular}{lcccccc}
			\hline
			Estimator & Bias & MC Std & As. Std. &  RMSE  & Length & Coverage\\\hline\hline
			$T=400$, $\text{$n_{IV}$}=1$\\
			\hline
			$\theta_{GMM,1}$  &     0.0062 &   0.1145 &   0.1096 &   0.1098  &  0.4298 &   0.9340\\
			$\theta_{TS2SLS,1}$  &     0.0036 &   0.1133 &   0.1087 &   0.1088 &   0.4262 &  0.9340\\
			$\theta_{TSGMM,1}$ &     0.0007 &   0.1065 &   0.1005 &   0.1005  &  0.3940 &   0.9300\\
			$\theta_{GMM,2}$  &     0.0011 &   0.0921 &   0.0896 &   0.0896  &  0.3510 &   0.9390\\
			$\theta_{TS2SLS,2}$  &    -0.0006 &   0.0867 &   0.0862 &   0.0862  &  0.3378 &   0.9500\\
			$\theta_{TSGMM,2}$ &    -0.0016 &  0.0865 &   0.0847 &   0.0847  &  0.3320 &   0.9440\\

			\hline
			$T=400$, $\text{$n_{IV}$}=4$\\
			\hline
			
			$\theta_{GMM,1}$  &    -0.0023   &  0.0987   &  0.0907   &  0.0908   &  0.3557   &  0.9130 \\
			$\theta_{TS2SLS,1}$  &     0.0028   &  0.0978   &  0.0967   &  0.0967   &  0.3789   &  0.9340 \\
			$\theta_{TSGMM,1}$   &    -0.0004   &  0.0939   &  0.0837   &  0.0837   &  0.3279   &  0.9180 \\
			$\theta_{GMM,2}$   &    -0.0050   &  0.0811   &  0.0765   &  0.0766   &  0.2998   &  0.9260 \\
			$\theta_{TS2SLS,2}$  &    -0.0013   &  0.0758   &  0.0767   &  0.0767   &  0.3006   &  0.9440 \\
			$\theta_{TSGMM,2}$   &    -0.0043   &  0.0776   &  0.0725   &  0.0726   &  0.2842   &  0.9270 \\
			
			\hline
			$T=800$, $\text{$n_{IV}$}=1$\\
			\hline
			$\theta_{GMM,1}$& 0.0018 & 0.0814 & 0.0784 & 0.0784 & 0.3074 & 0.9340\\
			$\theta_{TS2SLS,1}$ & 0.0002 & 0.0810 & 0.0782 & 0.0782 & 0.3065 & 0.9320\\
			$\theta_{TSGMM,1}$ & 0.0001 & 0.0741 & 0.0722 & 0.0722 & 0.2830 & 0.9400\\
			$\theta_{GMM,2}$& 0.0044 & 0.0651 & 0.0644 & 0.0646 & 0.2525 & 0.9420\\
			$\theta_{TS2SLS,2}$ & 0.0030 & 0.0628 & 0.0619 & 0.0620 & 0.2428 & 0.9430\\
			$\theta_{TSGMM,2}$ & 0.0028 & 0.0616 & 0.0611 & 0.0612 & 0.2397 & 0.9430\\
			
			\hline
			$T=800$, $\text{$n_{IV}$}=4$\\
			\hline
			$\theta_{GMM,1}$  &    -0.0046   &  0.0736   &  0.0671   &  0.0672   &  0.2630   &  0.9250\\
			$\theta_{TS2SLS,1}$  &    -0.0022    & 0.0731   &  0.0695  &   0.0696   &  0.2725   &  0.9450\\
			$\theta_{TSGMM,1}$ &    -0.0027  &   0.0698  &   0.0620   &  0.0621  &   0.2432   &  0.9190\\
			$\theta_{GMM,2}$  &    0.0003   &  0.0580   &  0.0557   &  0.0557   &  0.2182  &   0.9380\\
			$\theta_{TS2SLS,2}$  &     0.0014   &  0.0553   &  0.0546    & 0.0546   &  0.2140   &  0.9390\\
			$\theta_{TSGMM,2}$ &     0.0005   &  0.0554   &  0.0529   &  0.0529   &  0.2075  &   0.9350\\
			\hline
		\end{tabular}
	}
	\caption{Performance of the GMM, TS2SLS and TSGMM estimators under heteroskedasticity of type 1 (HET1) when the change point is known. We consider either 1 or 4 instruments with a sample size of either 400 or 800.}
	\label{tab:het1}
\end{table}

\newpage
\begin{table}[h]
	\scalebox{1}{
		\begin{tabular}{lcccccc}
			\hline
			Estimator & Bias & MC Std & As. Std. &  RMSE  & Length & Coverage\\\hline\hline
			$T=400$, $\text{$n_{IV}$}=1$\\
			\hline
			
			$\theta_{GMM,1}$  &    0.0062  &  0.0792  &  0.0768 &   0.0770  &  0.3009 &   0.9460\\
			$\theta_{TS2SLS,1}$  &    0.0047  &  0.0784  &  0.0762 &   0.0763  &  0.2987 &   0.9440\\
			$\theta_{TSGMM,1}$ &    0.0037 &   0.0750  &  0.0719  &  0.0720 &   0.2819  &  0.9420\\
			$\theta_{GMM,2}$  &    -0.0019  &  0.4586  &  0.4242  &  0.4242    &1.6629   & 0.9550\\
			$\theta_{TS2SLS,2}$  &    -0.0059  &  0.4558  &  0.4193  &  0.4194  &  1.6438  &  0.9530\\
			$\theta_{TSGMM,2}$ &    -0.0091  &  0.4572  &  0.4187&   0.4188  &  1.6411 &   0.9540\\
			
			\hline
			$T=400$, $\text{$n_{IV}$}=4$\\
			\hline
			$\theta_{GMM,1}$& -0.0027 & 0.0366 & 0.0348 & 0.0349 & 0.1366 & 0.9440\\
			$\theta_{TS2SLS,1}$ & -0.0017 & 0.0394 & 0.0375 & 0.0375 & 0.1470 & 0.9520\\
			$\theta_{TSGMM,1}$ & -0.0027 & 0.0350 & 0.0325 & 0.0326 & 0.1275 & 0.9280\\
			$\theta_{GMM,2}$& -0.0001 & 0.1893 & 0.1675 & 0.1675 & 0.6565 & 0.9390\\
			$\theta_{TS2SLS,2}$ & -0.0071 & 0.2341 & 0.2155 & 0.2157 & 0.8449 & 0.9600\\
			$\theta_{TSGMM,2}$ & 0.0002 & 0.1886 & 0.1657 & 0.1657 & 0.6494 & 0.9400\\
			
			\hline
			$T=800$, $\text{$n_{IV}$}=1$\\
			\hline
			$\theta_{GMM,1}$  &    -0.0091  &  0.3501  &  0.3164  &  0.3166 &   1.2403 &   0.9590\\
			$\theta_{TS2SLS,1}$  &    -0.0108  &  0.3488 &   0.3159  &  0.3161 &   1.2384   & 0.9580\\
			$\theta_{TSGMM,1}$ &    -0.0129   & 0.3471&    0.3113   & 0.3116&    1.2203  &  0.9580\\
			$\theta_{GMM,2}$  &     0.0034 &   0.0601 &   0.0588   & 0.0589&    0.2304&    0.9510\\
			$\theta_{TS2SLS,2}$ &     0.0021 &   0.0616 &   0.0595   & 0.0595  &  0.2332&    0.9480\\
			$\theta_{TSGMM,2}$ &     0.0022   & 0.0590 &   0.0572 &   0.0572 &   0.2241 &   0.9420\\

			\hline
			$T=800$, $\text{$n_{IV}$}=4$\\
			\hline
			$\theta_{GMM,1}$  &    -0.0001  &  0.1275 &   0.1220   & 0.1220  &  0.4780&    0.9590\\
			$\theta_{TS2SLS,1}$  &    -0.0052  &  0.1702  &  0.1581&    0.1582  &  0.6198 &   0.9700\\
			$\theta_{TSGMM,1}$ &     0.0010  &  0.1289 &   0.1200  &  0.1200  &  0.4702 &   0.9510\\
			$\theta_{GMM,2}$  &     0.0015  &  0.0287  &  0.0275  &  0.0276 &   0.1079  &  0.9410\\
			$\theta_{TS2SLS,2}$  &     0.0002 &   0.0313  &  0.0302 &   0.0302   & 0.1183 &   0.9490\\
			$\theta_{TSGMM,2}$ &     0.0012   & 0.0278  &  0.0267  &  0.0267 &   0.1047 &   0.9460\\
			
			\hline
		\end{tabular}
	}
	\caption{Performance of the GMM, TS2SLS and TSGMM estimators under heteroskedasticity of type 2 (HET2) when the change point is known. We consider either 1 or 4 instruments with a sample size of either 400 or 800.}
	\label{tab:het2}
\end{table}

\FloatBarrier
\newpage
\subsection{Unknown change point}
\FloatBarrier

\FloatBarrier
\begin{table}[h]
	\scalebox{.9}{
		\begin{tabular}{lcccccc}
			\hline
			Estimator & Bias & MC Std & As. Std. &  RMSE  & Length & Coverage\\\hline\hline
			$T=400$, $\text{$n_{IV}$}=1$\\
			\hline
			Estimated change location: 161.021 \\
			\hline
			$\theta_{GMM,1}$  &   0.0033 &  0.0819 &  0.0795 &  0.0796 &  0.3116 &  0.9460\\
			$\theta_{TS2SLS,1}$  &    0.0013 &  0.0810 &  0.0788 &  0.0789 &  0.3091 &  0.9410\\
			$\theta_{TSGMM,1}$ &    -0.0009 &  0.0751 &  0.0726 &  0.0726 &  0.2845 &  0.9370\\
			$\theta_{GMM,2}$  &    -0.0031 &  0.0655 &  0.0655 &  0.0656 &  0.2567 &  0.9490\\
			$\theta_{TS2SLS,2}$  &    -0.0044 &  0.0639 &  0.0653 &  0.0654 &  0.2558 &  0.9480\\
			$\theta_{TSGMM,2}$ &    -0.0051 &  0.0612 &  0.0619 &  0.0621 &  0.2426 &  0.9460\\
			\hline
			$T=400$, $\text{$n_{IV}$}=4$\\
			\hline
			Estimated change location: 160.912 \\
			\hline
			$\theta_{GMM,1}$  &    0.0042 &  0.0404 &  0.0397 &  0.0399 &  0.1557 &  0.9340\\
			$\theta_{TS2SLS,1}$  &    0.0065 &  0.0400 &  0.0408 &  0.0413 &  0.1599 &  0.9410\\
			$\theta_{TSGMM,1}$ &    0.0044 &  0.0384 &  0.0365 &  0.0367 &  0.1430 &  0.9260\\
			$\theta_{GMM,2}$  &    -0.0010 &  0.0330 &  0.0318 &  0.0318 &  0.1248 &  0.9350\\
			$\theta_{TS2SLS,2}$  &    -0.0008 &  0.0326 &  0.0322 &  0.0322 &  0.1263 &  0.9400\\
			$\theta_{TSGMM,2}$ &    -0.0009 &  0.0316 &  0.0300 &  0.0300 &  0.1175 &  0.9350\\
			
			\hline
			$T=800$, $\text{$n_{IV}$}=1$\\
			\hline
			Estimated change location: 321.038  \\
			\hline
			$\theta_{GMM,1}$  &    0.0049 &  0.0575 &  0.0566 &  0.0568 &  0.2218 &  0.9490\\
			$\theta_{TS2SLS,1}$  &    0.0037 &  0.0571 &  0.0564 &  0.0565 &  0.2209 &  0.9420\\
			$\theta_{TSGMM,1}$ &    0.0035 &  0.0513 &  0.0520 &  0.0521 &  0.2037 &  0.9500\\
			$\theta_{GMM,2}$  &    0.0035 &  0.0469 &  0.0459 &  0.0461 &  0.1801 &  0.9440\\
			$\theta_{TS2SLS,2}$  &    0.0022 &  0.0472 &  0.0458 &  0.0458 &  0.1795 &  0.9470\\
			$\theta_{TSGMM,2}$ &    0.0022 &  0.0442 &  0.0434 &  0.0435 &  0.1702 &  0.9400\\
			\hline
			$T=800$, $\text{$n_{IV}$}=4$\\
			\hline
			Estimated change location: 320.983 \\
			\hline
			$\theta_{GMM,1}$  &    -0.0016 &  0.0292 &  0.0275 &  0.0276 &  0.1078 &  0.9340\\
			$\theta_{TS2SLS,1}$  &    -0.0012 &  0.0289 &  0.0278 &  0.0278 &  0.1089 &  0.9400\\
			$\theta_{TSGMM,1}$ &    -0.0011 &  0.0272 &  0.0252 &  0.0252 &  0.0989 &  0.9340\\
			$\theta_{GMM,2}$  &    0.0004 &  0.0229 &  0.0226 &  0.0226 &  0.0885 &  0.9510\\
			$\theta_{TS2SLS,2}$  &    0.0004 &  0.0231 &  0.0227 &  0.0227 &  0.0891 &  0.9440\\
			$\theta_{TSGMM,2}$ &    0.0003 &  0.0219 &  0.0213 &  0.0213 &  0.0836 &  0.9340\\
			\hline
		\end{tabular}
	}
	\caption{Performance of the GMM, TS2SLS and TSGMM estimators under homoskedasticity (HOM) when the change point is estimated. We consider either 1 or 4 instruments with a sample size of either 400 or 800.}
	\label{tab:homb}
\end{table}

\newpage
\FloatBarrier
\begin{table}[h]
	\scalebox{.9}{
		\begin{tabular}{lcccccc}
			\hline
			Estimator & Bias & MC Std & As. Std. &  RMSE  & Length & Coverage\\\hline\hline
			$T=400$, $\text{$n_{IV}$}=1$\\
			\hline
			Estimated change point: 161.071  \\
			\hline
			$\theta_{GMM,1}$  &    0.0060 &  0.1147 &  0.1100 &  0.1101 &  0.4311 &  0.9380\\
			$\theta_{TS2SLS,1}$  &    0.0034 &  0.1136 &  0.1091 &  0.1091 &  0.4276 &  0.9370\\
			$\theta_{TSGMM,1}$ &    0.0006 &  0.1067 &  0.1008 &  0.1008 &  0.3950 &  0.9330\\
			$\theta_{GMM,2}$  &    -0.0029 &  0.0923 &  0.0901 &  0.0901 &  0.3531 &  0.9370\\
			$\theta_{TS2SLS,2}$  &    -0.0045 &  0.0869 &  0.0867 &  0.0868 &  0.3399 &  0.9480\\
			$\theta_{TSGMM,2}$ &    -0.0055 &  0.0866 &  0.0853 &  0.0855 &  0.3344 &  0.9470\\
			\hline
			
			$T=400$, $\text{$n_{IV}$}=4$\\
			\hline
			Estimated change point: 160.832 \\
			\hline
			
			$\theta_{GMM,1}$  &    0.0036 &  0.0990 &  0.0912 &  0.0913 &  0.3576 &  0.9170\\
			$\theta_{TS2SLS,1}$  &    0.0092 &  0.0978 &  0.0971 &  0.0976 &  0.3807 &  0.9350\\
			$\theta_{TSGMM,1}$ &    0.0054 &  0.0944 &  0.0843 &  0.0844 &  0.3303 &  0.9130\\
			$\theta_{GMM,2}$  &    -0.0048 &  0.0813 &  0.0766 &  0.0768 &  0.3004 &  0.9290\\
			$\theta_{TS2SLS,2}$  &    -0.0012 &  0.0760 &  0.0768 &  0.0768 &  0.3012 &  0.9470\\
			$\theta_{TSGMM,2}$ &    -0.0041 &  0.0779 &  0.0726 &  0.0727 &  0.2847 &  0.9270\\
			\hline
			$T=800$, $\text{$n_{IV}$}=1$\\
			\hline
			Estimated change point:  321.04 \\
			\hline
			
			$\theta_{GMM,1}$  &    0.0049 &  0.0813 &  0.0787 &  0.0789 &  0.3086 &  0.9360\\
			$\theta_{TS2SLS,1}$  &    0.0033 &  0.0808 &  0.0785 &  0.0785 &  0.3075 &  0.9340\\
			$\theta_{TSGMM,1}$ &   0.0032 &  0.0738 &  0.0726 &  0.0726 &  0.2845 &  0.9430\\
			$\theta_{GMM,2}$  &    0.0045 &  0.0651 &  0.0645 &  0.0646 &  0.2527 &  0.9430\\
			$\theta_{TS2SLS,2}$  &    0.0031 &  0.0629 &  0.0620 &  0.0621 &  0.2430 &  0.9430\\
			$\theta_{TSGMM,2}$ &    0.0029 &  0.0617 &  0.0612 &  0.0612 &  0.2398 &  0.9400\\
			\hline
			$T=800$, $\text{$n_{IV}$}=4$\\
			\hline
			Estimated change point:  320.92 \\
			\hline
			$\theta_{GMM,1}$  &    0.0010 &  0.0737 &  0.0673 &  0.0674 &  0.2640 &  0.9260\\
			$\theta_{TS2SLS,1}$  &    0.0037 &  0.0733 &  0.0697 &  0.0698 &  0.2734 &  0.9400\\
			$\theta_{TSGMM,1}$ &    0.0027 &  0.0699 &  0.0624 &  0.0624 &  0.2445 &  0.9170\\
			$\theta_{GMM,2}$  &    0.0004 &  0.0580 &  0.0558 &  0.0558 &  0.2186 &  0.9400\\
			$\theta_{TS2SLS,2}$  &    0.0016 &  0.0553 &  0.0547 &  0.0547 &  0.2144 &  0.9380\\
			$\theta_{TSGMM,2}$ &    0.0007 &  0.0554 &  0.0530 &  0.0530 &  0.2078 &  0.9340\\
			\hline
		\end{tabular}
	}
	\caption{Performance of the GMM, TS2SLS and TSGMM estimators under heteroskedasticity of type 1 (HET1) when the change point is estimated. We consider either 1 or 4 instruments with a sample size of either 400 or 800.}
	\label{tab:het1b}
\end{table}

\newpage
\begin{table}[h]
	\scalebox{.9}{
		\begin{tabular}{lcccccc}
			\hline
			Estimator & Bias & MC Std & As. Std. &  RMSE  & Length & Coverage\\\hline\hline
			$T=400$, $\text{$n_{IV}$}=1$\\
			\hline
			Estimated change location: 203.668 \\
			\hline
			
			$\theta_{GMM,1}$  &    0.0127 &  0.0812 &  0.0783 &  0.0794 &  0.3071 &  0.9440\\
			$\theta_{TS2SLS,1}$  &    0.0113 &  0.0804 &  0.0777 &  0.0785 &  0.3045 &  0.9440\\
			$\theta_{TSGMM,1}$ &   0.0104 &  0.0775 &  0.0736 &  0.0743 &  0.2885 &  0.9410\\
			$\theta_{GMM,2}$  &    -0.0019 &  0.4604 &  0.4261 &  0.4261  &  1.6702 &  0.9550\\
			$\theta_{TS2SLS,2}$  &    -0.0059 &  0.4576 &  0.4211 &  0.4211  &  1.6507 &  0.9530\\
			$\theta_{TSGMM,2}$ &    -0.0092 &  0.4587 &  0.4204 &  0.4205 &   1.6481 &  0.9560\\
			\hline
			$T=400$, $\text{$n_{IV}$}=4$\\
			\hline
			Estimated change location: 180.481 \\
			\hline
			$\theta_{GMM,1}$  &    0.0127 &  0.0812 &  0.0783 &  0.0794 &  0.3071 &  0.9440\\
			$\theta_{TS2SLS,1}$  &    0.0113 &  0.0804 &  0.0777 &  0.0785 &  0.3045 &  0.9440\\
			$\theta_{TSGMM,1}$ &    0.0104 &  0.0775 &  0.0736 &  0.0743 &  0.2885 &  0.9410\\
			$\theta_{GMM,2}$  &    -0.0019 &  0.4604 &  0.4261 &  0.4261  &  1.6702 &  0.9550\\
			$\theta_{TS2SLS,2}$  &    -0.0059 &  0.4576 &  0.4211 &  0.4211  &  1.6507 &  0.9530\\
			$\theta_{TSGMM,2}$ &    -0.0092 &  0.4587 &  0.4204 &  0.4205  &  1.6481 &  0.9560\\
			
			\hline
			$T=800$, $\text{$n_{IV}$}=1$\\
			\hline
			Estimated change location: 292.366 \\
			\hline
			$\theta_{GMM,1}$  &    -0.0152&    0.2785   & 0.2503 &   0.2508  &  0.9811    &0.9610\\
			$\theta_{TS2SLS,1}$  &    -0.0175  &  0.2784  &  0.2498 &   0.2505  &  0.9794   & 0.9590\\
			$\theta_{TSGMM,1}$ &    -0.0188  &  0.2738  &  0.2449 &   0.2457  &  0.9601&    0.9580\\
			$\theta_{GMM,2}$  &    -0.1168  &  0.1766  &  0.1528 &   0.1923 &   0.5990    &0.8420\\
			$\theta_{TS2SLS,2}$  &    -0.1179  &  0.1758  &  0.1519  &  0.1923 &   0.5956 &   0.8390\\
			$\theta_{TSGMM,2}$ &    -0.1179  &  0.1759  &  0.1517 &   0.1922  &  0.5948&    0.8380\\
			
			\hline
			$T=800$, $\text{$n_{IV}$}=4$\\
			\hline
			Estimated change location: 302.204 \\
			\hline
			$\theta_{GMM,1}$  &     0.0033   & 0.1274 &   0.1219 &   0.1219 &   0.4777&    0.9560\\
			$\theta_{TS2SLS,1}$  &    -0.0018  &  0.1696   & 0.1578 &   0.1578  &  0.6185&    0.9680\\
			$\theta_{TSGMM,1}$ &     0.0043 &   0.1287 &   0.1199 &   0.1200 &   0.4699 &   0.9480\\
			$\theta_{GMM,2}$  &     0.0015  &  0.0287  &  0.0276  &  0.0276 &   0.1080 &   0.9410\\
			$\theta_{TS2SLS,2}$  &     0.0002 &   0.0313  &  0.0302  &  0.0302 &   0.1185  &  0.9480\\
			$\theta_{TSGMM,2}$ &     0.0012 &   0.0278   & 0.0267  &  0.0268  &  0.1048 &   0.9470\\
			\hline
		\end{tabular}
	}
	\caption{Performance of the GMM, TS2SLS and TSGMM estimators under heteroskedasticity of type 2 (HET2) when the change point is estimated. We consider either 1 or 4 instruments with a sample size of either 400 or 800.}
	\label{tab:het2b}
\end{table}

\newpage
\begin{table}[h]
	\scalebox{.9}{
		\begin{tabular}{lccccccc}
			\hline
			Estimator & Bias & MC Std & As. Std. & RMSE  & Length & Coverage\\\hline\hline
			$T=400$, $\rho=-0.5$ and $\text{$n_{IV}$}=4$\\
			\hline
			\textbf{HOM} \\
			Estimated change location: 201.662  \\
			\hline
			$\theta_{GMM,1}$  &    -0.0006 &  0.0277 &  0.0274 &  0.0274 &  0.1072 &  0.9440\\
			$\theta_{TS2SLS,1}$  &    -0.0005 &  0.0276 &  0.0276 &  0.0276 &  0.1082 &  0.9420\\
			$\theta_{TSGMM,1}$ &    -0.0004 &  0.0275 &  0.0266 &  0.0266 &  0.1041 &  0.9330\\
			$\theta_{GMM,2}$  &    -0.0025 &  0.0607 &  0.0548 &  0.0548 &  0.2146 &  0.9230\\
			$\theta_{TS2SLS,2}$  &    0.0002 &  0.0585 &  0.0567 &  0.0567 &  0.2222 &  0.9340\\
			$\theta_{TSGMM,2}$ &    -0.0022 &  0.0561 &  0.0475 &  0.0475 &  0.1861 &  0.9080\\
			
			\hline
			\textbf{HET1}   \\
			Estimated change location: 198.848 \\
			\hline
			
			$\theta_{GMM,1}$  &    -0.0030 &  0.0706 &  0.0666 &  0.0667 &  0.2612 &  0.9210\\
			$\theta_{TS2SLS,1}$  &    0.0003 &  0.0698 &  0.0691 &  0.0691 &  0.2707 &  0.9450\\
			$\theta_{TSGMM,1}$ &    -0.0022 &  0.0703 &  0.0649 &  0.0649 &  0.2544 &  0.9190\\
			$\theta_{GMM,2}$  &    -0.0113 &  0.1470 &  0.1215 &  0.1220 &  0.4763 &  0.8850\\
			$\theta_{TS2SLS,2}$  &    0.0007 &  0.1451 &  0.1361 &  0.1361 &  0.5333 &  0.9250\\
			$\theta_{TSGMM,2}$ &    -0.0074 &  0.1361 &  0.1063 &  0.1066 &  0.4168 &  0.8660\\
			\hline
			\textbf{HET2}  \\
			Estimated change location: 335.398 \\
			\hline
			$\theta_{GMM,1}$  &    -0.0008 &  0.0281 &  0.0267 &  0.0267 &  0.1046 &  0.9310\\
			$\theta_{TS2SLS,1}$  &    -0.0006 &  0.0303 &  0.0286 &  0.0286 &  0.1120 &  0.9390\\
			$\theta_{TSGMM,1}$ &    -0.0010 &  0.0275 &  0.0262 &  0.0262 &  0.1028 &  0.9350\\
			$\theta_{GMM,2}$  &    -0.0083 &  0.6605 &  0.5624 &  0.5624  &  2.2044 &  0.9340\\
			$\theta_{TS2SLS,2}$  &    -0.0248 &  0.8108 &  0.7269 &  0.7274  &  2.8495 &  0.9530\\
			$\theta_{TSGMM,2}$ &    -0.0014 &  0.6695 &  0.5263 &  0.5263  &  2.0632 &  0.8950\\
			\hline
		\end{tabular}
	}
	\caption{Performance of the GMM, TS2SLS and TSGMM estimators under homoskedasticity (HOM) when a change point is estimated but there is no change point in the true DGP. We consider either 1 or 4 instruments with a sample size of either 400 or 800.}
	\label{tab:bob}
\end{table}
\FloatBarrier

\newpage
\FloatBarrier
\FloatBarrier

\newpage
\subsection{Detecting a change point}
\FloatBarrier
\begin{table}[h]
	\scalebox{.95}{
		\begin{tabular}{lcc}
			\hline
			Change size & Prob. of detecting change & Estimated change point \\
			\hline
			\hline
			\textbf{HOM} \\
			\hline
			1 & 1 & 160.912 \\
			0.5 & 1 & 160.705 \\
			0.3 & 1 & 160.728 \\
			0 & 0.07 & 196.174 \\
			\hline
			\textbf{HET1} \\
			\hline
			1 & 1 & 160.832 \\
			0.5 & 0.99 & 160.012 \\
			0.3 & 0.924 & 162.458 \\
			0 & 0.065 & 198.848\\
			\hline
			\textbf{HET2} \\
			\hline
			1 & 0.883 & 180.481 \\
			0.5 & 0.293 & 262.182 \\
			0.3 & 0.028 & 309.525 \\
			0 & 0.001 & 335.398\\
			\hline
		\end{tabular}
	}
	\caption{Probability of detecting a change point for different change sizes. We consider one external instrument and a sample size of $400$. Note that the change size equals zero implies the null hypothesis of no change.}
	\label{tab:ttest}
\end{table}
\FloatBarrier

\begin{table}[h]
	\scalebox{.9}{
		\begin{tabular}{lcccccc}
			\hline
			Estimator & Bias & MC Std & As. Std. & RMSE & Length  & Coverage \\
			\hline
			\hline
			\textbf{HOM} \\
			Estimated change point: 160.912\\
			Prob. of detecting the change: 1.000\\
			\hline
			$\theta_{GMM,1}$     &    0.0042 &  0.0404 &  0.0397 &  0.0399 &  0.1557 &  0.9340\\
			$\theta_{TS2SLS,1}$    &    0.0065 &  0.0400 &  0.0408 &  0.0413 &  0.1599 &  0.9410\\
			$\theta_{TSGMM,1}$     &   0.0044  &  0.0384 &  0.0365 &  0.0367 &  0.1430 &  0.9260\\
			$\theta_{GMM,2}$     &   -0.0010 &  0.0330 &  0.0318 &  0.0318 &  0.1248 &  0.9350\\
			$\theta_{TS2SLS,2}$    &  -0.0008  &  0.0326 &  0.0322 &  0.0322 &  0.1263 &  0.9400\\
			$\theta_{TSGMM,2}$     &   -0.0009 &  0.0316 &  0.0300 &  0.0300 &  0.1175 &  0.9350\\
			\hline
			\textbf{HET1} \\
			Estimated change point: 160.832 \\
			Prob. of detecting the change: 1.000  \\
			\hline
			$\theta_{GMM,1}$     &    0.0036 &  0.0990 &  0.0912 &  0.0913 &  0.3576 &  0.9170\\
			$\theta_{TS2SLS,1}$    &    0.0092 &  0.0978 &  0.0971 &  0.0976 &  0.3807 &  0.9350\\
			$\theta_{TSGMM,1}$     &    0.0054 &  0.0944 &  0.0843 &  0.0844 &  0.3303 &  0.9130\\
			$\theta_{GMM,2}$     &   -0.0048 &  0.0813 &  0.0766 &  0.0768 &  0.3004 &  0.9290\\
			$\theta_{TS2SLS,2}$    &   -0.0012 &  0.0760 &  0.0768 &  0.0768 &  0.3012 &  0.9470\\
			$\theta_{TSGMM,2}$     &   -0.0041 &  0.0779 &  0.0726 &  0.0727 &  0.2847 &  0.9270\\
			\hline
			\textbf{HET2} \\
			Estimated change point: 180.481 \\
			Prob. of detecting the change: 0.883 \\
			\hline
			$\theta_{GMM,1}$     &    0.0658  &  0.0542 &  0.0486 &  0.0986 &  0.1907 &  0.8480\\
			$\theta_{TS2SLS,1}$    &    0.0707  &  0.0589 &  0.0585 &  0.1056 &  0.2292 &  0.8670\\
			$\theta_{TSGMM,1}$     &    0.0659  &  0.0528 &  0.0468 &  0.0967 &  0.1834 &  0.8350\\
			$\theta_{GMM,2}$     &    -0.0439 &  0.3393 &  0.1593 &  0.1977 &  0.6242 &  0.8680\\
			$\theta_{TS2SLS,2}$    &    -0.0508 &  0.3476 &  0.2019 &  0.2336 &  0.7915 &  0.9110\\
			$\theta_{TSGMM,2}$     &    -0.0437 &  0.3389 &  0.1575 &  0.1959 &  0.6175 &  0.8680\\
			\hline
		\end{tabular}
	}
	\caption{Pre-testing for a change point. If a change point is detected, estimation is carried out across sub-samples as previously done; if no change point is detected, the parameters are estimated over the entire sample. We report a weighted average of the two specifications. We consider $T=400$ and $\text{$n_{IV}$}=4$.}
	\label{tab:avgspec}
\end{table}
\FloatBarrier

\section{Estimation of the NKPC}

\subsection{Data Plots}
\FloatBarrier

\FloatBarrier
\subsection{Estimation results in the benchmark model}
\FloatBarrier
\begin{table}[h]
	\scalebox{.9}{
		\begin{tabular}{ll|lll}
			\hline
			\multicolumn{5}{l}{\textbf{Panel A: Expected inflation}} \\ \hline
			&  & \multicolumn{3}{c}{Bai and Perron (2003) critical values}  \\
			& Test statistic & $1\%$ critical value & $5\%$ critical value & $10\%$ critical value \\
			\hline
			$F(1|0)$ &      3.49     &       3.46        &    2.90        &     2.61     \\
			$F(2|1)$ &       2.77    &        3.63       &      3.15       &     2.89    \\
			\hline
			Number of changes: &   & 2 & 2 & 2\\
			Break points: &  & 2001m9    &  2009m3  &  \\
			\hline \hline
			
			\multicolumn{5}{l}{\textbf{Panel B: Unemployment rate}} \\ \hline
			&  & \multicolumn{3}{c}{Bai and Perron (2003) critical values}  \\
			& Test statistic & $1\%$ critical value & $5\%$ critical value & $10\%$ critical value \\
			\hline
			$F(1|0)$ &     3.62       &      3.46        &     2.90      &   2.61       \\
			$F(2|1)$ &      5.39     &        3.63       &      3.15       &     2.89         \\
			\hline
			Number of changes: &   & 2 & 2 & 2\\
			Break points: &    &  2001m9   & 2011m10  &  \\
			\hline
			
		\end{tabular}
	}
	\caption{Breaks in the Reduced forms using \cite{baiperron2004} critical values. Trimming 0.2. $T=434$. }
	\label{tab:RF_changes}
\end{table}

\FloatBarrier
\begin{table}[h]
	\scalebox{.9}{
		\begin{tabular}{lllll}
			\hline
			\multicolumn{2}{c}{} & \multicolumn{3}{c}{Hall, Han and Boldea (2012) critical values}  \\
			& Test statistic & $1\%$ critical value & $5\%$ critical value & $10\%$ critical value \\
			\hline
			1983m9-2001m9: & & & & \\
			$F(1|0)$ &     36.86      &        24.45      &    20.08       &    17.95     \\
			Break point: & 1992m4  &      &     &  \\
			\hline
			2001m10-2009m3: & & & & \\
			$F(1|0)$ &     21.22      &       24.45      &    20.08       &    17.95         \\
			Break point: & 2006m8 &      &     &  \\
			\hline
			2009m4-2020m4: & & & & \\
			$F(1|0)$ &      5.20    &         24.45      &    20.08       &    17.95      \\
			Break point: & - &      &     &  \\
			\hline
		\end{tabular}
	}
	\caption{Breaks in the structural equation over each stable RF segment using \cite{HHB} critical values. Trimming 0.15.}
	\label{tab:sf_change}
\end{table}
\FloatBarrier
\newpage

\begin{table}[h]
	\centering
	\scalebox{.9}{
		\begin{tabular}{l|ccc|ccc||c}
			\cline{2-8}
			&    \multicolumn{3}{c|}{1983m9-1992m4} & \multicolumn{3}{c||}{1992m5-2001m9} & 1983m9-2001m9\\ \cline{2-8}
			& TS2SLS  & GMM & TSGMM & TS2SLS & GMM & TSGMM & GMM\\
			& (1) & (2) & (3) & (4) & (5) & (6) & (7) \\\hline
			
			$\beta_0$   & -0.0024       & -0.0030       & -0.0022       & 0.0035       & -0.0001       & -0.0004       & 0.0001        \\
			&   (0.0256)      & (0.0264)      & (0.0237)      & (0.0199)     & (0.0201)      & (0.0198)      & (0.0161)      \\

			$\gamma_1$ & 1.4323*** & 1.4751*** & 1.4748*** & 0.5369       & 0.6529*** & 0.6512*** & 1.2028***       \\
			&  (0.2472)      & (0.1190)      & (0.1097)      & (0.6315)     & (0.1266)      & (0.1218)      & (0.0976)      \\

			$\gamma_2$ & -0.5522** & -0.5861*** & -0.5888*** & -0.1292     & -0.2287** & -0.2273** & -0.3989***      \\
			& (0.2302)      & (0.1039)      & (0.0931)      & (0.3610)     & (0.1003)      & (0.0989)      & (0.0854)      \\
			
			$\kappa_\Psi$ & 0.0150        & 0.0179\textcolor{blue}{*} & 0.0175\textcolor{blue}{*} & 0.0138       & 0.0267\textcolor{blue}{**} & 0.0266\textcolor{blue}{***} & 0.0111\textcolor{blue}{*}       \\
			& (0.0262)      & (0.0131)      & (0.0119)      & (0.0370)     & (0.0119)      & (0.0113)      & (0.0082)      \\
			
			$\gamma_f$   & -0.2618       & -0.2818*  & -0.2814** & 0.2859      & -0.0817       & -0.0598       & -0.1915       \\
			& (0.3514)      & (0.1793)      & (0.1616)      & (1.3275)     & (0.4235)      & (0.3828)      & (0.1664)      \\
			
			$\kappa_u$  & 0.0150        & 0.0151        & 0.0143        & -0.0280\textcolor{red}{*} & -0.0434\textcolor{red}{***} & -0.0427\textcolor{red}{***} & -0.0036      \\
			& (0.0130)      & (0.0111)      & (0.0096)      & (0.0209)     & (0.0167)      & (0.0156)      & (0.0085)      \\
			\hline
			$N$                 &     & $104$ &   &   & $114$ &  & $218$\\\hline
		\end{tabular}
	}
	\caption{*$p < 0.10$, **$p < 0.05$, ***$p < 0.01$. Black stars indicate bi-directional testing. Blue stars indicate right-tailed (positive) tests, and Red stars left-tailed (negative) tests.}
	\label{tab:1983-2001}
\end{table}
\FloatBarrier
\normalsize

\subsection{Robustness checks}
\subsubsection{Weak identification concerns}
\begin{table}[h]
	\centering
	\scalebox{.9}{
		\begin{tabular}{lcc}
			\hline
			Sub-sample &  KP statistic & N. Obs.\\
			\hline
			1983m9-2001m10	 & 	7.959		 & 	216\\
			2001m11-2009m3	 & 	16.96		 & 	89\\	
			2009m4-2020m4 & 6.087   & 129\\
			\hline
		\end{tabular}
	}
	\caption{\cite{kleibergen2006} KP statistics reported over the stable RF subperiods. We test the null of rank deficiency (rank equal to 1 for the (2,4)-matrix of RF coefficients) and the associated critical value are from the Chi2(3): 7.81 at 5\% and 6.25 at 10\%. }
	\label{tab:kp}
\end{table}

\subsubsection{Robustness checks: alternative instruments sets}

In this section, we present the results obtained with alternate instrument sets: 2 instruments (one lag of expected inflation and unemployment rate); 5 instruments (one lag of expected inflation, unemployment rate, real exchange rate, terms of trade, short-long term interest rate spread). We follow exactly the same steps as described in Section \ref{subsection:estim procedure}.

\newpage
\FloatBarrier
$\bullet$ \textbf{Results obtained under just-identification with 2 instruments:}

\begin{table}[h]
	\scalebox{.9}{
		\begin{tabular}{ll|lll}
			\hline
			\multicolumn{5}{l}{\textbf{Panel A: Expected inflation}} \\ \hline
			&  & \multicolumn{3}{c}{Bai and Perron (2003) critical values}  \\
			& Test statistic & $1\%$ critical value & $5\%$ critical value & $10\%$ critical value \\
			\hline
			$F(1|0)$ &2.88&	3.97&	3.26	&2.90\\
			$F(2|1)$ &      3.31	&4.17	&3.57&	3.25      \\
			\hline
			Number of changes: &   & 0 & 0 & 2\\
			Break points: &  &   2001m10  &  2009m3  &  \\
			\hline \hline
			
			\multicolumn{5}{l}{\textbf{Panel B: Unemployment rate}} \\ \hline
			&  & \multicolumn{3}{c}{Bai and Perron (2003) critical values}  \\
			& Test statistic & $1\%$ critical value & $5\%$ critical value & $10\%$ critical value \\
			\hline
			$F(1|0)$ &      3.53&	3.97	&3.26&	2.9         \\
			$F(2|1)$ &      4.02&	4.17	&3.57	&3.25        \\
			\hline
			Number of changes: &  & 0 & 2 &2 \\
			Break points: &    &  2001m10     &  2008m12   &  \\
			\hline
		\end{tabular}
	}
	\caption{Breaks in the Reduced forms using \cite{baiperron2004} critical values. Trimming 0.2. $T=434$. }
	\label{tab:rf_2IV}
\end{table}

\FloatBarrier
\begin{table}[h]
	\scalebox{.9}{
		\begin{tabular}{lllll}
			\hline
			\multicolumn{2}{c}{} & \multicolumn{3}{c}{\cite{baiperron2004} critical values}  \\
			& Test statistic & $1\%$ critical value & $5\%$ critical value & $10\%$ critical value \\
			\hline
			1983m9-2001m10: & & & & \\
			$F(1|0)$ &    35.43        &       24.50      &    20.08        &   17.95     \\
			Break point: & 1992m4  &      &     &  \\
			\hline
			2001m11-2009m3: & & & & \\
			$F(1|0)$ &       17.91     &       24.50      &    20.08        &   17.95     \\
			Break point: & 2006m9 &      &     &  \\
			\hline
			2009m3-2020m4: & & & & \\
			$F(1|0)$ &      4.35       &       24.50      &    20.08        &   17.95     \\
			Break point: & - &      &     &  \\
			\hline
		\end{tabular}
	}
	\caption{Breaks in the structural equation over each stable RF segment. Trimming is 0.15. Note: the change point estimated at 2006m9 is at the margin of significance; it is included in all our results.}
	\label{tab:sf_change 2IV}
\end{table}

\newpage

\begin{table}[h]
	\centering
	\begin{tabular}{lcc}
		\hline
		Sub-sample &  F statistic & N. Obs.\\
		\hline
		1983m9-2001m10	&	9.501	&	216\\
		
		\hline
	\end{tabular}
	\caption{\cite{kleibergen2006} KP statistics reported over the stable RF subperiods. We test the null of rank deficiency (rank equal to 1 for the (2,2)-matrix of RF coefficients) and the associated critical value are from the Chi2(1): 3.84 at 5\% and 2.71 at 10\%. }
	\label{tab:KP 2IV}
\end{table}
\FloatBarrier
\begin{table}[h]
	\centering
	\scalebox{.9}{
		\begin{tabular}{l|ccc|ccc||c}
			\hline
			&    \multicolumn{3}{c|}{1983m9-1992m4}  &  \multicolumn{3}{c||}{1992m5-2001m10}  & 1983m9-2001m10\\\hline
			& TS2SLS  & GMM & TSGMM & TS2SLS & GMM & TSGMM & GMM\\
			& (1) & (2) & (3) & (4) & (5) & (6) & (7) \\\hline
			$\beta_0$   & -0.0022       & -0.0050       & -0.0050       & 0.0041       & 0.0011       & 0.0011       & -0.0001       \\
			& (0.0256)      & (0.0264)      & (0.0242)      & (0.0189)     & (0.0232)     & (0.0230)     & (0.0163)      \\
			$\gamma_1$  & 1.4328***     & 1.4539***     & 1.4539***     & 0.5631       & 0.6641***    & 0.6641***    & 1.2018***     \\
			& (0.2510)      & (0.1234)      & (0.1158)      & (0.4831)     & (0.2320)     & (0.2297)     & (0.0983)      \\
			$\gamma_2$ & -0.5543**     & -0.5786***    & -0.5786***    & -0.1353      & -0.2296      & -0.2296      & -0.4048***    \\
			& (0.2295)      & (0.1049)      & (0.0960)      & (0.2706)     & (0.1589)     & (0.1568)     & (0.0854)      \\
			$\kappa_\Psi$  & 0.0146        & 0.0161        & 0.0161\textcolor{blue}{*}   & 0.0141       & 0.0303       & 0.0303       & 0.0102       \\
			& (0.0262)      & (0.0131)      & (0.0124)      & (0.0283)     & (0.0457)     & (0.0449)     & (0.0083)      \\
			$\gamma_f$     & -0.2624       & -0.2432       & -0.2432       & 0.2187       & -0.2295      & -0.2295      & -0.1843       \\
			& (0.3696)      & (0.1822)      & (0.1666)      & (1.0502)     & (2.0702)     & (2.0304)     & (0.1716)      \\
			$\kappa_u$ & 0.0138        & 0.0140        & 0.0140        & -0.0278\textcolor{red}{*}     & -0.0477      & -0.0477      & -0.0050      \\
			& (0.0127)      & (0.0110)      & (0.0097)      & (0.0181)     & (0.0725)     & (0.0714)     & (0.0088)      \\
			\hline
			$N$                 &     & $102$ &   &   & $114$ &  & $216$\\\hline
			
		\end{tabular}
	}
	\caption{*$p < 0.10$, **$p < 0.05$, ***$p < 0.01$. Black stars indicate bi-directional testing. Blue stars indicate right-tailed (positive) tests, and Red stars left-tailed (negative) tests.}
	\label{tab:1983-2001 2IV}
\end{table}
\FloatBarrier

\newpage

$\bullet$ \textbf{Results obtained with 5 instruments:}

\begin{table}[h]
	\scalebox{.9}{
		\begin{tabular}{lllll}
			\hline
			\multicolumn{5}{l}{\textbf{Panel A: Expected inflation}} \\ \hline
			&  & \multicolumn{3}{c}{Bai and Perron (2003) critical values}  \\
			& Test statistic & $1\%$ critical value & $5\%$ critical value & $10\%$ critical value \\
			\hline
			$F(1|0)$ &      3.20    &         3.30       &     2.77    &        2.49 \\
			$F(2|1)$ &     2.88        &     3.52     &       2.99       &     2.76 \\
			\hline
			Number of changes: &   & 0 & 1 & 2\\
			Break points: &  & 2009m3    &    &  \\
			\hline \hline
			
			\multicolumn{5}{l}{\textbf{Panel B: Unemployment rate}} \\ \hline
			&  & \multicolumn{3}{c}{Bai and Perron (2003) critical values}  \\
			& Test statistic & $1\%$ critical value & $5\%$ critical value & $10\%$ critical value \\
			\hline
			$F(1|0)$ &    3.47      &       3.30     &       2.77    &        2.49    \\
			$F(2|1)$ &    4.75      &       3.52       &     2.99     &       2.76    \\
			\hline
			Number of changes: &   & 2 & 2 & 2\\
			Break points: &    &   2001m10   &  2008m12  &  \\
			\hline
		\end{tabular}
	}
	\caption{Breaks in the Reduced forms. Trimming 0.2. $T=434$. }
	\label{tab:rf change 5iv}
\end{table}
\FloatBarrier

\begin{table}[h]
	\scalebox{.9}{
		\begin{tabular}{lllll}
			\hline
			\multicolumn{2}{c}{} & \multicolumn{3}{c}{\cite{baiperron1998} critical values}  \\
			& Test statistic & $1\%$ critical value & $5\%$ critical value & $10\%$ critical value \\
			\hline
			1983m9-2001m10: & & & & \\
			$F(1|0)$ &     36.81      &        24.45     &      20.08     &       17.97  \\
			Break point: & 1992m6 &      &     &  \\
			\hline
			2001m10-2009m3: & & & & \\
			$F(1|0)$ &     20.31      &        24.45     &      20.08     &       17.97  \\
			Break point: & 2006m9 &      &     &  \\
			\hline
			2009m4-2020m4: & & & & \\
			$F(1|0)$ &     4.90      &        24.45     &      20.08     &       17.97  \\
			Break point: & - &      &     &  \\
			\hline
		\end{tabular}
	}
	\caption{Breaks in the equation of interest over each stable RF segment. Trimming 0.15.}
	\label{tab:sf_change 5iv}
\end{table}
\FloatBarrier
\newpage
\begin{table}[h]
	\centering
	\begin{tabular}{lcc}
		\hline
		Sub-sample &  F statistic & N. Obs.\\
		\hline
		1983m9 - 2001m10 & 4.26  & 216\\
		\hline
	\end{tabular}
	\caption{\cite{kleibergen2006} KP statistic reported over the first stable RF period. We test the null of rank deficiency (rank equal to 1 for the (2,3)-matrix of RF coefficients) and the associated critical value are from the Chi2(4): 9.49 at 5\% and 7.78 at 10\%.}
	\label{tab:kp_ex}
\end{table}

\FloatBarrier
\begin{table}[h]
	\centering
	\scalebox{.85}{
		\begin{tabular}{l|ccc|ccc||c}
			\hline
			&    \multicolumn{3}{c|}{1983m9-1992m6}  &  \multicolumn{3}{c||}{1992m7-2001m10}  & 1983m9-2001m10\\\hline
			& TS2SLS  & GMM & TSGMM & TS2SLS & GMM & TSGMM & GMM\\
			& (1) & (2) & (3) & (4) & (5) & (6) & (7) \\\hline
			$\beta_0$  &  -0.0051	&	-0.0045	&	0.0000	&	0.0051	&	0.0022	&	0.0022	&	0.0002	\\
			&   (0.0254)	&	(0.0261)	&	(0.0221)	&	(0.0238)	&	(0.0202)	&	(0.0164)	&	(0.0161)	\\
			
			$\gamma_1$ &  1.4289	*** &	1.4779	*** &	1.4779	*** &	0.5117	&	0.6402***	&	0.6207***	&	1.2038***	\\
			&  (0.2515)	&	(0.1167)	&	(0.0965)	&	(0.8955)	&	(0.1244)	&	(0.1039)	&	(0.0973)	\\
			
			$\gamma_2$ &  -0.5522	** &	-0.5951	*** &	-0.5934	*** &	-0.1149	&	-0.2221**	&	-0.2356***	&	-0.4006***	\\
			&   (0.2398)	&	(0.1015)	&	(0.0824)	&	(0.5249)	&	(0.0995)	&	(0.0816)	&	(0.0850)	\\
			
			$\kappa_\Psi$  &  0.0136	&	0.0163	&	0.0178\textcolor{blue}{*} &	0.0135	&	0.0262\textcolor{blue}{**} &	0.0267\textcolor{blue}{***} &	0.0109\textcolor{blue}{*} \\	
			&  (0.0255)	&	(0.0131)	&	(0.0110)	&	(0.0459)	&	(0.0113)	&	(0.0093)	&	(0.0081)	\\
			
			$\gamma_f$  &  -0.2576	&	-0.2992	&	-0.2812	&	0.3596	&	-0.0481	&	-0.0360	&	-0.1786	\\
			&  (0.3473)	&	(0.1775)	&	(0.1254)	&	(1.9307)	&	(0.3953)	&	(0.2906)	&	(0.1595)	\\
			
			$\kappa_u$  &  0.0136	&	0.0135	&	0.0125	&	-0.0254	&	-0.0418	\textcolor{red}{**} &	-0.0445	\textcolor{red}{***} &	-0.0036	\\
			&  (0.0123)	&	(0.0112)	&	(0.0091)	&	(0.0264)	&	(0.0167)	&	(0.0129)	&	(0.0084)	\\
			\hline
			$N$                 &     & $106$ &   &   & $112$ &  & $218$\\\hline
			
		\end{tabular}
	}
	\caption{*$p < 0.10$, **$p < 0.05$, ***$p < 0.01$. Black stars indicate bi-directional testing. Blue stars right-tailed (positive) tests, and red stars (negative) left-tailed tests. One lag of expected inflation is included as instrument.}
	\label{tab:1983-2001_ex}
\end{table}
\FloatBarrier
\newpage
\section{Proof of theoretical results}
\textbf{$\bullet$ Definition 1}:

Recall that $\Pi_z= [I_{p_1},  O_{(q-p_1)\times p_1}]$ and  $\Pi^a \deff [\Pi_z, \Pi^{0}]$ is the matrix that augments $
\Pi^{0}$ with zeros and ones such that $Z_t'\Pi^a = [Z_t'\Pi_z, Z_t'\Pi^0] =[Z_{1t}', Z_t'\Pi^0] $. Let $Q \deff Q_1+Q_2$. Then:

$\bullet$ $V_{TS2SLS,vec} = \begin{bmatrix}D_1' \Omega_1 D_1 & D_1' \Omega_{12} D_2 \\
		D_2' \Omega_{12}' D_1 & D_2' \Omega_2 D_2
	\end{bmatrix},$  $A_i = \Pi^{a'} Q_i \Pi^a$, $D_i \,= \, A_i^{-1} (\Pi^{a'} M_i')$,  $M_1'= (I, Q_2 Q^{-1}, - Q_1 Q^{-1})$,
	$M_2' = (I, Q_1 Q^{-1}, - Q_2 Q^{-1})$, $a_i=\theta_{x,i}^0 \otimes I_q$,
	\begin{displaymath}\Omega_i = \left(
		\begin{array}{ccc}
			S_{u,i} & S_{uv,i}' a_i  & O \\
			a_i' S_{uv,i} & a_i'S_{v,i} a_i & O \\
			O & O &  a_i' (S_v-S_{v,i}) a_i   \\
		\end{array}
		\right),  \
		\Omega_{12} = \left(
		\begin{array}{ccc}
			O & O & S_{uv,1}'a_2 \\
			O & O & a_1'  S_{v,1}  a_2 \\
		    a_1	S_{uv,2}  & a_1'  S_{v,2} a_2  & O  \\
		\end{array}
		\right).
	\end{displaymath}
	
		\vspace{0.1in}
	
$\bullet$  $ V_{GMM,vec} = \diag(V_{GMM,1}, V_{GMM,2})$ and $V_{GMM,i} \, = \, \left[\, \Pi^{a'} \, Q_i \, (S_{u,i})^{-1} \, Q_i \, \Pi^{a}\, \right]^{-1}$.

$\bullet$ 	 $ V_{TSGMM,vec}^a = [\Gamma' \mathcal S^{-1} \Gamma]^{-1}$ with:
	
	\begin{align*}
		\mathcal S = \begin{bmatrix}
			S_{u,1} & O & S_{uv,1}' & O \\
			O & S_{u,2} & O & S_{uv,2}' \\
			S_{uv,1} & O & S_{v,1} & O \\
			O &S_{uv,2} & O & S_{v,2}
		\end{bmatrix},
		\quad \Gamma & = - \begin{bmatrix} Q_1\Pi^a  & O & O  \\
			O & Q_2 \Pi^a & O  \\
			O & O & \begin{bmatrix}I_{p_2} \otimes Q_{1} \\ I_{p_2} \otimes Q_2 \\
			\end{bmatrix}
		\end{bmatrix} = -\diag(\Gamma_1, \Gamma_2),
	\end{align*}
	with $\Gamma_1 = \diag(Q_1\Pi^a, Q_2 \Pi^a)$ and $\Gamma_2 = \begin{bmatrix}I_{p_2} \otimes Q_{1} \\ I_{p_2} \otimes Q_2
	\end{bmatrix}$.

	\begin{eqnarray*}
	\bullet	&& V_{TSGMM,vec} = (V_{GMM,vec}^{-1} +  \mathcal G'\mathcal G)^{-1}, \qquad \mathcal G =  \mathcal M_{\mathcal J} \mathcal E^{-1/2} \H, \qquad \mathcal M_{\mathcal J} = I - \mathcal J (\mathcal J' \mathcal J)^{-1} \mathcal J',  \\
		&& \mathcal J =\mathcal E^{-1/2} \ \Gamma_2, \mathcal E =  \diag(S_{v,1} - S_{uv,1} S_{u,1}^{-1} S_{uv,1}' \ , \ S_{v,2} - S_{uv,2} S_{u,2}^{-1} S_{uv,2}'),
		\\
		&&\mathcal H  = [\diag \ ( S_{uv,1} \ S_{u,1}^{-1},  S_{uv,2} \ S_{u,2}^{-1}) ] \ \Gamma_1.
	\end{eqnarray*}
- \textit{Additional notation:}  Let $\theta^0(t) \deff \theta_1^0 \mathbf{1}[t\leq T^0] + \theta_2^0 \mathbf{1}[t> T^0]$,  $\widetilde u_t \deff y_t- \hat W_t '\theta^0(t) = u_t + (W_t - \hat W_t)' \theta^0(t) = u_t + v_t '\theta_x^0(t) + Z_t'(\Pi_T - \hat \Pi)$, $\theta_x^0(t) \deff \theta_{x,1}^0 \mathbf{1}[t\leq T^0] + \theta_{x,2}^0 \mathbf{1}[t> T^0]$,   $\Pi_T \deff \Pi^0/r_T$, $\Pi_T^{a} \deff (\Pi_z,\Pi_T)$, and $\hat \Pi^a \deff (\Pi_z, \hat \Pi)$ such that $\hat W_t' = [Z_{1t}', \hat X_t'] = Z_t' [\Pi_z ,\hat \Pi] = Z_t'\hat \Pi^{a}$.  Also, for simplicity, let $\overline R_T = \diag(I_{p_1},r_T I_{p_2})$, $\hat A_1 (r) \deff \overline R_T \ T^{-1} \sum_{t=1}^{[Tr]} \hat W_t \hat W_t' \ \overline R_T $,  $\hat A_{2} (r) \deff \hat A_1(1)- \hat A_{1} (r)$, $\xi_1(r) \deff \overline R_T T^{-1/2} \sum_{t=1}^{[Tr]} \hat W_t \widetilde u_t$,  $\xi_2(r) \deff \xi_1(1)-\xi_1(r)$, and recall that $\sum_{1} \deff \sum_{t=1}^{T_1^0}$, $\sum_2 \deff \sum_{t=T_1^0+1}^{T}$.

\textbf{$\bullet$ Proof of Theorem \ref{theo:consistent}:}\\
	For simplicity, in each of parts (i),(ii), (iii), we write $\hat{\theta}_{vec}$ for each of
	 $\hat{\theta}_{TS2SLS,vec}(\lambda_0)$, $\hat{\theta}_{GMM,vec}(\lambda_0)$, and $\hat{\theta}_{TSGMM,vec}(\lambda_0)$, and  $\hat{\theta}_{vec} \deff \vec(\hat{\theta}_1,\hat \theta_2)$, where $\hat \theta_i$ are the $p\times 1$ estimators in each of the two samples.\\
\textit{(i). Asymptotic properties of $\hat{\theta}_{vec}=\hat{\theta}_{TS2SLS,vec}(\lambda_0)$}. This proof generalizes \cite{HHB}, Theorem 3, to near-weak instruments. Since $\hat \Pi = \Pi_T + (T^{-1} \sum_{t=1}^T Z_t Z_t')^{-1} T^{-1} \sum_{t=1}^T Z_t v_t'$, and $T^{-1} \sum_{t=1}^T Z_t Z_t' \inp Q$, $T^{-1} \sum_{t=1}^T Z_t v_t \inp 0$ by Assumptions \ref{ass:reg1}-\ref{ass:reg-identif} and a WLLN for NED processes, $\Hat \Pi \inp \Pi_T$, and $\overline R_T \Hat \Pi \inp \Pi^a$. With the additional notation above, $T^{1/2} \overline R_T^{-1} (\hat \theta_i -\theta_i^0) = \hat A_i^{-1} (\lambda_0) \ \xi_i (\lambda_0)$, where:
\begin{align*}
	\hat A_i^{-1} (\lambda_0) &= \overline R_T \ T^{-1} \textstyle \sum_{i} \hat W_t \hat W_t' \ \overline R_T  = \overline R_T \ \hat \Pi^{a'} (T^{-1} \textstyle \sum_{i} Z_t Z_t' ) \overline R_T \hat \Pi^a \\
	&= \overline R_T \ \Pi_T^{a'}  (T^{-1} \textstyle \sum_{i} Z_t Z_t' ) \overline R_T \Pi_T^{a'} +\op(1) = \Pi^{a'} Q_i \Pi^a +\op(1) = A_i+\op(1).
\end{align*}
On the other hand,
$\xi_i (\lambda_0) = R_T T^{-1/2} \sum_{1} \hat W_t \widetilde u_t = (\Pi^a +\op(1))' T^{-1/2} \sum_{i} Z_t \widetilde u_t$. Note that, for $i=1,2$, and $j=1,2, j\neq i$:
\begin{align*}
&\textstyle T^{-1/2} \sum_{i} Z_t \widetilde u_t \\
&= \textstyle 	T^{-1/2} \sum_{i} Z_t u_t + T^{-1/2} \sum_{i} Z_t v_t'\theta_{x,i}^0 - T^{-1} \sum_{i} Z_t Z_t' (\sum_{t=1}^T Z_t Z_t')^{-1} (T^{-1/2}\sum_{t=1}^T Z_t v_t'\theta_{x,i}^0) \\
& =\textstyle  T^{-1/2} \sum_{i} Z_t u_t +\textstyle T^{-1/2} \sum_{i} Z_t v_t'\theta_{x,1}^0 - (Q_i^{-1} Q +\op(1)) \textstyle T^{-1/2} \sum_{i} Z_t v_t'\theta_{x,1}^0 \\
&- (Q_i^{-1} Q +\op(1)) \textstyle T^{-1/2} \sum_{j} Z_t v_t'\theta_{x,1}^0 \\
& = \Psi^{u}_i + (I-Q_i Q^{-1}) \Psi_1^v \theta_{x,i}^0 - (Q_i Q^{-1}) \Psi_2^v \theta_{x,i}^0 + \op(1) \\
& = M_i'\vec (\Psi^{u}_i, \Psi_1^v \theta_i, \Psi_2^v \theta_i),
\end{align*}
where $\Psi^{u}_i = T^{-1/2} \sum_{1} Z_t u_t, \Psi^{v}_i = T^{-1/2} \sum_{i} Z_t  v_t', i=1,2$.
Note that the terms $\Psi_1^v, \Psi_2^v$ are present because of the full-sample first-stage. Under Assumptions \ref{ass:reg1}-\ref{ass:reg-identif}, by the CLT for NED processes (Theorem 2.11 in \cite{wooldridge1988}),  $\vec (\Psi^{u}_i, \Psi_1^v, \Psi_2^v ) \ind \mathcal{N}(0, \Omega_i)$, where $\Omega_i$ is defined above. Thus, $\textstyle T^{-1/2} \sum_{i} Z_t \widetilde u_t \ind \mathcal{N}(0, M_i' \Omega_i M_i)$, so
$
\Lambda_T [\hat \theta_i - \theta_i^0] \ind \mathcal{N}\left(0, D_i' \Omega_1 D_i\right).
$
Moreover, because of the full-sample first-stage, asymptotically, $\Lambda_T [\hat \theta_1 - \theta_1^0] \not \perp \Lambda_T [\hat \theta_2 - \theta_2^0]$, and
$
\acov \left\{\Lambda_T [\hat \theta_1 - \theta_1^0], \Lambda_T [\hat \theta_2 - \theta_2^0]\right\}   = D_1' \Omega_{12} D_2.
$
Putting these together, $[I_2 \otimes \Lambda_T] \ [\hat \theta_{vec} - \theta_{vec}^0]
\ind \mathcal{N} \left( 0, V_{TS2SLS,vec}\right)$.

\textit{(ii). Asymptotic properties of $\hat{\theta}_{vec}=\hat{\theta}_{GMM,vec}(\lambda_0)$}. Let $\overline W^0 \deff \diag(\overline Z_1^0, \overline Z_2^0)$, where $\overline Z_1^0, \overline Z_2^0$ are the first $T_1^0\times q$, respectively the last $(T-T_1^0) \times q$ blocks of $Z$, the matrix with rows $Z_t'$ stacked in order. Also let $\overline W \deff \vec (\overline W_1^0, \overline W_2^0)$ be the partitions of $W$ (the matrix with rows $W_t'$) at $\hat T$, respectively $T^0$, and similarly partition $y=\vec (\overline y_1^0, \overline y_2^0)$ and  $u= \vec (\overline U_1^0, \overline U_2^0)$. Then for the weighting matrices $\hat S_{u,i}^{-1} \inp S_{u,i}$,
\begin{align*}
	\hat \theta_{GMM,i} &= (\overline W_i^{0'} \overline Z_i^0 \, \hat S_{u,i}^{-1}  \,  \overline Z_i^{0'} \overline W_i^0)^{-1} \,  \overline W_i^{0'} \overline Z_i^0 \,  \hat S_{u,i}^{-1} \, \overline Z_i^{0'} \overline y_i \\&= \theta_i^0+ ( \overline W_i^{0'} \overline Z_i^0 \, \hat S_{u,i}^{-1}  \,  \overline Z_i^{0'} \overline W_i^0)^{-1} \,  \overline W_i^{0'} \overline Z_i^0 \,  \hat S_{u,i}^{-1} \, \overline Z_i^{0'} \overline U_i^0.
\end{align*}
The asymptotic distributions for the these split sample estimators follow from the WLLN and CLT for NED processes, under Assumptions \ref{ass:reg1}0\ref{ass:reg-identif}. In particular, $ \overline R_T T^{-1} \overline W_i^{0'} \overline Z_i^0  \inp \Pi^{a'} Q_i $, $ T^{-1/2} \overline Z_i^{0'} \overline U_i^0 \ind \mathcal{N}(0, S_{u,i})$, for $i=1,2$, and since $\hat S_{u,i} \inp S_{u,i}$, $\Lambda_T [\hat \theta_{GMM,i} - \theta_i^0] \ind \mathcal{N} (0, (\Pi^{a'} Q_i S_{u,i}^{-1} Q_i \Pi^a)^{-1})$, and they are asymptotically independent by Assumption \ref{ass:reg1}.

\textit{(ii). Asymptotic properties of $\hat{\theta}_{vec}=\hat{\theta}_{TSGMM,vec}(\lambda_0)$ and  $\hat{\Pi}_{vec}=\hat{\Pi}_{TSGMM,vec}(\lambda_0)$}. Let $\breve \Lambda_T \deff \ \diag(I_2 \otimes \Lambda_T, T^{1/2} I_{qp2})$,  $\hat \beta  \deff \begin{bmatrix} \hat \theta \\
	\hat \Pi_{TSGMM,vec}  \end{bmatrix}$ and $\beta_T^0 \deff \begin{bmatrix} \theta_{vec}^0  \\
	\Pi_{vec}^0/r_T \end{bmatrix}$.  Then, the sample moment conditions for the estimator $\hat \beta$ can be rewritten as:
\begin{align*}
	\breve g_T(\theta_{vec}, \Pi) = \begin{cases} g_{T,1}(\theta_{vec}) \deff T^{-1} \overline Z^{0'}(y - \overline W^0 \theta_{vec}) & \\
		g_{T,2}(\Pi_{vec,T}) \deff T^{-1} \begin{bmatrix} \left(I_{p_2} \otimes \overline Z_1^{0'} \right)\left( X_{vec,1}^0 -\left(I_{p_2} \otimes \overline Z_1^0 \right) \Pi_{vec,T} \right) \\
			\left(I_{p_2} \otimes \overline Z_2^{0'} \right)\left(X_{vec,2}^0 - \left(I_{p_2} \otimes \overline  Z_2^0 \right) \Pi_{vec,T}\right)  \\
		\end{bmatrix}  & \\
	\end{cases},
\end{align*}
where $\theta_{vec}\deff\vec(\theta_1,\theta_2)$, $\Pi_{vec,T} \deff\vec(\Pi/r_T)$  for any $\Pi$,   $\Pi_{vec,T}^0 \deff\vec(\Pi^0/r_T)$, and $X_{vec,i}^0\deff \vec(\overline X_i^0)$, for $i=1,2$, with $\overline X^0\deff\vec(\overline X_1^0, \overline X_2^0)$ being the partition of $X$
at the true change $T^0$. Let $\overline{\mathcal Z_i}^{0'} \deff I_{p_2} \otimes \overline Z_i^{0'}
$, $\overline{\mathcal Z}^{0'} \deff
\diag (\overline{\mathcal Z}_1^{0'},  \overline{\mathcal Z}_2^{0'})$, $\mathcal X_{vec}^0 \deff  \begin{bmatrix}X_{vec,1}^0\\ X_{vec,2}^0\end{bmatrix}$, $\mathcal Z_{vec}^0 \deff \begin{bmatrix}\overline{\mathcal Z}_1^0\\ \overline{\mathcal Z}_2^0\end{bmatrix}$.  This notation is employed to stack the moment conditions for each column of $\Pi$ to arrive at the usual formula for a GMM estimator:
\begin{align*}
	g_{T,2}(\Pi_{vec,T}) &= T^{-1} \overline{\mathcal X}^{0'}
	\left(\mathcal X_{vec}^0 -
	\mathcal Z_{vec}^0 \Pi_{vec,T}\right) \\
	g_{T}(\beta) = & T^{-1}\begin{cases} \overline Z^{0'}(y - \overline W^0 \theta_{vec})  &  \\
		\overline{\mathcal Z}^{0}
		\left(\mathcal X_{vec}^0 -
		\mathcal Z_{vec}^0 \Pi_{vec,T}\right) &  \\
	\end{cases} =T^{-1} \begin{bmatrix} \overline Z^{0} & O \\
		O & \overline{\mathcal Z}^{0} \\
	\end{bmatrix}'\left(\begin{bmatrix}y \\
		\mathcal X_{vec}^0  \\
	\end{bmatrix} - \begin{bmatrix}\overline W^0 & O \\
		O & \mathcal Z_{vec}^0 \\
	\end{bmatrix} \beta\right).
\end{align*}

Hence, the estimation of $\beta_T^0$ is now written as a usual GMM problem, therefore, by usual GMM asymptotics,
\begin{align}\label{eq:GMMaug0}
	\breve \Lambda_T( \hat \beta - \beta_{T}^0) \ind \mathcal{N} (0 ,  [\Gamma' \mathcal S^{-1} \Gamma]^{-1}).
\end{align}
Below we justify this result. First,
\begin{align*}
	\Gamma & = \plim \frac{\partial \breve g_T (\beta_T^0)}
	{\partial \beta'} (T^{-1/2} \breve \Lambda_T^{-1}) = - T^{-1} \plim \begin{bmatrix} \overline Z^{0} & O \\
		O & \overline{\mathcal Z}^{0} \\
	\end{bmatrix}'\begin{bmatrix}\overline W^0 & O \\
		O & \mathcal Z_{vec}^0 \\
	\end{bmatrix} (T^{-1/2} \breve \Lambda_T^{-1}) \\
	& =- \begin{bmatrix}
		\plim [(T^{-1} \overline Z^{0'} \overline W^0)(I_2 \otimes  {\overline R}_T)] & O \\
		O & \plim (T^{-1}
		\overline{\mathcal Z}^{0'} \mathcal Z_{vec}^0) \\
	\end{bmatrix}
	= -\diag (\Gamma_1,\Gamma_2),
\end{align*}
and the latter equality holds because under Assumptions \ref{ass:reg1}-\ref{ass:reg-IV},
\begin{align*}
	\Gamma_1 & =  -\diag [\ \plim (T^{-1} \overline Z_1^{0'} \overline W_1^0 \ \overline R_T), \plim (T^{-1} \overline Z_2^{0'} \overline W_2^0 \ \overline R_T)\ ] \\
	&= -\diag[\ \plim (T^{-1} \overline Z_1^{0'} \overline Z_1^0) \Pi^a, \plim (T^{-1} \overline Z_2^{0'} \overline Z_2^0) \Pi^a \ ] = -\diag(Q_i \Pi^a), \\
	\Gamma_2 & =  -\plim  (T^{-1} \overline{\mathcal Z}^{0'} \mathcal Z_{vec}^0) = -\begin{bmatrix}  \plim  (T^{-1} \overline{\mathcal Z}_{1}^{0'} \overline{\mathcal Z}_{1}^0) \\
		\plim  (T^{-1} \overline{\mathcal Z}_{2}^{0'} \overline{\mathcal Z}_{2}^0) \\
	\end{bmatrix} =- \begin{bmatrix}I_{p_2} \otimes Q_{1} \\
		I_{p_2} \otimes Q_2 \\
	\end{bmatrix}.
\end{align*}
Let	$\mathcal V_{vec}^0$ be defined as $\mathcal X_{vec}^0$ with $X_t$ replaced by $v_t$, and let $\mathcal S_{1,1}$ be a $q\times q$ matrix, and $\mathcal S_{2,2}$ is $(p_2 q) \times (p_2 q)$ matrix. Then:
\begin{align*}\mathcal S &= \avar \, \left \{ T^{-1/2} \begin{bmatrix} \overline Z^{0} & O \\
	O & \overline{\mathcal Z}^{0} \\
\end{bmatrix}'\begin{bmatrix}U \\
	\mathcal V_{vec}^0 \end{bmatrix}\right\} = \avar \, \begin{bmatrix}  T^{-1/2} \overline Z^{0'}u \\
	T^{-1/2} \overline{\mathcal Z}^{0'} \mathcal V_{vec}^0 \\
\end{bmatrix} \deff \begin{bmatrix} \mathcal S_{1,1} & \mathcal S_{1,2} \\
	\mathcal S_{1,2}' & \mathcal S_{2,2} \end{bmatrix}, \end{align*}
where
\begin{align*}
	\mathcal S_{1,1} &=  \avar ( T^{-1/2} \overline Z^{0'}u ) = \diag(S_{u,1},S_{u,2}),\\
	\mathcal S_{1,2} &= \acov ( T^{-1/2} \overline Z^{0'}u, T^{-1/2} \overline{\mathcal Z}^{0'} \mathcal V_{vec}^0) \\
	& =  \diag \left[ \acov \left( \textstyle T^{-1/2} \sum_1 Z_t u_t, \textstyle \sum_1 Z_t \otimes v_t\right), \acov\left( T^{-1/2} \textstyle\sum_2 Z_t u_t, \textstyle\sum_2 Z_t \otimes v_t\right)\right] \\
	& = \diag (S_{uv,1}',S_{uv,2}'),\\
	\mathcal S_{2,2} & = \avar ( T^{-1/2} \overline{\mathcal Z}^{0'} \mathcal V_{vec}^0) = \avar \begin{bmatrix} T^{-1/2} \overline{\mathcal Z}_{1}^{0'} V_{vec,1}^0  \\
		T^{-1/2} \overline{\mathcal Z}_{2}^{0'} V_{vec,2}^0   \\
	\end{bmatrix} = \diag[ \avar \ ( T^{-1/2} (I_{p_2} \otimes \overline Z_i^{0'}) V_{vec,i}^0)] \\
	& = \diag\left[ \avar  \left(T^{-1/2} \textstyle \sum_1 Z_t \otimes v_t\right), \avar \left(T^{-1/2} \textstyle\sum_2 Z_t \otimes v_t\right)\right] = \diag (S_{v,1},S_{v,2}) . \end{align*}
This completes the proof of \eqref{eq:GMMaug0}. To get from this result a formula for the asymptotic variance of the TSGMM estimator as a function of that of the split-sample GMM estimator, we use the partitioned inverse formula in Abadir and Magnus (2005), pp. 106. For any $A,B, C,D$ matrices, for which $A,D$ and $E = D- C A^{-1} B$ and $F =A-BD^{-1} C$ are nonsingular,
\begin{align*}
	\begin{bmatrix} A & B \\
		C & D \\
	\end{bmatrix}^{-1} & = \begin{bmatrix} A^{-1} + A^{-1} B E^{-1} C A^{-1} & -A^{-1} B E^{-1} \\
		-E^{-1} C A^{-1}  & E^{-1}\\
	\end{bmatrix}  = \begin{bmatrix} F^{-1} & -F^{-1} B D^{-1} \\
		-D^{-1} C F^{-1}  & D^{-1} +D^{-1}C F^{-1} B D^{-1}\\
	\end{bmatrix}.
\end{align*}
We use the first formula for $\mathcal S^{-1}$, and the second for $(\Gamma' \mathcal S^{-1} \Gamma)^{-1}$. Let $\mathcal E =  \mathcal S_{2,2} - \mathcal S_{2,1} \S_{1,1}^{-1} \mathcal S_{1,2}$.
\begin{align*}
	\mathcal S^{-1} & = \begin{bmatrix} \mathcal S_{1,1} & \mathcal S_{1,2} \\
		\mathcal S_{1,2}' & \mathcal S_{2,2} \\
	\end{bmatrix}^{-1} = \begin{bmatrix} \S_{1,1}^{-1} + \S_{1,1}^{-1} \, \S_{1,2} \, \mathcal E^{-1} \, \S_{1,2}' \, \S_{1,1}^{-1}  & - \S_{1,1}^{-1}\ \S_{1,2} \ \mathcal E^{-1} \ \\
		-\mathcal E^{-1} \ \S_{1,2}' \ \S_{1,1}^{-1}   & \mathcal E^{-1} \\
	\end{bmatrix},\\
	\Gamma' \S^{-1} \Gamma  & = \begin{bmatrix} \Gamma_1' ( \S_{1,1}^{-1} + \S_{1,1}^{-1} \, \S_{1,2} \, \mathcal E^{-1} \, \S_{1,2}' \, \S_{1,1}^{-1}) \Gamma_1  & - \Gamma_1' \S_{1,1}^{-1}\ \S_{1,2} \ \mathcal E^{-1} \Gamma_2\ \\
		-\Gamma_2' \mathcal E^{-1} \ \S_{1,2}' \ \S_{1,1}^{-1} \Gamma_1   & \Gamma_2' \mathcal E^{-1} \Gamma_2
	\end{bmatrix}
\end{align*}
Then, according to the second formula above, $V_{TSGMM,vec} = \mathcal F^{-1}$, where
\begin{align*}
	\mathcal F &\deff \Gamma_1' ( \S_{1,1}^{-1} + \S_{1,1}^{-1} \, \S_{1,2} \, \mathcal E^{-1} \, \S_{1,2}' \, \S_{1,1}^{-1}) \Gamma_1 -  (\Gamma_1' \S_{1,1}^{-1}\ \S_{1,2} \ \mathcal E^{-1} \Gamma_2 ) (\Gamma_2' \mathcal E^{-1} \Gamma_2)^{-1}\Gamma_2' \mathcal E^{-1} \ \S_{1,2}' \ \S_{1,1}^{-1} \Gamma_1.
\end{align*}
Recalling that $\mathcal J \deff \mathcal E^{-1/2}\Gamma_2$, $\H  \deff \S_{1,2}' \ \S_{1,1}^{-1} \Gamma_1$, it follows that:
\begin{align*}
	\mathcal F = \Gamma_1'  \S_{1,1}^{-1} \Gamma_1 + \H' \mathcal E^{-1/2} (I- \mathcal J (\mathcal J' \mathcal J)^{-1} \mathcal J')  \mathcal E^{-1/2} \H =  \Gamma_1'  \S_{1,1}^{-1} \Gamma_1 + \H' \mathcal E^{-1/2} \mathcal M_{\mathcal J} \mathcal E^{-1/2} \H.
\end{align*}
Note that from the above, $\Gamma_1' \S_{1,1} \Gamma_1 = V_{GMM,vec}^{-1}$, completing the proof. $\blacksquare$

\textbf{$\bullet$ Proof of Theorem \ref{theo:eff}:}

(i) Recall $\mathcal G \deff \mathcal M_{\mathcal J} \mathcal E^{-1/2} \H$. Then $\mathcal A \deff (\mathcal M_{\mathcal J} \mathcal E^{-1/2} \H)' \mathcal M_{\mathcal J} \mathcal E^{-1/2} \H  = \mathcal G'\mathcal G$ is psd by construction, so $V_{TSGMM,vec} \leq V_{GMM,vec}$, and its rank is equal to the rank of $\mathcal G$. Since $\mathcal E$ is a Schur complement derived from the variance matrix $\mathcal S$ with is pd by Assumption \ref{ass:reg-change}, $\mathcal E^{-1/2}$ is $2p_2q \times 2 p_2 q$ and full rank. $\Gamma_2$ is of dimension $2 p_2q \times p_2 q$ and of full column rank $p_2 q$ by Assumption \ref{ass:reg-change}. Therefore, $\mathcal J$ is $2p_2 q \times p_2 q$ and of rank $p_2 q$,  $\mathcal M_{\mathcal J}$ is $2p_2q \times 2p_2 q$ and of rank $p_2 q$, hence $ \mathcal B= \mathcal E^{-1/2}\mathcal M_{\mathcal J} \mathcal E^{-1/2}$ is $2p_2q \times 2p_2 q$ and of rank $p_2q$. We can therefore write $\mathcal B = \mathcal C' \Lambda \mathcal C$, where $\mathcal C$ is the $2p_2 \times 2p_2$ full rank eigenvector matrix, and $\Lambda$ is the $2p_2 \times 2 p_2$ diagonal matrix of eigenvalues, with the first $p_2$ diagonal elements strictly positive, and the rest equal to zero. Now  $\H = \S_{1,2}' \ \S_{1,1}^{-1} \Gamma_1 $ where $\Gamma_1$ is $2q \times 2 p$ and full column rank $2p$ and $\S_{1,1}$ is $2q \times 2q$ is full rank $2q$ by Assumptions \ref{ass:reg1}-\ref{ass:reg-change}. If $\S_{1,2}$, of dimension $2p_2q \times 2q$, is also full column rank $2 q$, then $\H$ has rank equal to $2p$. It follows that $\mathcal D = \mathcal C \H$ is of rank $2p$, and
$\mathcal A = \mathcal D'  \Lambda \mathcal D $ is of rank $min(p_2q, 2p)$. So if $p_2q \geq 2p$, it is also of rank $2p$, implying that $V_{TSGMM,vec} < V_{GMM,vec}$ in this case.

(ii) Under Assumptions \ref{ass:reg-secondmom}-\ref{ass:cond-hom}, one can show that $M_1' \Omega_1 M_1 = \lambda^0 c Q$,
where $c= \Phi_u + 2(1-\lambda^0) \Phi_{uv}' \theta_{x,1}^0 + (1-\lambda^0) \theta_{y,1}^{0'}  \Phi_v \theta_{x,1}^0$. So $\Pi^{a'} M_1' \Omega_1 M_1\Pi^{a} = \lambda^0 c (\Pi^{a'}Q \Pi^a \deff \lambda^0)  \deff \lambda^0 c A$.  Also, $A_{i}=\Pi^{a'} Q_1 \Pi^{a} = \lambda^0 \Pi^{a'} Q \Pi^a= \lambda_0 A$, so $V_{TS2SLS,1}^{-1} = \lambda^0 (\Pi^{a'} Q \Pi^a )/ c = \lambda^0 A/c$. On the other hand, $V_{GMM,1}^{-1} = \lambda^0 A/\Phi_u$, so we can compare the two by comparing $c$ with $\Phi_u$. If
$
c-\Phi_u = 2 (1-\lambda^0)  [ \Phi_{uv}' \theta_{x,1}^0+\theta_{y,1}^{0'} \Phi_v \theta_{x,1}^0  ]\leq 0$, then $V_{2SLS,1}^{-1} \geq V_{GMM,1}^{-1}$, which implies that $V_{TS2SLS,1} \leq V_{GMM,1}$. The proof for $V_{TS2SLS,2} \leq V_{GMM,2}$ is similar.

We  also have:
\begin{equation*}
\displaystyle V_{TS2SLS,1} =\frac{c}{\lambda^0} A^{-1} = \frac{\Phi_u}{\lambda^0} + \frac{1-\lambda^0}{\lambda^0} [ 2  \Phi_{uv}' \theta_{x,1}^0+ \theta_{y,1}^{0'} \Phi_v \theta_{x,1}^0 ].
\end{equation*}
Recall that  $Q_i = \lambda_i^0 Q$, with $\lambda_1^0 \deff \lambda^0$, $\lambda_2^0 \deff 1-\lambda^0$. Also, $S_{j,i} = \lambda_i^0 \ (\Phi_{j} \otimes Q) $, for $j\in\{u,v,uv\}$. Also,  $\mathcal E_i  = \lambda_i^0 \ (\Phi_v-\Phi_{uv} \Phi_u^{-1} \Phi_{uv}') \ \otimes Q \deff  \lambda_i^0 \ (\overline{\mathcal  E} \otimes Q )$. Hence:
\begin{align*}
	\mathcal J &=-  \begin{bmatrix} \sqrt{\lambda^0} \  \overline{\mathcal  E}^{-1/2} \otimes Q^{1/2} \\
		\sqrt{1-\lambda^0} \   \overline{\mathcal  E}^{-1/2} \otimes Q^{1/2} \end{bmatrix} \\
	\mathcal J' \mathcal J & =  \overline{\mathcal  E}^{-1} \otimes Q, (\mathcal J' \mathcal J)^{-1} = \overline{\mathcal  E} \otimes Q^{-1} \\
	\mathcal J (\mathcal J' \mathcal J)^{-1} \mathcal J' &=  \begin{bmatrix} \lambda^0  \ I_{p_2 q} & \sqrt{\lambda^0 (1-\lambda^0)} \ I_{p_2 q} \\
		\sqrt{\lambda^0 (1-\lambda^0)}  \ I_{p_2 q} &  (1-\lambda^0)\ I_{p_2 q}
	\end{bmatrix} \\
	\mathcal M_{\mathcal J} &= \begin{bmatrix} (1- \lambda^0) \ I_{p_2 q} & - \sqrt{\lambda^0 (1-\lambda^0)} \ I_{p_2 q} \\
		- \sqrt{\lambda^0 (1-\lambda^0)}  \ I_{p_2 q} &  \lambda^0\ I_{p_2 q}
	\end{bmatrix}  \\
	\mathcal E_i^{-1/2} \Gamma_i  & = \sqrt{\lambda_i^0} \ (\overline{\mathcal  E}^{-1/2} \otimes Q^{-1/2} )S_{uv,i}' \ S_{u,i}^{-1} \ Q_{i}\Pi^a \\
	& = \sqrt{\lambda_i^0} \ (\overline{\mathcal  E}^{-1/2} \otimes Q^{-1/2} )  [(\Phi_{uv}' \ \Phi_{u}^{-1}) \otimes I_q ]\ ( Q \ \Pi^a) \\
	& =  \sqrt{\lambda_i^0} (\overline{\mathcal  E}^{-1/2}\Phi_{uv}' \ \Phi_{u}^{-1}) \otimes Q^{1/2} \Pi^a.
\end{align*}
Also, let
$	\mathcal G = \mathcal M_{\mathcal J} \mathcal E^{-1/2} \H \deff \begin{bmatrix} \mathcal G_{11} & \mathcal G_{12} \\
		\mathcal G_{21} & \mathcal G_{22}. \end{bmatrix}
$
Then:
\begin{align*}
	\mathcal G_{11} &=  \sqrt{\lambda^0} (1- \lambda^0)(\overline{\mathcal  E}^{-1/2}\Phi_{uv}' \ \Phi_{u}^{-1}) \otimes (Q^{1/2} \Pi^a)    \\
	\ \mathcal G_{12} &=-\sqrt{\lambda^0} (1- \lambda^0)(\overline{\mathcal  E}^{-1/2}\Phi_{uv}' \ \Phi_{u}^{-1}) \otimes (Q^{1/2} \Pi^a)  \\
	\ \mathcal G_{21} &=-\lambda^0 \sqrt{1- \lambda^0}(\overline{\mathcal  E}^{-1/2}\Phi_{uv}' \ \Phi_{u}^{-1}) \otimes (Q^{1/2} \Pi^a)  \\
	\mathcal G_{22} &=  \lambda^0 \sqrt{1- \lambda^0}(\overline{\mathcal  E}^{-1/2}\Phi_{uv}' \ \Phi_{u}^{-1}) \otimes (Q^{1/2} \Pi^a).
\end{align*}
Letting $\gamma \deff (\overline{\mathcal  E}^{-1/2}\Phi_{uv}' \ \Phi_{u}^{-1})$, and $\delta\deff \gamma' \gamma \deff \Phi_{u}^{-2} \Phi_{uv}'\overline{\mathcal  E}^{-1}\Phi_{uv}$,  it follows that:
\begin{align*}
	-\mathcal G &= \sqrt{\lambda^0 (1-\lambda^0)} \begin{bmatrix}
		\sqrt{1-\lambda^0} \ \gamma & -\sqrt{1-\lambda^0} \ \gamma \\
		- \sqrt{\lambda^0} \ \gamma & \sqrt{\lambda^0}  \ \gamma
	\end{bmatrix} \otimes  (Q^{1/2} \Pi^a) \\
	\mathcal G' \mathcal G & = \lambda^0 (1-\lambda^0) \delta \begin{bmatrix}
		1 \  & -1 \  \\
		- 1 \  & 1
	\end{bmatrix} \otimes  (\Pi^{a'} Q \Pi^a) = \lambda^0 (1-\lambda^0) \delta \begin{bmatrix}
		1 \  & -1 \  \\
		- 1 \  & 1
	\end{bmatrix} \otimes  A.
\end{align*}

Recall that  $V_{TSGMM,vec} = \mathcal F^{-1}$, with
\begin{align*}
\mathcal F &=\Gamma_1'  \S_{1,1}^{-1} \Gamma_1 + \H' \mathcal E^{-1/2} \mathcal M_{\mathcal J} \mathcal E^{-1/2} \H  = \diag( \lambda_i^0 \Phi_u^{-1} A) + \mathcal G' \mathcal G \\
& =  \begin{bmatrix}
\lambda^0 [\Phi_u^{-1}+ (1-\lambda^0) \delta] & - \lambda^0 (1-\lambda^0) \delta \\
- \lambda^0 (1-\lambda^0) \delta & (1-\lambda^0) [\Phi_u^{-1}+ \lambda^0 \delta]
\end{bmatrix} \otimes A \deff \overline{\mathcal F} \otimes A, \\
 V_{TSGMM,vec} &= \mathcal F^{-1} =\overline{\mathcal F}^{-1} \otimes A^{-1}.
\end{align*}
Now,
\begin{align*}
\det \overline{\mathcal F} &= \lambda^0 (1-\lambda^0)\{ [\Phi_u^{-1}+ (1-\lambda^0) \delta][\Phi_u^{-1}+ \lambda^0 \delta]- \lambda^0 (1-\lambda^0) \delta^2\} \\
& = \lambda^0 (1-\lambda^0) (\Phi_u^{-2} + \Phi_u^{-1} \delta) = \lambda^0 (1-\lambda^0) \Phi_u^{-2} (1 + \Phi_u \delta) \\
\overline{\mathcal F}^{-1} &= \frac{\Phi_u^2}{\lambda^0 (1-\lambda^0) (1 + \Phi_u \delta) } \begin{bmatrix}
(1-\lambda^0) [\Phi_u^{-1}+ \lambda^0 \delta] &  \lambda^0 (1-\lambda^0) \delta \\
 \lambda^0 (1-\lambda^0) \delta & \lambda^0 [\Phi_u^{-1}+ (1-\lambda^0) \delta]
\end{bmatrix} \\
& = \frac{\Phi_u^2}{(1 + \Phi_u \delta) } \begin{bmatrix}
\frac{1}{\lambda^0} \ [\Phi_u^{-1}+ \lambda^0 \delta] &   \delta \\
  \delta & \frac{1}{1-\lambda^0} [\Phi_u^{-1}+ (1-\lambda^0) \delta]
\end{bmatrix}.
\end{align*}
\begin{align*}
V_{TSGMM,1} &= \frac{\Phi_u(1+ \lambda^0 \Phi_u \delta)}{\lambda^0(1 + \Phi_u \delta)} A^{-1} = \left(\frac{\Phi_u}{\lambda^0} + \frac{\Phi_u(1+ \lambda^0 \Phi_u \delta)}{\lambda^0(1 + \Phi_u \delta)} - \frac{\Phi_u}{\lambda^0} \right)  A^{-1}\\
& = \left(\frac{\Phi_u}{\lambda^0} + \frac{\Phi_u(1+ \lambda^0 \Phi_u \delta - 1 - \Phi_u \delta)}{\lambda^0(1 + \Phi_u \delta)} \right)A^{-1}= \left(\frac{\Phi_u}{\lambda^0} - \frac{1-\lambda^0}{\lambda^0}\frac{\Phi_u^2 \delta}{1+\Phi_u \delta} \right)A^{-1}\\
V_{TS2SLS,1} & = \left(\frac{\Phi_u}{\lambda^0} + \frac{1-\lambda^0}{\lambda^0} ( 2  \Phi_{uv}' \theta_{x,1}^0+ \theta_{y,1}^{0'} \Phi_v \theta_{x,1}^0 ) \right)A^{-1}\\
V_{TS2SLS,1} - V_{TSGMM,1} &= \frac{1-\lambda^0}{\lambda^0} \left( 2  \Phi_{uv}' \theta_{x,1}^0+ \theta_{y,1}^{0'} \Phi_v \theta_{x,1}^0  + \frac{\Phi_u^2 \delta}{1+\Phi_u \delta} \right)  A^{-1}\geq 0 \\
 \Leftrightarrow \ & 2  \Phi_{uv}' \theta_{x,1}^0+ \theta_{y,1}^{0'} \Phi_v \theta_{x,1}^0 \geq \ - \frac{\Phi_u^2 \delta}{1+\Phi_u \delta}.
\end{align*}
The proof for $V_{TSGMM,2}$ and $V_{TS2SLS,2}$ is similar and therefore omitted. $\blacksquare$

\textbf{$\bullet$ Proof of Theorem \ref{theo:change-order}}:\\
(i) \textbf{Asymptotic properties of $\hat{\lambda}$}\\

- We start by proving the following Lemma.
\begin{lemma}\label{lemma:1}
Under the assumptions maintained in Theorem \ref{theo:change-order}, (i) $\hat A_1(r) = O_p(1)$, uniformly in $r\in [0,1]$ (\ur \ henceforth); (ii) $\xi_1(r) = O_p(1)$ \ur.
\end{lemma}
\textbf{Proof of Lemma \ref{lemma:1}:}

\emph{Part (i).} Note that $\hat A_1(r) = \overline R_T \  T^{-1} \sum_{t=1}^{[Tr]} \hat W_t \hat W_t' \ \overline R_T = \overline R_T \hat \Pi^{a'}   \hat Q_1(r) \hat \Pi^a \overline R_T$, and $\hat Q_1(r)=T^{-1}\sum_{t=1}^{[T\lambda]} Z_t Z_t' \inp Q_1(r)$, \ur, respectively  $\hat{\Pi} =\Pi_T + \Op(T^{-1/2})$. It follows that $ \overline R_T \hat \Pi^a = \Pi^a+ \op(1)=\Op(1)$, independently of $r$. So
$
\hat A_1(r)  =\Op(1) \times \Op(1) \times \Op(1) = \Op(1) \, u.r.
$

\emph{Part (ii).} By a standard CLT, $T^{1/2}(\hat \Pi^{a} - \Pi_T^a)=\Op(1)$. Also, by the FCLT in Wooldridge and White (1988), Theorem 2.11, $ \Psi^{uv}_1(r) = T^{-1/2}\sum_{t=1}^{[Tr]} Z_t (u_t + v_t' \theta_x^0(t)) = \Op(1)$ u.r. Hence:
\begin{align*}
\xi_1(r) &= \Pi^{a'} \Psi^{uv}_1(r)-\Pi^{a'} \hat Q_1(r) \  [T^{1/2}(\hat \Pi - \Pi_T)] + \op(1)\\
& = \Op(1) + \Op(1) \times \Op(1) + \op(1) = \Op(1)\, u.r. \square
\end{align*}

We now return to the proof of the main result.

Assume throughout that $\hat T < T^0$. The proof for $\hat T \geq T^0$ is similar and omitted.

\emph{Consistency of $\hat \lambda$.} Let $\hat u_t \deff y_t-W_t'\hat \theta_1$ for $t \in \{1,\ldots, \hat T\}$, $\hat u_t \deff
y_t-W_t' \hat \theta_2$ otherwise, and $d_t \deff \hat u_t - \widetilde u_t$. As before, we show consistency by contradiction, in two steps.
By definition, $\sum_{t=1}^T \hat{u}_t^2 \leq \sum_{t=1}^T \widetilde u_t^2 $, hence $
2 \sum_{t=1}^T \widetilde u_t d_t +  \sum_{t=1}^T d_t^2 \leq 0.
$ In step 1, we show that:
\begin{equation}\label{contrad3}
T^{-1} \sum_{t=1}^T d_t^2 = \Op(1) \mbox{ and } T^{-1/2} \sum_{t=1}^T \widetilde u_t d_t = \Op(1),
\end{equation}
implying that $T^{-1} \sum_{t=1}^T d_t^2 =\Op(1)$ and $2 T^{-1} \sum_{t=1}^T \widetilde u_t d_t=\op(1)$.  So $\plim \left( T^{-1} \sum_{t=1}^T d_t^2 \right)\leq 0.$\footnote{Note that if $r_T = T^{1/2}$,
then $\sum_{t=1}^T d_t^2$ and $2 \sum_{t=1}^T \widetilde u_t d_t$ are always of the same order, and the proof argument cannot be applied.}
But because $T^{-1} \sum_{t=1}^T d_t^2 \geq 0$, $\plim \left(T^{-1} \sum_{t=1}^T d_t^2 \right)= 0$. In step 2, if $\hat{\lambda} \not \inp \lambda^0$, then with positive probability, $T^{-1} \sum_{t=1}^T d_t^2 > 0$,
contradicting $\plim \left(T^{-1}\sum_{t=1}^T d_t^2 \right) = 0$, so $\hat{\lambda} \inp \lambda^0$.

\textbf{Step 1.}
Start by noting:
\begin{align*}
d_t = \hat u_t - \widetilde u_t = \begin{cases} y_t- \hat W_t' \hat \theta_1 - y_t + \hat W_t' \theta^0(t) , &  t \leq \hat T \\
  y_t- \hat W_t' \hat \theta_2 - y_t + \hat W_t' \theta^0(t) , &  t > \hat T  \end{cases} =  \begin{cases} \hat W_t' (\theta_1^0 - \hat \theta_1), &  t \leq \hat T \\
   \hat W_t (\theta_1^0 - \hat \theta_2), &  \hat T +1 \leq  t \leq T^0 \\
    \hat W_t  (\theta_2^0 - \hat \theta_2), &  t > T^0  \end{cases}.
\end{align*}
It follows that, for $\xi_{\Delta} \deff \xi_1 (\lambda^0) - \xi_1(\hat \lambda)$,
\begin{align*}
\sum_{t=1}^T \widetilde u_t d_t & =  (\theta_1^0 - \hat \theta_1)' \Lambda_T \ \xi_1(\hat \lambda)   +  (\theta_1^0 - \hat \theta_2)' \Lambda_T \
\xi_{\Delta}  +  (\theta_2^0 - \hat \theta_2)' \Lambda_T  \ \xi_2(\hat \lambda).
\end{align*}
By Lemma \ref{lemma:1}, $\xi_1(\hat \lambda)  = \Op(1), \xi_{\Delta} = \Op(1), \xi_2(\hat \lambda) =\Op(1)$. Also by Lemma \ref{lemma:1}, letting $\hat A_{\Delta} \deff \hat A_1(\hat \lambda) - \hat A_1$,
\begin{align*}
\Lambda_T [\hat \theta_1 -\theta_1^0] &= \hat A_1^{-1} (\hat \lambda) \  \xi_1(\hat \lambda) = \Op(1) \\
\Lambda_T[\hat \theta_2 -\theta_2^0] & = \hat A_2^{-1} (\hat \lambda) \  \xi_2(\hat \lambda) + \hat A_1(\hat \lambda)\hat A_{\Delta} \left( \Lambda_T \theta_{\Delta} \right) = \Op(1) + \Op(T^{1/2} ) = \Op( T^{1/2}).
\end{align*}
Hence, $\Lambda_T \ [\hat \theta_2 -\theta_1^0] =\Lambda_T \ [\hat \theta_2 -\theta_2^0] + \Lambda_T \ [ \theta_2^0-\theta_1^0] = \Op (T^{1/2})$. Adding these together, $
\sum_{t=1}^T \widetilde u_t d_t = \Op (T^{1/2}).
$ Next, note that,
\begin{align}\label{parti7} \nonumber
\sum_{t=1}^T d_t^2 & = \sum_{t=1}^{\hat T} d_t^2 + \sum_{\hat T+1 }^{T^0} d_t^2  + \sum_{t= T^0+1 }^{T} d_t^2 \\ \nonumber
&= (\theta_1^0 - \hat \theta_1)' \Lambda_T \ \hat A_1(\hat \lambda) \ \Lambda_T (\theta_1^0 - \hat \theta_1)
+ (\theta_1^0 - \hat \theta_2)' \Lambda_T  \  \hat A_{\Delta} \ \Lambda_T (\theta_1^0 - \hat \theta_2)
\\
 &+ (\theta_2^0 - \hat \theta_2)' \Lambda_T\  \hat A_2(\hat \lambda) \ \Lambda_T (\theta_2^0 - \hat \theta_2) =  \Op(1)  + \Op(T)+ \Op(1) = \Op(T).
\end{align}

\textbf{Step 2.} If $\hat \lambda \not \inp \lambda^0$, then there exists $\eta \in (0,1)$, such that
with positive probability $\epsilon$, $T^0 - \hat T = [ T\lambda^0] - [T \hat \lambda]  \geq T \eta$. Then, it can be shown that with probability $\epsilon$,
\begin{align*}
T^{-1} \sum_{t=1}^T d_t^2 & \geq T^{-1}\left(  \sum_{t=T^0-T\eta+1}^{T^0} d_t^2\right) =
(\theta_1^0 - \hat \theta_2)' \overline R_T^{-1} [\hat A_1(\lambda^0)-\hat A_1(\eta)] \overline R_T^{-1} (\theta_1^0 - \hat \theta_2)\\
& = (\theta_{z,1}^0 - \hat \theta_{z,2})' [\hat A_{z,1}(\lambda^0)-\hat A_{z,2}(\eta)] (\theta_{z,1}^0 - \hat \theta_{z,2}) + \Op(r_T^{-1}) >C + \op(1),
\end{align*}
for some $C>0$, where $\theta_{z,i}^0, \hat \theta_{z,i}$ are the true and estimated parameters corresponding to the exogenous regressors, and $\hat A_{z,1}(\hat \lambda) = T^{-1} \sum_{t=1}^{[T\hat \lambda]} Z_{1t} Z_{1t}'$. Thus, with positive probability, $T^{-1} \sum_{t=1}^T d_t^2 >0$, reaching a contradiction and completing the consistency proof.

\emph{Rate of convergence of $\hat \lambda$.} From above, any change point estimator $\hat T=[T\hat \lambda]$ is such that $T^0-\hat T \leq \epsilon T$, for some chosen $\epsilon>0$. Assume that for chosen $C>0$, $T^0-\hat T > C$. Define $SSR_1$,
$SSR_2$ and $SSR_3$ as the 2SLS sum of squared residuals in the equation of interest, obtained with change points $\hat T$, $T^0$ and $(\hat T, T^0)$ respectively. It is sufficient to show that if $C < T^0-\hat T \leq \epsilon T$ for some large but fixed $C$ and small but fixed $\epsilon$, because then it follows that $\plim T^{-1} (SSR_1-SSR_2) >0$, which cannot hold by definition of the sum of squared residuals. So, $T^0-\hat T \leq C $, and by symmetry of the argument, if $\hat T\geq T^0$, $\hat T-T^0\leq C$, establishing the desired convergence rate for the change-fraction estimator.

We now show that $\plim  T^{-1} (SSR_1-SSR_2) >0$. Denote by $(\hat \theta_1, \hat \theta_2)$ the 2SLS estimators based on sample partition $(1,\hat T,T)$, $(\hat \theta_1, \hat \theta_{\Delta}, \widetilde \theta_2)$ the ones based on $(1, \hat T,T^0,T)$, and $(\widetilde \theta_1, \widetilde \theta_2)$ the ones based on $(1,T^0,T)$, all using the full-sample first-stage predictor $\hat W_t$. Also let the true and estimated parameters corresponding to $Z_{1t}$ be denoted  by subscript $z$, and, as before, the outer products of the exogenous regressors $Z_{1t}$ have subscript $z$ for $\hat A_{1}(\cdot), \hat A_2(\cdot), \hat A_{\Delta}(\cdot)$.  Then, it can be shown that:
\begin{align*}
SSR_1 - SSR_3 &= (\widetilde \theta_2 - \hat \theta_{\Delta})' \Lambda_T \hat A_{\Delta} \Lambda_T (\widetilde \theta_2 - \hat \theta_{\Delta}) - (\widetilde \theta_2 - \hat \theta_{\Delta})' \Lambda_T \ \hat A_{\Delta} \hat A_2^{-1}(\hat \lambda) \hat A_{\Delta} \Lambda_T (\widetilde \theta_2 - \hat \theta_{\Delta}) \\
& \deff N_1 - N_2 \\
SSR_2 - SSR_3 &= (\hat \theta_1 - \hat \theta_{\Delta})' \ \Lambda_T^{-1} \hat A_{\Delta} \ \Lambda_T^{-1} (\hat \theta_1 - \hat \theta_{\Delta})- (\hat \theta_1 - \hat \theta_{\Delta})'  \Lambda_T^{-1}\    \hat A_{\Delta} \hat A_1^{-1} (\hat \lambda)\hat A_{\Delta} \ \Lambda_T^{-1} (\hat \theta_1 - \hat \theta_{\Delta})\\
& \deff N_3 - N_4 \\
N_{z,1} &\deff (\widetilde \theta_{z,2} - \hat \theta_{z,\Delta})' T^{1/2} \hat A_{z,\Delta} T^{1/2} (\widetilde \theta_{z,2} - \hat \theta_{z,\Delta}) \\
N_{z,2} & \deff(\widetilde \theta_{z,2} - \hat \theta_{z,\Delta})'  \ T^{1/2} \ \hat A_{z,\Delta} \hat A_{z,2}^{-1}(\hat \lambda) \hat A_{z,\Delta} \  T^{1/2} (\widetilde \theta_{z,2} - \hat \theta_{z,\Delta}) \\
N_{z,3} &\deff (\hat \theta_{z,1} - \hat \theta_{z,\Delta})' \ T^{1/2}  \hat A_{z,\Delta} \  T^{1/2} (\hat \theta_{z,1} - \hat \theta_{z,\Delta}) \\
N_{z,4} &\deff (\hat \theta_{z,1} - \hat \theta_{z,\Delta})' \  T^{1/2}\    \hat A_{z,\Delta} \hat A_{z,1}^{-1} (\hat \lambda)\hat A_{z,\Delta} \ T^{1/2} (\hat \theta_{z,1} - \hat \theta_{z,\Delta}).
\end{align*}
It can be shown that $N_{z,i}$ are the dominant terms in $N_i$, and so we only analyze the order of $N_{z,i}$, in turn.
Since $\hat A_{z,\Delta}$ contains $T^0- \hat T \leq \epsilon T$ terms, $ \hat A_{\Delta} = \Op(\epsilon)$, while $ A_2(\hat \lambda) = \Op(1)$, by Lemma \ref{lemma:1}. It follows that $T^{-1} N_{z,1} >> T^{-1} N_{z,2}$ (meaning that $T^{-1} N_{z,1}$ is of a strictly larger probability order than $T^{-1} N_{z,2}$ when $\epsilon$ tends to zero, see also derivations below for $T^{-1} N_{z,1}$). Because $\widetilde \theta_{z,2}$ is estimating $\theta_{z,2}^0$ with observations only in the second regime, $[T^0+1,T]$, it can be shown that $\widetilde \theta_{z,2} -\theta_{z,2}^0= \Op(T^{-1/2} )$; on the other hand, $\hat \theta_{z,\Delta}$ is estimating $\theta_{z,1}^0$ in sub-sample $[\hat T+1, T^0]$, so for large enough C, it can be shown that $\hat \theta_{z,\Delta} -\theta_{z,1}^0 =\Op(T^{-1/2})$, hence $T^{1/2}(\widetilde \theta_{z,2} -\hat \theta_{z,\Delta}) =  T^{1/2} \theta_{z,\Delta}^0 + \Op(1)$, where recall $\theta_{z,\Delta}^0 = \theta_{z,1}^0-\theta_{z,2}^0$. Also, by Assumption \ref{ass:reg-IV}, for large enough $C$, $\hat A_{z,\Delta} \inp A^*$, a p.d. matrix of constants, and so:
\begin{equation}\label{eq:nz1}
N_{z,1}= T \theta_{z,\Delta}^{0'} A^* \theta_{z,\Delta}^0 + \op(T)
\end{equation}
for some $B>0$. For $N_{z,3}$, it can be shown that $\hat \theta_{z,1}-\theta_{z,1}^0 = \Op(T^{-1/2})$, and since $\hat \theta_{z,\Delta} -\theta_{z,1}^0 =\Op(T^{-1/2})$, $\hat \theta_{z,1} -\hat \theta_{z,\Delta} = \Op(T^{-1/2})$,  $N_{z,3}= \op(1)$. Similarly to $T^{-1} N_{z,1}>> T^{-1} N_{z,2}$, it can be shown that $T^{-1} N_{z,3} >> T^{-1} N_{z,4}$. It follows that $T^{-1} N_{z,1}>> T^{-1} N_{z,j}$, for $j=2,3,4$ for chosen small $\epsilon$ and large $C$, and also that $T^{-1} (SSR_1- SSR_2) = T^{-1} N_{z,1} +\op(1)$, hence
$
\plim T^{-1} (SSR_1- SSR_2) = (\theta_{z,\Delta}^0)' A^* \theta_{z,\Delta}^0 + \op(1) \geq B + \op(1),
$
for some $B>0$, by Assumption \ref{ass:reg-IV}.

(ii) \textbf{Asymptotic properties of parameter estimators evaluated at the estimated change point}:

\emph{Part (a). TS2SLS estimator $\hat \theta =\vec(\hat \theta_1, \hat \theta_2)=\hat \theta_{TS2SLS,vec}(\hat \lambda)$.} According to the additional notation before the proof of Theorem 1, $ \Lambda_T [\hat \theta_1 - \theta_1^0] = \hat A_1^{-1}(\hat \lambda) \xi_1(\hat \lambda)$, and
analyze first $\hat A_1^{-1}(\hat \lambda)$, with the knowledge that $\hat T - T^0 = \Op(1)$. Since $\hat A_{\Delta} = \Op[(\hat T-T^0)/T] = \Op(T^{-1}) = \op(1)$,
$
\hat A_1(\hat \lambda) =   \hat A_1(\lambda^0) - \hat A_{\Delta} = \hat A_1(\lambda^0) + \op(1).
$
Similarly, $\hat A_2(\hat \lambda) = \hat A_2(\lambda^0) + \op(1)$. Also, $\hat A_i(\lambda^0) \inp A_i$, where recall that  $A_i=\Pi^{a'} Q_i\Pi^a $ and $Q_i = Q_i (\lambda^0)$,   so:
\begin{align*}
\Lambda_T [\hat \theta_1 - \theta_1^0] &= A_1^{-1} \xi_1(\hat \lambda) + \op(1) = A_1^{-1} \Pi^{a'} \widetilde \Psi_1(\hat \lambda) + \op(1) \\
\Lambda_T [\hat \theta_2 - \theta_2^0] & =A_2^{-1} \xi_2(\hat \lambda) + \op(1) = A_2^{-1} \Pi^{a'} \widetilde \Psi_2(\hat \lambda) + \op(1),
\end{align*}
where  $\widetilde \Psi_1(\lambda) =  T^{-1/2} \sum_{t=1}^{[Tr]} Z_t \widetilde u_t$, $\widetilde \Psi_2(\lambda) =  T^{-1/2} \sum_{t=[Tr]+1}^{1} Z_t \widetilde u_t$, and $\widetilde \Psi_{\Delta} = T^{-1/2} \sum_{\hat T+1}^{T^0} Z_t \widetilde u_t$.
It remains to analyze the asymptotic distributions of $\widetilde \Psi_i (\hat \lambda)$. Note that $\widetilde \Psi_1 (\hat \lambda) = \widetilde \Psi_1 (\lambda^0)- \widetilde \Psi_{\Delta}$, and $\widetilde \Psi_{\Delta} = \Op(T^{-1/2})$ because it contains a finite number of terms. Thus, $\widetilde \Psi_i (\hat \lambda) = \widetilde \Psi_i(\lambda^0)+ \op(1)$ which shows that the asymptotic distribution of $\vec(\hat \theta_1, \hat \theta_2)$ is as if the change point was known.

\emph{Part (b): Split-sample GMM estimator: $\hat \theta =\vec(\hat \theta_1, \hat \theta_2) = \hat \theta_{GMM,vec}(\hat \lambda)$.} Define the partitions as in the Proof of Theorem 1 (ii) but at $\hat T$ instead of $T^0$: i.e. $\overline W = \diag(\overline W_1, \overline W_2)$, where $\overline W_1, \overline W_2$ are the first $\hat T\times q$, respectively the last $(T-\hat T) \times q$ blocks of $W$, and similarly for $\overline Z =\diag(\overline Z_1, \overline Z_2)$, $\overline y =\diag(\overline y_1, \overline y_2)$,  $\overline U=\diag(\overline U_1, \overline U_2)$.  Then, for the weighting matrices $\hat S_{u,i}^{-1} \inp S_{u,i}$,
\begin{align*}
\hat \theta_{GMM,i} &= ( \overline W_i' \overline Z_i \, \hat S_{u,i}^{-1}  \,  \overline Z_i' \overline W_i)^{-1} \,  \overline W_i' \overline Z_i \,  \hat S_{u,i}^{-1} \, \overline Z_i' \overline y_i\\
&= \theta_i^0 + ( \overline W_i' \overline Z_i \, \hat S_{u,i}^{-1}  \,  \overline Z_i' \overline W_i)^{-1} \,  \overline W_i' \overline Z_i \,  \hat S_{u,i}^{-1} \, \overline Z_i' \overline U_i.
\end{align*}
It can be shown that $T^{-1/2} \overline Z_1' \overline U_1 =   T^{-1/2} \overline Z_1^{0'} \overline U_1^0+ \op(1)$, because $\hat T - T^0 = \Op(1)$. Also, $\overline R_T T^{-1} \overline W_1' \overline Z_1 = \overline R_T T^{-1} \overline W_1^{0'} \overline Z_1^0+ \op(\overline R_T)$. It follows that $\hat \theta_{GMM,1}$ is asymptotically equivalent to the estimator using $T^0$ instead of $\hat T$, and similarly for $\hat \theta_{GMM,2}$, as well as for their joint distribution.

\emph{Part (c): Asymptotic normality of TSGMM.} Let $\breve \Lambda_T \deff \ \diag(I_2 \otimes \Lambda_T, T^{1/2} I_{qp2})$,  $\hat \beta  \deff \begin{bmatrix} \hat \theta_{TSGMM,vec} \\
\hat \Pi_{TSGMM,vec}  \end{bmatrix}$ and $\beta_T^0 \deff \begin{bmatrix} \theta_{vec}^0  \\
\Pi_{vec}^0/r_T \end{bmatrix}$.  Since this estimator is based on the same
quantities as $\hat \theta_{GMM,vec}$ and $\hat \theta_{TS2SLS,vec}$, by similar arguments, it can be shown that the asymptotic distribution of $\breve
\Lambda_T (\hat \beta- \beta_T^0)$ is as if the change point $T^0$ was known. $\blacksquare$

\textbf{$\bullet$ Proof of Theorem \ref{theo:wald}:}

\emph{Part (i). }Let $\theta^0$ be the common value of $(\theta_i^0)$ under the null hypothesis, for $i=1,2$, and so $\theta_x^0$ is the common value of $(\theta_{x,i}^0)$. The asymptotic distribution of the Wald test is determined by that of $\Lambda_T (\hat \theta_i (\lambda) - \theta^0) =
\hat A_i^{-1} (\lambda) \, \xi_i(\lambda)$. From Lemma \ref{lemma:1}, the proof of Theorem \ref{theo:change-order} and Assumption \ref{ass:reg-secondmom all candidates}, $\hat A_i (\lambda) \inp \lambda \Pi^{a'} Q \Pi^a.$ Also, under Assumptions \ref{ass:reg}-\ref{ass:reg-secondmom all candidates}, by the FCLT,
\begin{align} \nonumber
\xi_1(\lambda) & = \op(1) + \Pi^{a'} \Psi_1^{uv}(\lambda) -\Pi^{a'} [\hat Q_1(\lambda) \hat Q_1^{-1}(1)]  \Psi^{v'} \theta_x^0 \\ \label{eq:wald3}
& = \Pi^{a'} \Psi_1^{uv}(\lambda) - \Pi^{a'} \lambda \ \Psi^{v'} \theta_x^0+ \op(1) \Rightarrow \Pi^{a'} [P^{1/2} \mathcal B_q(\lambda) - \lambda (P^*)^{1/2} \mathcal B_q^*(1)],
\end{align}
where $(P^{*})^{1/2} \mathcal B^*_q(\lambda)$ and $P^{1/2} \mathcal B_q(\lambda)$ are two (dependent) $q \times 1$ Brownian motions generated by partial sums of $(Z_t v_t' \theta_x^0)$, respectively $[Z_t (u_t+v_t' \theta_x^0)]$, and for $a= \theta_x^0 \otimes I_q$,  $P = S_u +a'\  S_{uv}' +  S_{uv} a + a'  S_v a $, $P^* = a' S_v a$.
Let $G^* = [\Pi^{a'} Q \Pi^a]^{-1} \,  [\Pi^{a'} P^* \Pi^a] \,  [\Pi^{a'} Q  \Pi^a]^{-1}$, $G = [\Pi^{a'} Q  \Pi^a]^{-1} \, [\Pi^{a'} P  \Pi^a] \, [\Pi^{a'} Q  \Pi^a]^{-1} $, $G^{1/2} G^{1/2'} = G$, and
$(G^*)^{1/2} (G^*)^{1/2'} = G^*$, with $(G^*)^{1/2} =[\Pi' Q \Pi]^{-1} \,  \Pi^{a'} (P^*)^{1/2}$ and $G^{1/2} =  [\Pi^{a'} Q  \Pi^a]^{-1} \,  \Pi^{a'} P^{1/2}$. Then from \eqref{eq:wald3}, letting $\mathcal{R}_p^{\Lambda} = \Lambda_T \times [ (1,-1) \otimes I_p]  = (1,-1) \otimes \Lambda_T$,
\begin{align*}
\Lambda_T (\hat \theta_1 (\lambda) - \theta^0) &\Rightarrow  G^{1/2} \frac{\mathcal B_q(\lambda)}{\lambda} -  (G^{*})^{1/2}  \mathcal B_q^*(1) \\
\Lambda_T (\hat \theta_2 (\lambda) - \theta^0) & \Rightarrow  G^{1/2} \frac{\mathcal B_q(1) - \mathcal B_q(\lambda)}{1-\lambda} - (G^{*})^{1/2}  \mathcal B_q^* (1) \\
\mathcal{R}_p^{\Lambda}   \hat \theta_{vec} (\lambda) & =\Lambda_T (\hat \theta_1 (\lambda) - \theta^0) -                                                                                \Lambda_T (\hat \theta_2 (\lambda) - \theta^0) \\
&\Rightarrow G^{1/2} \frac{\mathcal B_q(\lambda)}{\lambda} -  G^{1/2}\frac{\mathcal B_q(1) - \mathcal B_q(\lambda)}{1-\lambda} = G^{1/2} \frac{\mathcal B_q(\lambda) - \lambda \mathcal B_q(1)}{\lambda(1-\lambda)}.
\end{align*}

Since $\Lambda_T^{-1} \hat H_{i}(\lambda) \Lambda_T^{-1} \inp \lambda_i \Pi^{a'} P \Pi^a$, for $\lambda_1=\lambda$, $\lambda_2=1-\lambda$,   $\Lambda_T^{-1} \hat G_{i} (\lambda)\Lambda_T^{-1} \inp \lambda_i^{-1} G$. Hence,
\begin{align*}
& \mathcal R_p^{\Lambda}  \hat G_{T} (\lambda) \mathcal R_p^{\Lambda'}  = \mathcal R_p \ \diag (\lambda^{-1} G, (1-\lambda)^{-1} G) \ \mathcal R_p' = \frac{G}{\lambda(1-\lambda)}.
\end{align*}
It can be shown that:
\begin{align*}
&  \hat{\theta}'(\lambda)\,  \mathcal R_p^\prime\,  [\, \mathcal R_p \, \hat G_{T} (\lambda) \, \mathcal R_p^\prime \, ]^{-1} \mathcal R_p \ \hat{\theta}(\lambda)   =  \hat{\theta}'(\lambda) \mathcal R_p^{\Lambda \prime}\,  [\, \mathcal R_p^{\Lambda} \, \hat G_{iT} (\lambda) \, \mathcal R_p^{\Lambda \prime} \, ]^{-1} \mathcal R_p^{\Lambda} \ \hat{\theta}(\lambda)  \\ &  \Rightarrow \frac {\mathcal{BB}_q^{'} (\lambda)\,[ G^{1/2'} G^{-1} G^{1/2}]\, \mathcal{BB}_q (\lambda)}{\lambda(1-\lambda)} = \frac{\mathcal{BB}_p^{'} (\lambda) \mathcal{BB}_p (\lambda) }{\lambda(1-\lambda)},
\end{align*}
because $G^{1/2'} (G)^{-1} G^{1/2}$ is a projection matrix of rank $p$, selecting only the first $p$ elements of $\mathcal{BB}_q(\lambda) = \mathcal{B}_g(\lambda) - \lambda \mathcal{B}_q(1)$ (for an extensive proof, see \cite{HHB}, Supplemental Appendix, pp. 23-27).

\emph{Part (ii).} The proof for $p_1=0$ can be derived in a similar fashion to $p_1\neq 0$ and is omitted for simplicity. We now write the proof for $p_1 \neq 0$, where $\theta_{z,\Delta}^0 \neq\ 0$.
If $\lambda \leq \lambda^0$, uniformly in $\lambda$ (u. $\lambda$),
\begin{align*}
\Lambda_T [\hat \theta_1(\lambda) -\theta_1^0] &= \hat A_1^{-1} (\lambda) \   \xi_1 (\lambda) = \Op(1) \\
\Lambda_T [\hat \theta_2(\lambda) -\theta_2^0 ] & = \hat A_2^{-1} (\lambda) \ \xi_2(\lambda) + \hat A_2^{-1}(\lambda) \hat A_{\Delta} \ \Lambda_T\ \theta_{\Delta}^0 \\& = \Op(1)+ A_2^{-1}(\lambda)\ [ A_2(\lambda)-A_2] \ \Lambda_T\ \theta_{\Delta}^0  = \Op(1) + \{ I - A_2^{-1}(\lambda) A_2 \} \Lambda_T\ \theta_{\Delta}^0
\end{align*}
It follows that u. $\lambda$,
$
\mathcal R_p^{\Lambda}  \hat \theta_{vec}(\lambda) = \Lambda_T  \hat \theta_1(\lambda) - \Lambda_T \hat \theta_2(\lambda) =\Op(1) +  [A_2^{-1}(\lambda) A_2] \ \Lambda_T \  \theta_{\Delta}^0.
$
Since $\mathcal {R}_p^{\Lambda} \hat   G (\lambda) \mathcal R_p^{\Lambda'} = \Lambda_T \hat G_{1} (\lambda)\Lambda_T + \Lambda_T \hat G_{2}(\lambda) \Lambda_T  \inp G_1(\lambda) + G_2(\lambda) \deff G(\lambda)$,
\begin{align*}
Wald_T(\lambda) & = O_p(1) + \theta_{\Delta}^{0'} \Lambda_T \left  \{ A_2 A_2^{-1}(\lambda) G^{-1}(\lambda) A_2^{-1}(\lambda) A_2 \right \} \Lambda_T \theta_{\Delta} \\
& = \Op(r_T) + T \theta_{z,\Delta}^{0'} \left  \{ A_2 A_2^{-1}(\lambda) G^{-1}(\lambda) A_2^{-1}(\lambda) A_2 \right \}_z \theta_{z,\Delta}^0
\end{align*}
 where $\{ \cdot \}_z$ selects the first $p_1 \times p_1$ elements of the matrix, and the latter term dominates because it is $\Op(T)$, so u. $\lambda \leq \lambda^0$,
\begin{align*}
T^{-1} Wald_T(\lambda) \inp \theta_{z,\Delta}^{0'} \left  \{ A_2 A_2^{-1}(\lambda) G^{-1}(\lambda) A_2^{-1}(\lambda) A_2 \right \}_z \theta_{z,\Delta}^0.
\end{align*}
Similarly, it can be shown that u. $\lambda > \lambda^0$, \begin{align*}
T^{-1} Wald_T(\lambda) \inp \theta_{z,\Delta}^{0'} \left  \{ A_1 A_1^{-1}(\lambda) G^{-1}(\lambda) A_1^{-1}(\lambda) A_1 \right \}_z \theta_{z,\Delta}^0.
\end{align*}
Under Assumption \ref{ass:reg-secondmom all candidates}, $A_{i} = \lambda_i A$, $H_{i} = \lambda_i \Pi'(S_{u} + a_i' S_{uv} + S_{uv}' a_i  + a_i' S_{v} a_i)\Pi \deff \lambda_i \mathbf H_i$, and so $G^{-1} (\lambda) = \lambda (1-\lambda)\ A  \ (\lambda \mathbf H_2 + (1-\lambda) \mathbf H_1)^{-1}  A$,
  \begin{align*}
T^{-1} Wald_T(\lambda) \inp  \begin{cases} (1-\lambda^0)^2 \ \frac{\lambda}{1-\lambda} \ \theta_{z,\Delta}^{0'}A \ \{[(1-\lambda) \mathbf H_1 + \lambda \mathbf H_2]^{-1}\}_z \ A \ \theta_{z,\Delta}^0& \lambda \leq \lambda^0 \\
(\lambda^{0})^2  \ \frac{1-\lambda}{\lambda}  \  \theta_{z,\Delta}^{0'}  A \ \{[(1-\lambda) \mathbf H_1 + \lambda \mathbf H_2]^{-1}\}_z \ A \ \theta_{z,\Delta}^0 & \lambda > \lambda^0.
\end{cases}
\end{align*}
If $\theta_{x,1}^0=\theta_{x,2}^0$, then $H_1=H_2 \deff H$, and so:
\begin{align*}
T^{-1} Wald_T(\lambda) \inp  \begin{cases} (1-\lambda^0)^2 \ \frac{\lambda}{1-\lambda} \ \theta_{z,\Delta}^{0'}A \ \{H^{-1}\}_z \ A \ \theta_{z,\Delta}^0& \lambda \leq \lambda^0 \\
(\lambda^{0})^2  \ \frac{1-\lambda}{\lambda}  \  \theta_{z,\Delta}^{0'}  A \ \{H^{-1}\}_z \ A \ \theta_{z,\Delta}^0 & \lambda > \lambda^0.
\end{cases}
\end{align*}

It follows that under the Assumption \ref{ass:reg-secondmom all candidates} and no change in the endogenous parameters, $Wald_T(\lambda)$ is asymptotically maximized at $\lambda^0$, thus, by continuous mapping theorem, $\hat \lambda^W \inp \lambda^0$. However, if any of these conditions are violated, it's not clear that $\hat \lambda^W \inp \lambda^0$. If Assumption \ref{ass:reg-secondmom all candidates} doesn't hold, and for some $\lambda^* < \lambda^0$, $Q_2(\lambda)=e \  Q_1(\lambda)$, for a scalar $e$, then it can be shown that for $0<e<1$, $Wald_T(\lambda^*)> Wald_T(\lambda^0) + \op(1)$, so $\hat \lambda^W \not \inp \lambda^0. \blacksquare$

\textbf{$\bullet$ Proof of Theorem \ref{theo:common}:}

\emph{Part (i).} Let $\theta^0$ be the common value of $\theta_i^0$ under the null hypothesis. By arguments similar to the Proof of Theorem \ref{theo:change-order}, the distribution of the split-sample 2SLS estimators $\hat \theta_i^c$ is as if the change point $\nu^0$ was known, so
$
\Lambda_{iT} [\hat \theta_i^c -\theta^0] \ind \mathcal{N}(0, G_i^{c}),
$
where recall that $G_i^c \deff (A_i^c)^{-1} \, B_i^c \, (A_i^c)^{-1}$, $A_i^c \deff \Pi_i^{a'} Q_i (\nu^0) \Pi_i^a$, and $B_i^c \deff \ \Pi_i^{a'} S_{u,i}(\nu^0) \Pi_i^a$, and $\Lambda_{iT} = T^{1/2} \diag(I_{p_1}, r_{iT}^{-1} I_{p_2})$. Moreover, $\hat \theta_1^c \perp \hat \theta_2^c$ asymptotically, because they are constructed with asymptotically independent sub-samples in the first-stage. Also, by construction, $\Lambda_{iT}^{-1}  \hat B_{i}^c \Lambda_{iT}^{-1} \inp B_i^c$, and since $\hat A_i^c \deff \Lambda_{iT}^{-1} \left(\sum_{i} \hat W_t \hat W_t' \right) \Lambda_{iT}^{-1} \inp A_i$, it follows that  $\Lambda_{iT} \ \hat G^c_{i} \ \Lambda_{iT} \inp (A_i^c)^{-1} B_i^c (A_i^c)^{-1} \deff G_i^c$.

Case (a): $r_T=r_{1T}= r_{2T}$. Then $\Lambda_{1T}= \Lambda_{2T} \deff \Lambda_T$, and
\begin{align*}
&  \Lambda_T \ \mathcal{R}_p \, \hat \theta_{vec}^c \ind \mathcal{N}(0, G_1^c + G_2^c)\\
&  \Lambda_T  \ \mathcal{R}_p  \hat G^c \mathcal{R}_p' \ \Lambda_T  \inp G_1^c + G_2^c \\
 & Wald_T^c =   [ \Lambda_T  \mathcal{R}_p \, \hat \theta_{vec}^c] ' \  [\Lambda_T  \ \mathcal{R}_p  \hat G^c \mathcal{R}_p' \ \Lambda_T ]^{-1} \  [ \Lambda_T \mathcal{R}_p \, \hat \theta_{vec}^c] \inp \chi^2_p
\end{align*}

Case (b): Wlog, $r_T = r_{1T}< r_{2T}$. Then:
 \begin{align*}
& \Lambda_{2T}\ \mathcal{R}_p \, \hat \theta_{vec}^c  =[ \Lambda_{2T} \Lambda_{1T}^{-1}] \ \Lambda_{1T} (\hat \theta_1^c- \theta^0) -  \Lambda_{2T} \ (\hat \theta_2^c- \theta^0) \\
& = \begin{bmatrix}
T^{1/2} (\hat \theta_{z,1}^c - \theta_{z}^0) \\
 [r_{2T}^{-1} r_{1T}] \  T^{1/2} r_{1T}^{-1} (\hat \theta_{y,1}-\theta_{y}^0) \end{bmatrix} - \begin{bmatrix}
T^{1/2}  (\hat \theta_{z,2} - \theta_{z}^0)  \\
 T^{1/2} r_{2T}^{-1} (\hat \theta_{y,2}-\theta_{y}^0)
\end{bmatrix}= \begin{bmatrix}
T^{1/2} (\hat \theta_{z,1} - \theta_{z}^0) \\
\op(1)
\end{bmatrix} - \begin{bmatrix}
 T^{1/2}  (\hat \theta_{z,2} - \theta_{z}^0) \\
 T^{1/2} r_{2T}^{-1} (\hat \theta_{y,2}-\theta_{y}^0)
\end{bmatrix}\\
& \ind \mathcal{N}(0,\mbox{ diag }(\{G_1^c\}_z, O_{p_2}) + G_2^c) = \mathcal{N}(0, G_{1}^{z,c} + G_2^c) \\
& \Lambda_{2T} \  \mathcal{R}_p  \ \hat G^c \ \mathcal{R}_p'\Lambda_{2T} = \Lambda_{2T} \ [\hat G^c_{1} + \hat G^c_{2}] \  \Lambda_{2T}  = [\Lambda_{2T} \Lambda_{1T}^{-1}] \ [ \Lambda_{1T} \ \hat G^c_{1} \ \Lambda_{1T}] \  [\Lambda_{1T}^{-1} \Lambda_{2T}]   \\
&+ \Lambda_{2T}\ \hat G^c_{2} \  \Lambda_{2T} \inp G_1^{z,c} + G_2^c \\
&Wald_T^c = [ \Lambda_{2T}\ \mathcal{R}_p \, \hat \theta_{vec}]' \ [\Lambda_{2T} \  \mathcal{R}_p  \  \hat G^c \  \mathcal{R}_p'\Lambda_{2T}]^{-1} \ [ \Lambda_{2T}\ \mathcal{R}_p \,  \hat \theta_{vec}] \inp \chi^2_p.
\end{align*}

\emph{Part (ii).} Assume again for simplicity that $p_1 \neq 0$; for $p_1=0$ the proof follows along the same lines.

Case (a): $r_T=r_{1T}= r_{2T}$. Then $\Lambda_{1T}=\Lambda_{2T} = \Lambda_T$, and
$
\Lambda_T \mathcal R_p \ \hat \theta^c_{vec} =  \Lambda_T [\hat \theta^c_1 - \hat\theta^c_2] =\Op(1) +   \ \Lambda_T \theta_{\Delta}^0.
$
Also,  $\Lambda_T \mathcal {R}_p \ \hat G^c \  \mathcal R_p' \Lambda_T \inp G^c$, hence $ \  Wald_T^c = \theta_{\Delta}^{0'} \Lambda_T (G^c)^{-1} \Lambda_T \theta_{\Delta}^0 + \op(T)$, so $T^{-1} Wald_T^{c}  \inp \theta_{z,\Delta}^{0'} \left\{(G^c)^{-1}\right\}_z \theta_{z, \Delta}^0$.

Case (b): Wlog, $ r_T=r_{1T} < r_{2T}$. Then:
 \begin{align*}
& \Lambda_{2T} \ \mathcal{R}_p \, \hat  \theta_{vec}^c  =[ \Lambda_{2T} \  \Lambda_{1T}^{-1}] \  \Lambda_{1T} (\hat \theta_1^c- \theta_1^0) -  \Lambda_{2T} \ (\hat \theta_2^c- \theta_2^0) + \Lambda_{2T} \theta_{\Delta}^0 \\
 &= \begin{bmatrix} T^{1/2} \theta_{z,\Delta}^0 +\Op(1)\\
\op(T^{1/2})
\end{bmatrix} \\
& \Lambda_T \hat G^c \Lambda_T\inp \mbox{diag}[G_{1}^{z,c}, G_2^c] \  \Leftrightarrow  \Lambda_T \mathcal {R}_p \hat G^c  \mathcal R_p' \Lambda_T \inp G_1^{z,c}+ G_2^c,
\end{align*}
so $T^{-1}\  Wald_T^c \inp  \theta_{z,\Delta}^{0'} \ \left\{(G_{1}^{z,c}+ G_2^c)^{-1} \right\}_z\ \theta_{z,\Delta}^0. \blacksquare$

\vspace{0.1in}

\textbf{$\bullet$ Proof of Theorem \ref{theo:eff-ls}: Efficiency of first-stage estimators}

\textit{Part (i). } The distribution of $\hat \Pi_{OLS}$ is obtained by usual OLS asymptotics; for $\hat \Pi_{TSGMM}$ with the optimal weighting matrix $(\hat S_v^a)^{-1} \inp \diag(S_{v,1}, S_{v,2})$, we have:
\begin{align*}
T^{1/2}(\hat \Pi_{TSGMM}-\Pi) & = [(T^{-1} W'\overline W^0) (\hat S_v^a)^{-1} (T^{-1} \overline W^{0'} W)]^{-1}(T^{-1} W'\overline W^0) (\hat S_v^a)^{-1} T^{-1/2} \overline W^0 v \\
T^{-1/2} \overline W^0 v & \ind \mathcal N(0, \diag(S_{v,1}, S_{v,2})), \  T^{-1} \overline W^{0'} W \inp \vec(Q_1, Q_2), \\
 \Rightarrow  T^{1/2}(\hat \Pi_{GMM}-\Pi) &\ind \mathcal N(0, (Q_1 S_{v,1}^{-1} Q_1 + Q_2 S_{v,2}^{-1} Q_2)^{-1} ).
\end{align*}
The distribution of $\hat \Pi_{TSGMM}$ follows from Theorem \ref{theo:consistent}, by similar arguments as for $V_{TSGMM,vec}$.

\textit{Part (ii). } $V_{OLS, \Pi} = (Q_1+Q_2)^{-1} (S_{v,1}+S_{v,2}) (Q_1+Q_2)^{-1}$ and $V_{GMM,\Pi} = (Q_1 S_{v,1}^{-1} Q_1 +Q_2 S_{v,2}^{-1} Q_2)^{-1}$. Using similar arguments as in Antoine and Boldea (2018), Proof of Theorem 2, but replacing $\mu_i$ with $Q_i a$, for any $q \times 1$ vector $a$, it follows that $V_{GMM,\Pi} \leq V_{OLS, \Pi}$, with equality iff $S_{v,1}^{-1} Q_1 a = S_{v,2}^{-1} Q_2 a$ for all $a$. Similarly to Theorem \ref{theo:eff}(i),
$V_{TSGMM,\Pi} \leq V_{GMM,\Pi}$, because:
\begin{align*}
\mathcal H_* &= \mathcal S_{1,2} \mathcal S_{2,2}^{-1} \Gamma_2 = - \vec(S_{uv,i}' S_{v,i}^{-1} Q_i) \deff - \vec( \mathbf \Gamma_{i*}),\\
 \mathcal E_* & = \diag( S_{u,i} -S_{uv,i}' S_{v,i}^{-1} S_{uv,i}) \deff \diag( \mathcal E_{i*}^{-1/2}).
\end{align*}
Hence, $\mathcal E_{i*}, \mathbf \Gamma_{i*}$ play the role of $\mathcal E_i, \mathbf \Gamma_i$ in the proof of Theorem \ref{theo:eff}(i), and so $V_{TSGMM,\Pi} \leq V_{GMM,\Pi}$. \\
 \textit{Part (iii). } From the above, it follows that even under Assumptions \ref{ass:reg-secondmom}-\ref{ass:cond-hom}, $V_{TSGMM,\Pi} \leq V_{GMM,\Pi}$. However, from (ii), $V_{GMM,\Pi}=V_{OLS,\Pi}$ iff $ S_{v,1}^{-1} Q_1 a = S_{v,2}^{-1} Q_2 a$ for all vectors $a$. Under Assumption \ref{ass:reg-secondmom}, $S_{v,i}^{-1} Q_i = (\lambda_i^0 S_v)^{-1} \lambda_i^0 Q = S_v^{-1} Q$, so $V_{GMM,\Pi}=V_{OLS,\Pi}$. Also, under Assumption \ref{ass:cond-hom}, $S_{v,i} = \Phi_v Q_i$, so $ S_{v,i}^{-1} Q_i = \Phi_v I_q$. $\blacksquare$.

\newpage

\appendix

\setcounter{page}{1}
\setcounter{table}{0}

\begin{center}
	\LARGE \textbf{Supplementary Appendix} \\
\end{center}

\normalsize

\hspace{.5cm}

In this Supplementary appendix, we report additional empirical results associated with the estimation of the NKPC model highlighted in Section \ref{section:NKPC} of the main paper.

\section{Additional results for the Benchmark model}

\begin{table}[h]
	\centering
	\scalebox{.9}{
		\begin{tabular}{l|ccc|ccc||c|}
			\cline{2-8}
			&    \multicolumn{3}{c|}{2001m11-2006m8} & \multicolumn{3}{c||}{2006m9-2009m3} & 2001m11-2009m3\\ \cline{2-8}
			& TS2SLS  & GMM & TSGMM & TS2SLS & GMM & TSGMM & GMM\\
			& (1) & (2) & (3) & (4) & (5) & (6) & (7) \\\hline
			$\beta_0$ &  0.0514		 &0.0545	 &	0.0495	 &	-0.0053	 &	-0.0685	 &	-0.0454	 &	-0.0230\\
			&(0.0577)	 &	(0.0453)	 &	(0.0412)	 &	(0.4383)		 &(0.1601)	 &	(0.1455)	 &	(0.0420)\\

			$\gamma_1$ &0.8883**	 &0.8359*** &	0.8436***	 &1.1517*** &	1.0462***	 &1.0110*** &	1.0568***\\
			&(0.4037)	 &	(0.1366)	 &	(0.1217)	 &	(0.5065)		 &(0.1635)	 &	(0.1508)		 &(0.0805)\\
			
			$\gamma_2$ &-0.6068* &	-0.5899*** &	-0.5552***	 &-0.5973	 &	-0.6304*** &	-0.6430*** &	-0.4347***\\
			&(0.3570)	 &	(0.1277)	 &	(0.1064)		 &(1.1428)	 &	(0.1393)	 &	(0.1354)	 &	(0.0897)\\\
			
			$\kappa_\Psi$ &-0.0700	 &	-0.0597	 &	-0.0526	 &	0.0248		 &0.1512	 &	0.1665\textcolor{blue}{*} &	0.0448\textcolor{blue}{**}\\
			&(0.1144)	 &	(0.0484)	 &	(0.0444)	 &	(0.5529)	 &	(0.1259)		 &(0.1188)	 &	(0.0248)\\
			
			$\gamma_f$ &-0.2514	 &	-0.2500	 &	-0.2254	 &	0.5099	 &	0.2415	 &	0.2813	 &	0.1620\\
			&(0.4843)	 &	(0.2473)		 &(0.2362)	 &	(0.8966)	 &	(0.2614)	 &	(0.2560)		 &(0.1365)\\
			$\kappa_u$   &-0.0284	 &	-0.0276\textcolor{red}{*} &	-0.0255\textcolor{red}{*} &	0.0234	 &	0.0353	 &	0.0523		 &-0.0113\\
			&(0.0369)	 &	(0.0200)	 &	(0.0182)	 &	(0.1211)	 &	(0.1252)	 &	(0.1152)	 &	(0.0253)\\
			\hline
			$N$                 &     & $58$ &   &   & $31$ &  & $89$\\\hline
			
		\end{tabular}
	}
	\caption{*$p < 0.10$, **$p < 0.05$, ***$p < 0.01$. Black stars indicate bi-directional testing. Blue stars right-tailed (positive) tests, and red stars (negative) left-tailed tests.}
	\label{tab:2001-2009}
\end{table}

\newpage

\begin{table}[h]
	\centering
	\scalebox{.9}{
		\begin{tabular}{l|ccc|}
			\cline{2-4}
			&    \multicolumn{3}{c|}{2009m4-2020m4} \\ \cline{2-4}
			& TS2SLS  & GMM & TSGMM \\
			& (1) & (2) & (3)  \\\hline
			$\beta_0$     & 0.0017          & -0.0068         & -0.0068         \\
			& (0.0265)        & (0.0263)        & (0.0262)        \\
			$\gamma_1$    & 1.2495***  & 1.2711***  & 1.2711 ***         \\
			& (0.0741)        & (0.0733)        & (0.0736)        \\
			$\gamma_2$    & -0.4993*** & -0.5514*** & -0.5514***         \\
			& (0.0863)        & (0.0797)        & (0.0825)        \\
			$\kappa_\Psi$  & -0.0103         & -0.0125         & -0.0125         \\
			& (0.0159)        & (0.0161)        & (0.0159)        \\
			$\gamma_f$    & 0.0175          & -0.0401         & -0.0401         \\
			& (0.1957)        & (0.1883)        & (0.1931)        \\
			$\kappa_u$    & -0.0002         & -0.0008         & -0.0008         \\
			& (0.0020)        & (0.0019)        & (0.0019)        \\
			$N$                &      & $129$    & \\\hline
			
		\end{tabular}
	}
	\caption{*$p < 0.10$, **$p < 0.05$, ***$p < 0.01$. Black stars indicate bi-directional testing. Blue stars right-tailed (positive) tests, and red stars (negative) left-tailed tests.}
	\label{tab:2009-2020}
\end{table}
\end{document}